\documentclass[%
 reprint,
 amsmath,amssymb,
 aps,
]{revtex4-2}

\usepackage{bm}
\usepackage{svg}
\usepackage{comment}
\usepackage{upgreek}
\usepackage{graphicx}
\usepackage{dcolumn}
\usepackage{braket}
\usepackage{bm}
\usepackage{amsmath}
\usepackage[font=small,labelfont=bf]{caption}

\usepackage{hyperref}

\begin{document}

\preprint{APS/123-QED}

\title[Article Title]{A diamond nanophotonic interface with an optically accessible \\ deterministic electronuclear spin register}


\author{Ryan A. Parker$^{1,*}$}
\author{Jes\'{u}s Arjona Mart\'{i}nez$^{1,*}$}
\author{Kevin C. Chen$^{2}$}
\author{Alexander M. Stramma$^{1}$}
\author{Isaac B. Harris$^{2}$}
\author{Cathryn P. Michaels$^{1}$}
\author{Matthew E. Trusheim$^{2}$}
\author{Martin Hayhurst Appel$^{1}$}
\author{Carola M. Purser$^{1}$}
\author{William G. Roth$^{1}$}
\author{Dirk Englund$^{2,\dagger}$}
\author{Mete Atatüre$^{1,\dagger}$}
\affiliation{%
 ${}^{1}$Cavendish Laboratory, University of Cambridge, JJ Thomson Avenue, Cambridge CB3 0HE, United Kingdom
}%
\affiliation{%
 ${}^{2}$Department of Electrical Engineering and Computer Science, Massachusetts Institute of Technology, Cambridge, Massachusetts 02139, USA
}

\date{\today}

\begin{abstract}

A contemporary challenge for the scalability of quantum networks is developing quantum nodes with simultaneous high photonic efficiency and long-lived qubits. Here, we present a fibre-packaged nanophotonic diamond waveguide hosting a tin-vacancy centre with a spin-1/2 $^{117}$Sn nucleus. The  interaction between the electronic and  nuclear spins results in a signature $452(7)$ MHz hyperfine splitting. This exceeds the natural optical linewidth by a factor of 16, enabling direct optical nuclear-spin initialisation with 98.6(3)\% fidelity and single-shot readout with 80(1)\% fidelity. The waveguide-to-fibre extraction efficiency of our device of 57(6)\% enables the practical detection of 5-photon events. Combining the photonic performance with the optically initialised nuclear spin, we demonstrate a spin-gated single-photon nonlinearity with 11(1)\% contrast in the absence of an external magnetic field. These capabilities position our nanophotonic interface as a versatile quantum node in the pursuit of scalable quantum networks.

\end{abstract}

\maketitle

Scalable quantum networking motivates the search for a platform with simultaneous access to a high-quality high-efficiency optical interface and to long-lived memory qubits \cite{kimble2008quantum, wehner2018quantuminternet}. Commonly, the choice of the material platform presents a trade-off between these independent requirements. Self-assembled quantum dots stand out for their photonic quality and integration maturity \cite{tomm2021brightsource, uppu2020scalable, papon2022independent}, which has enabled demonstrations of near-deterministic spin-photon interactions \cite{javadi2018spin, appel2021spinphoton}, remote entanglement \cite{delteil2016generation, stockill2017phase} and cluster-state generation \cite{istrati2020clusterstates, cogan2023clusterstate}. The intrinsically limited spin coherence \cite{zaporski2023idealrefocusing} and the lack of an easily accessible long-lived register \cite{denning2019magnon} has hindered their potential for long-distance quantum networking. This is in stark contrast to the nitrogen vacancy centre in diamond, which sustains an impressive degree of spin coherence \cite{bar2013solid, balasubramanian2009ultralong} and provides access to long-lived nuclear memories \cite{bradley2019tenqubit}. However, spectral instability and low-rate coherent-photon emission \cite{johnson2015tunable} remains an obstacle for scaling up quantum networks \cite{pompili2021realization, hermans2022qubit}.

The negatively charged group-IV colour centres in diamond (SiV, GeV, SnV) span this gap, as they possess both high-quality optical properties \cite{knall2022efficientsinglephotons, bhaskar2017quantum, martinez2022photonic} and long spin-coherence times \cite{sukachev2017silicon, debroux2021quantum}. Long-lived nuclear memories can be realised through hyperfine coupling to stochastically present $^{13}\text{C}$ \cite{nguyen2019integrated}. Alternatively, the nuclear host spin may be employed, as demonstrated for SiV \cite{stas2022sivnuclear}, and has the advantage of a stronger and deterministic coupling. 

The recent success of group-IV colour centres lies in their ease of integration into photonic waveguides \cite{bhaskar2017quantum, rugar2020narrow, bradac2019quantum, koch2022super, martinez2022photonic} and nanocavities \cite{nguyen2019integrated, evans2018photon, sipahigil2016integrated, rugar2020narrow}. This is attributed to their centro-symmetry and therefore insensitivity to surface electrical noise. Building upon the pioneering role of SiV, the first centre to be investigated in this category, the SnV centre stands out, as it possesses an intrinsically higher quantum efficiency (80\% vs. 10\% for SiV) \cite{iwasaki2017tin, sukachev2017silicon}, robustness of the optical-cyclicity against an applied external magnetic field \cite{debroux2021quantum}, and intrinsically long coherence times above 1 K \cite{trusheim2020transformlimited}. In particular, the high quenum efficiency and large Debye-Waller factor relaxes the requirement of large Purcell enhancement to ameliorate the imperfect photonic efficiency and enables a waveguide-based route for a high-efficiency spin-photon interface. For a sufficiently well-coupled spin, a waveguide can also serve as a photonic gate within an all optical quantum circuit \cite{le2022dynamical}. Relative to nanocavities, the waveguide architecture offers a broadband platform with superior scalability \cite{wan2020large}, reduced fabrication complexity and control overhead \cite{douglas2015quantum}, and improved outcoupling to optical fibres \cite{knall2022efficientsinglephotons}. For scalable quantum communication, the remaining challenges for the SnV centre are to demonstrate integration into such a high photonic-efficiency platform and to establish access to long-lived nuclear memories.

\begin{figure*}
    \centering
    \includegraphics[width=1.0\textwidth]{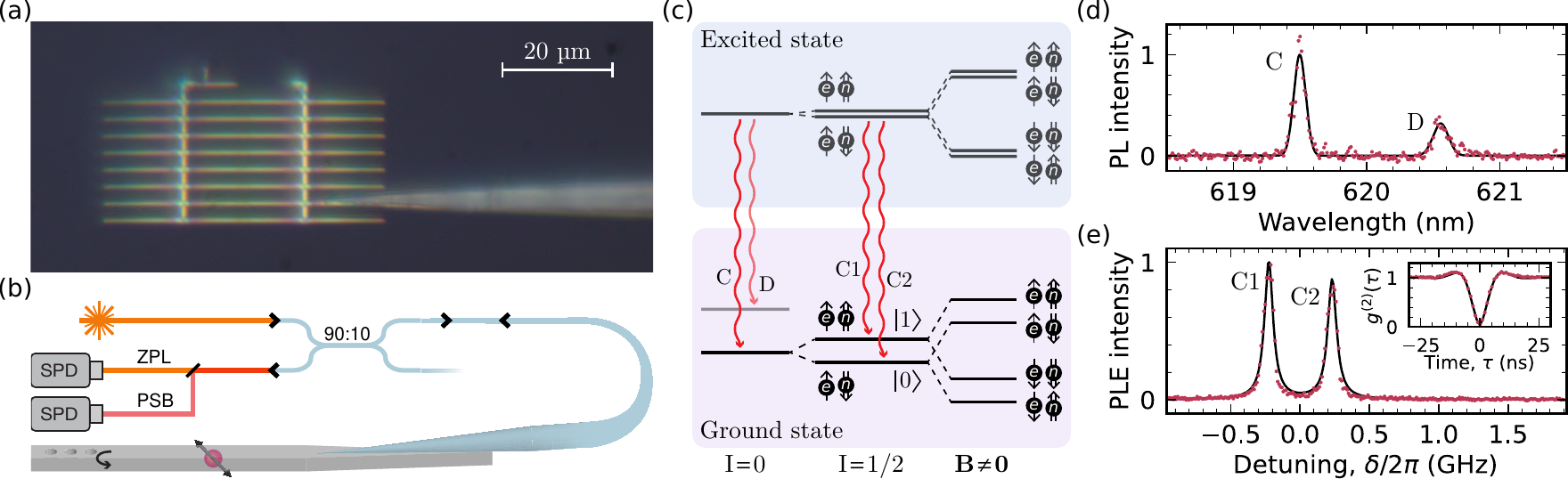}
    \caption{ \textbf{A nanophotonic quantum device hosting an electronuclear spin register.}
        (a) Microscope photograph of the packaged device. A UV-cured optical adhesive permanently fixes the tapered fibre after it is contacted onto a diamond microchiplet. (b) Experimental setup. We excite the emitter near resonance and collect both the phonon-sideband fluorescence (PSB) and the resonant reflection (ZPL) into single-photon detectors (SPD). (c) Electronic structure of the SnV. The hyperfine coupling between the electron and $^{117}\text{Sn}$ nucleus results in a splitting of the optical transition at zero magnetic field. A non-zero magnetic field further splits the two electronic states. (d) Photoluminescence (PL) under 130 nW of 532 nm off-resonant excitation. An inhomogeneous distribution FWHM of 90(2) GHz is extracted from Gaussian fits (solid line). (e) Photoluminescence under resonant excitation (PLE) at $\bm{B}=\bm{0}$. A homogeneous linewidth FWHM of 70(5) MHz is extracted from Lorentzian fits (solid line). Inset: second-order resonant autocorrelation function with a Bloch equation-based fit \cite{kim2020analytical} (solid line).
    }
    \label{fig:1}
\end{figure*}

In this work, we present a fibre-packaged nanophotonic waveguide device with efficient optical coupling and direct optical access to a deterministically present nuclear spin. Isotopically filtered $^{117}$Sn implantation results in $^{117}$SnV centres with optically addressable hyperfine transitions separated by 452(7) MHz. As this is 16 times greater than the optical linewidth, it enables direct optical access to the hyperfine-split states and therefore initialisation and single-shot readout of the nuclear spin. The nuclear spin can thus serve as an ancillary memory to the electron or a stand-alone qubit. Interfacing the waveguide with an adiabatically tapered optical fibre results in a waveguide-to-fibre extraction efficiency of at least 57(6)\%, enabling practical detection of up to five consecutive photons which will enable multidimensional photonic cluster state generation \cite{michaels2021multidimensional}. We further show a single-photon nonlinearity conditional on the electronuclear state, which we employ as a spin-gated photonic switch. The realised integration of the $^{117}$SnV centre into a fibre-packaged and high-efficiency photonic device, combined with the demonstrated coherent spin control \cite{debroux2021quantum} and the intrinsically high single-photon quality \cite{martinez2022photonic}, realises a scalable solid-state platform to advance the pursuit of quantum networks.

\section*{Results}
\subsection*{A fibre-packaged \\ nanophotonic device}

Our quantum device is a diamond microchiplet coupled to a tapered optical fibre, as shown in Fig. \ref{fig:1}(a). The quantum microchiplets are formed through isotopically selective ion implantation and subsequent top-down nanofabrication with the incorporation of a Bragg reflector in the centre of the waveguide to reflect all light towards a single adiabatically tapered port \cite{mouradian2017rectangular} (SI I). A conically tapered fibre is adhered to the adiabatically tapered port of the waveguides to form a packaged device (see Methods). Figure \ref{fig:1}(b) illustrates this setup and the subsequent collection optics. A single, fibre-based, input and output port is employed for all measurements in this work, with no tuning requirements owing to the broadband nature of the platform. A 90:10 fibre-based beamsplitter separates the two counter-propagating modes. From reflection measurements, we extract a waveguide to fibre coupling efficiency greater than 57(6)\% at 620 nm (SI II), with no degradation of the coupling efficiency when the device is handled or cooled to cryogenic temperatures.

Figure \ref{fig:1}(c) highlights the relevant optical transitions of the SnV centre. In the absence of magnetic field, the ground-state consists of two orbital levels each with degenerate spin sublevels \cite{hepp2014electronic}. The two transitions to a shared excited state are visible in Fig. \ref{fig:1}(d), where the $^{117}$SnV centre is excited with off-resonant light. For the $^{117}$SnV centre, the electron hyperfine interaction with the $I = 1/2$ nucleus results in an optically resolvable splitting of the ground state into aligned and anti-aligned electronuclear spin levels. Figure \ref{fig:1}(e) presents a direct measurement of the zero-field substructure, where the phonon sideband (PSB) fluorescence is collected while  the excitation wavelength is scanned through the two transitions using a chirped optical resonance excitation (CORE) scheme \cite{martinez2022photonic} (SI III). This scheme minimises optical pumping by repeatedly scanning the laser at a constant rate over the transitions, which is $\sim$\,0.2~MHz/ns for our system. The inset shows a second order autocorrelation function ($g^{(2)}(0) = 0.052(4)$) obtained by resonantly exciting the two transitions simultaneously, verifying that this sub-structure results from a single $^{117}$SnV centre \cite{kim2020analytical}.

\begin{figure}
    \centering
    \includegraphics[width=\columnwidth]{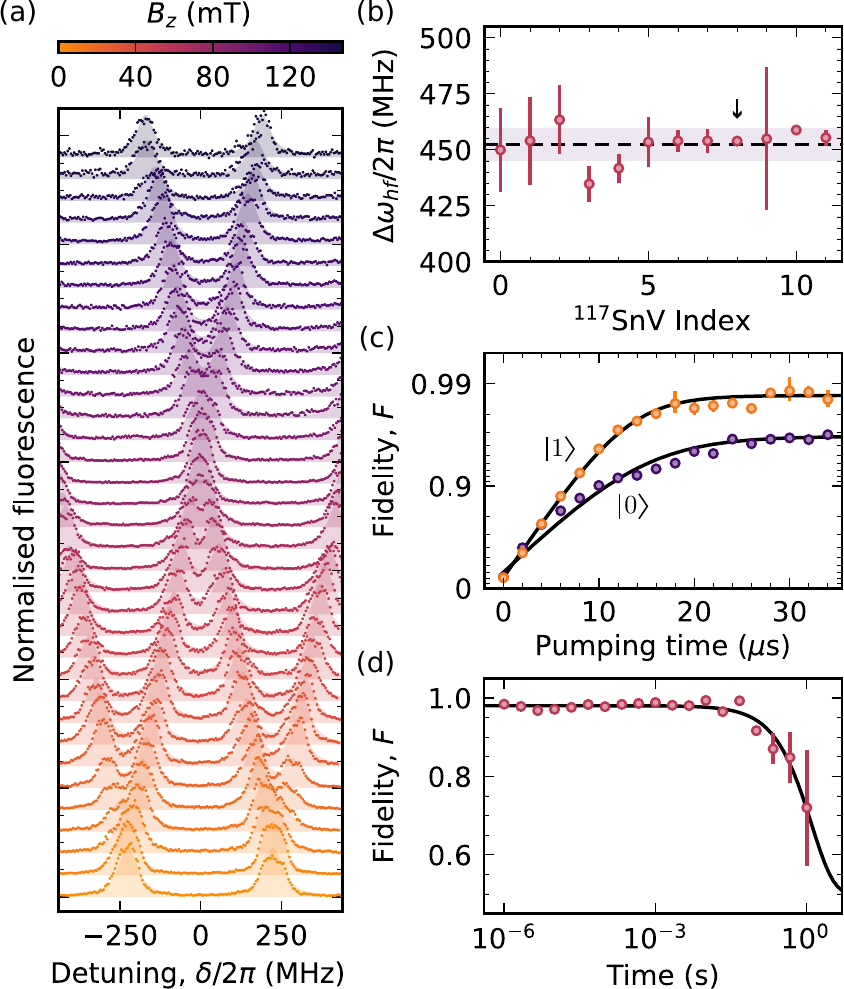}
    \caption{ \textbf{Accessing the $^{117}$SnV electronuclear spin manifold.}
        (a) Repeated CORE scans as a function of magnetic field along the [111] axis. The magnetic field is varied from 0 mT to 147 mT in steps of 4.3 mT. (b) Hyperfine splitting of the optical transitions at $\bm{B}=\bm{0}$ for multiple emitters in the waveguide. The emitter employed for all other measurements is highlighted with an arrow. The shaded region corresponds to the standard deviation. (c) Initialisation fidelity as a function of optical pumping time for the $\ket{0}$ (purple) and $\ket{1}$ (orange) states. (d) State fidelity as a function of time after initialisation. 
    }
    \label{fig:2}
\end{figure}

\subsection*{Optical access to the $^{117}$SnV}

When a magnetic field $B_{\text{z}}$ is present along the symmetry axis of the defect, as in Fig. \ref{fig:2}(a), the remaining spin-degeneracy is split by the electronic Zeeman effect \cite{hepp2014electronic}. Due to the forbidden nature of spin flips in electric dipole transitions \cite{hepp2014electronic}, the four visible transitions correspond to the allowed electron and nuclear-spin preserving transitions. As the hyperfine coupling is much stronger than the nuclear Zeeman effect, the nuclear quantisation axis is strongly pinned by the electronic quantisation axis, and the ground and excited states are each described by the Hamiltonian

\begin{equation}
H = \gamma_S^* B_z S_z + \gamma_I B_z I_z + A_\parallel S_z I_z,
\end{equation}
where $\gamma_S^*$ is an effective electron gyromagnetic ratio, $\bm{S}$ and $\bm{I}$ are the electron and nuclear spin operators and $A_\parallel$ is the parallel component of the hyperfine tensor (SI IV). The four visible spin-conserving transitions occur at detunings $\delta \in \{\pm \Delta A_\parallel/4 \pm \Delta (\gamma_S^*/2) B_z\}$ where $\Delta$ refers to the difference between ground and excited state values of the respective parameter. A fit to the spectra in Fig. \ref{fig:2}(a) yields a splitting rate of $\Delta \gamma_S^* / 2 \pi = 5.41(6)$ GHz/T and a zero-field splitting that is $15.8$-times greater than the lifetime-limited optical linewidth $\gamma_0/2 \pi = 28.6(1)$ MHz (SI V).

Figure \ref{fig:2}(b) highlights that the optical splitting is intrinsic to the $^{117}$Sn nucleus. Across all 12 $^{117}$SnV centres measured in our device, we extract a zero-field hyperfine splitting of $|\Delta A_\parallel|/4 \pi = 452(7)$ MHz. This consistency is in sharp contrast with $^{13}\text{C}$-based memories, whose couplings are stochastic \cite{bradley2019tenqubit}. Due to the split-vacancy structure of the defect, we expect this optical splitting to be dominated by the ground-state hyperfine tensor. In the even-parity ground state the hyperfine interaction should be dominated by the Fermi contact interaction and thus be primarily isotropic \cite{mcconnell1958isotropic}, whereas in the odd-parity excited state the Fermi contact interaction vanishes and the dipolar contribution should result in a much smaller interaction \cite{hepp2014electronic}. As the hyperfine coupling is proportional to the nuclear gyromagnetic ratio, our measurements should extend with little modification to the other ($I =  1/2$) spin-active tin isotopes $^{115}\text{Sn}$ and $^{119}\text{Sn}$, for which we predict optical splittings of approximately 415 MHz and 475 MHz respectively \cite{laaksonen1995absolute, lassigne1981heteronuclear}. Such a large ground-state hyperfine coupling should enable two-qubit gates at a $\sim$\,$20$ MHz rate, as previously achieved with SiV centres \cite{stas2022sivnuclear}, but without the observed infidelities or overhead associated with indirect driving of nearby transitions. Previous reports of nuclear couplings to tin-vacancy centres likely correspond to nearby $^{13}\text{C}$ whose couplings have been predicted in a matching range \cite{debroux2021quantum, defo2021carbon13snv}.

\begin{figure}
    \centering
    \includegraphics[width=\columnwidth]{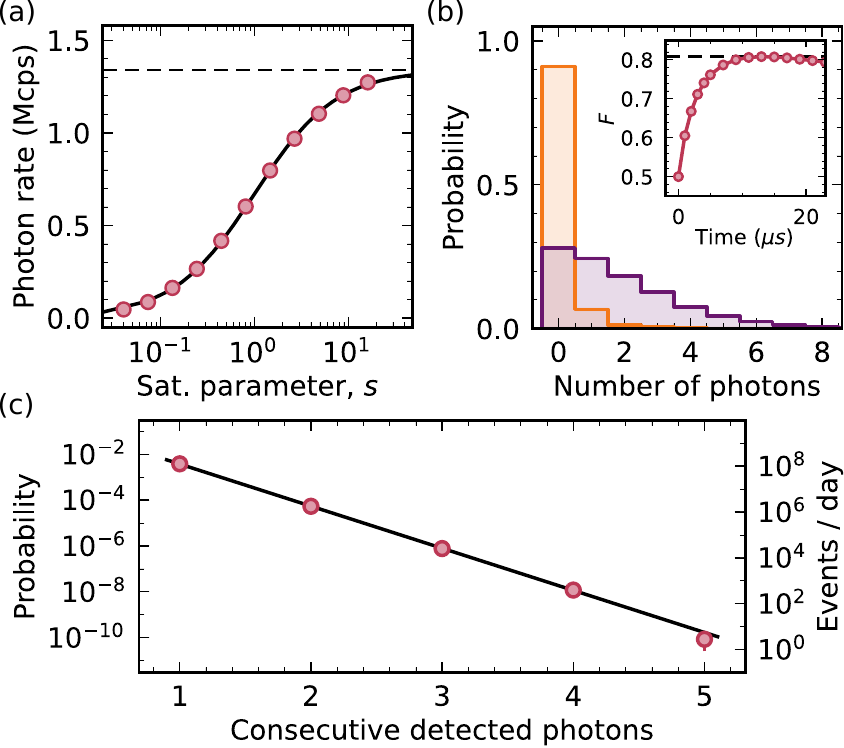}
    \caption{  \textbf{An efficient optical interface.}
        (a) PSB fluorescence as a function of the saturation parameter, $s = P/P_\text{sat}$. (b) Histogram of photon counts per readout pulse after the state is initialised into the $\ket{0}$ (purple) or $\ket{1}$ (orange) state. Inset: fidelity as a function of readout time. (c) Probability of detecting N consecutive photons at a $0.38$ MHz generation rate.
    }
    \label{fig:3}
\end{figure}

In Fig. \ref{fig:2}(c) we utilise the optical accessibility to the hyperfine manifold to perform state initialisation at a magnetic field of $\bm{B} = \bm{0}$. By pumping on the C1 (C2) transition we initialise the system into the $\ket{0}$ ($\ket{1}$) state with a fidelity of 96.9(3)\% (98.6(3)\%) as derived from the polarisation asymmetry in a subsequent CORE scan. Direct optical pumping presents lower control overhead than typical nuclear initialisation mechanisms based on electron initialisation followed by a SWAP gate \cite{stas2022sivnuclear}, measurement-based initialisation \cite{waldherr2014quantum} or dynamic nuclear polarisation \cite{scheuer2016optically}. This initialisation protocol is potentially extendable to the other solid-state emitters with a comparable ratio of hyperfine coupling to optical linewidth such as the V$^{4+}$ centre in SiC \cite{wolfowicz2020vanadium} and the rare-earth ions in wide-bandgap semiconductors  \cite{hedges2010efficient, liu2021heralded, saglamyurek2011broadband, zhong2017nanophotonic}.

Figure \ref{fig:2}(d) shows a measurement of the longitudinal relaxation time, where the $^{117}$SnV centre is initialised into the $\ket{1}$ state and the polarisation asymmetry is measured after a time delay. The time constant of the decay $T_1 = 1.25(10)$ s provides a lower bound on the intrinsic nuclear polarisation lifetime, as residual laser leakage may lead to additional depolarisation processes. This result highlights that the nuclear spin can act as a long-lived nuclear memory where the large hyperfine splitting suppresses flip-flop interactions with the surrounding nuclear bath \cite{merkulov2002electron, gillard2021fundamental, seo2016quantum}.

\begin{figure}
    \centering
    \includegraphics[width=\columnwidth]{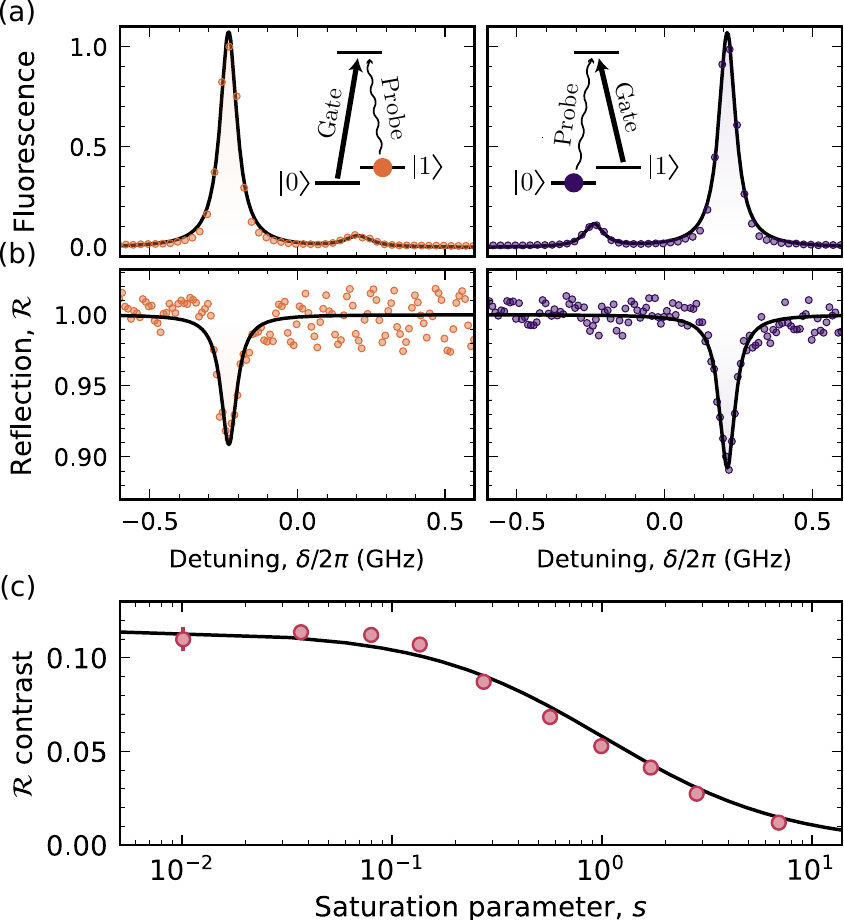}
    \caption{ \textbf{A spin-gated optical switch.}
        (a) PSB fluorescence as a function of probe detuning after the emitter has been pumped by a gate pulse into the $\ket{0}$ (purple) and $\ket{1}$ (orange) states. (b). Reflection intensity as a function of detuning. (c) Contrast of the resonant reflection dip as a function of the saturation parameter. The reflection contrast presented is the mean of the contrast of the two transitions.
    }
    \label{fig:4}
\end{figure}

\subsection*{An efficient optical interface}

Having identified an optically resolvable nuclear sub-structure, we now evaluate the optical coupling of the centre to our nanostructured waveguide. Figure \ref{fig:3}(a) shows a saturation curve obtained by conducting a CORE scan over the optical transitions and summing the fluorescence amplitude of the two hyperfine-split transitions. From a fit of the form $I_\infty/(1 + P_\text{sat}/P)$ \cite{radulaski2017scalable} we extract a power at saturation $P_\text{sat} = 120(30)$ pW at the input of the adiabatically tapered fibre. This corresponds to the $^{117}$SnV centre being saturated by a flux of $2.1(5)$ photons per lifetime, confirming an efficient light-matter interface. The maximum resonance fluorescence rate $I_\infty = 1.34(5)$ Mcps is the highest achieved for SnV centres and extends beyond previous reports for SnV centres in bulk diamond ($\sim$\,25 kcps) \cite{gorlitz2022coherence} and nanopillars (81 kcps) \cite{trusheim2020transformlimited} by over an order of magnitude.

Leveraging this efficient interface, we now demonstrate single-shot readout of the nuclear spin through a pump-probe measurement and counting photon arrival events. Figure \ref{fig:3}(b) shows histograms of the number of photons collected in a readout pulse for the case where the qubit was initialised into the $\ket{0}$ (purple) or $\ket{1}$ (orange) state. The two distributions have an average photon number of $\langle n \rangle_{\ket{0}} = 1.83$ and $\langle n \rangle_{\ket{1}} = 0.13$. We obtain a fidelity of $F = 80(1)\%$, using the optimal one-photon threshold and a readout time of $15$ $\upmu$s (inset).

Our high photon-collection efficiency ultimately manifests itself in the detection of multi-photon coincidences, as quantified in Fig. \ref{fig:3}(c). We highlight the detection of multiple 5-photon coincidence events in a 24 hour time period. The high overall detection efficiency of 1.40(5)\% thus brings to light the possibility of generating complex entanglement resources, such as cluster states, which are only distinct from GHZ states when consisting of at least 4 photons \cite{istrati2020sequential}. When paired with high-purity ZPL photon collection efficiencies recently demonstrated for SnV centres in the waveguide architecture used in this work \cite{martinez2022photonic}, the path to photonic cluster-states becomes accessible in the near-term. Improvements in collection efficiency could be realised through the use of superconducting nanowire single photon detectors and improvements to the adiabatic mode transfer through optimised fabrication protocols, further enhancing the complexity of the entanglement resources achievable with the $^{117}$SnV centre (SI II).

\section*{A spin-gated optical switch}

We now combine our controllable electronuclear state with our high-efficiency light-matter interface. Figures \ref{fig:4}(a) and \ref{fig:4}(b) show respectively the PSB fluorescence and the resonant reflection collected from a CORE scan across the two hyperfine-split transitions with the electronuclear register initialised into the $\ket{0}$ (purple) or $\ket{1}$ (orange) state, and corrected for spectral diffusion (SI III). In Fig. \ref{fig:4}(b), we observe a decrease in the reflectivity, with up to 11(1)\% contrast,
as the probe-field is brought on resonance with the two transitions. This is a characteristic signature of high-efficiency coupling in waveguide quantum electrodynamics \cite{grangier2000implementations}. Importantly, this nonlinearity is conditional on the spin-state of the electronuclear manifold and thus acts a spin-gated photonic switch. The saturability of the nonlinear response, shown in Fig. \ref{fig:4}(c), is a signature of the nonclassical character of the nonlinearity. This opens up a pathway for memory-enhanced quantum communication \cite{bhaskar2020experimental} or photon-photon spin-mediated interactions \cite{le2022dynamical}, and Fock-state readout using the $^{117}$SnV centre \cite{le2022dynamical}.

A fit to the reflection contrast can be employed to characterise the efficiency of the spin-photon interface. Specifically, for a two-level system in a broadband cavity the reflectivity coefficient for near-resonant light is given by 

\begin{equation}
r(\omega) = 1 - \frac{2 f}{1 + C/(1 + 2 \rm{i} \delta/\gamma_\text{h})},
\end{equation}
where $C$ is the cooperativity, $\gamma_\text{h}$ the homogeneous linewidth of the $^{117}$SnV centre, $\delta$ the detuning of the probe field, and $f$ the fraction of the cavity photons that exit toward the input port of the cavity (SI VI). The cooperativity is related to $\beta_\text{tot}$, defined as the ratio of the ZPL emission into the waveguide mode to the total population decay rate. From a fit of the measured power reflectivity, $\mathcal{R} = |r|^2$, we extract $\beta_{\text{tot}} > 6.2(7)\%$.

The $\beta$-factor measured in Fig. \ref{fig:4}(c) can be decomposed into the product of the intrinsic $\beta$-factor of the waveguide ($\beta_{\text{wg}}$), the internal quantum efficiency ($\eta_{\text{QE}}$), the Debye-Waller factor ($\eta_\text{DW}$) and the ZPL orbital branching ratio ($\eta_{\text{orb}}$) as $\beta_{\text{tot}} = \beta_{\text{wg}} \times \eta_{\text{QE}} \times \eta_{\text{DW}} \times \eta_{\text{orb}}$. A higher $\beta$-factor could be realised through: improved centre positioning, as can be achieved through shallow implantation and subsequent overgrowth \cite{rugar2020generation}; improved emission dipole alignment, achievable through the use of \textless111\textgreater-oriented diamond in fabrication; or Purcell enhancement, as has been previously achieved for the SnV centre in nanocavities \cite{rugar2021quantum}. This decomposition further constrains the quantum efficiency of the SnV centre to be $\eta_{\text{QE}} > 51(8)\%$. This is compatible with theoretical calculations (91\%) \cite{thiering2018abinitio} and previous rate-equation measurements (79\%) and represents an intrinsic advantage of SnV over SiV, for which the quantum efficiency is in the range of 5 to 10\% \cite{neu2012photophysics, sukachev2017silicon}. A large quantum efficiency results in a higher cooperativity factor for a given scattering loss rate, fixed by fabrication imperfections, and thus lessens the challenge of efficient photonic outcoupling present for critically coupled nanocavities \cite{knall2022efficientsinglephotons}.

\section*{Discussion}

In this work, we have introduced a versatile quantum device, consisting of a fibre-packaged waveguide with access to a nuclear spin that serves to address three key challenges for optically interconnected qubit systems. First, as a quantum node, the high waveguide-to-fibre extraction efficiency of 57(6)\% should enable high repetition rate remote entanglement based on single-photon protocols when combined with previous demonstrations of quantum control \cite{debroux2021quantum} and interference of resonant photons \cite{martinez2022photonic}. This entanglement generation rate would be limited by initialisation and readout overheads in the current device, which could be lowered through fast-feedback control logic \cite{stas2022sivnuclear}. For entanglement distribution over more than two nodes, an ancillary register is necessary \cite{baier2021realization}. The immediate next steps thus involve accessing the coherence of the nuclear state through microwave \cite{stas2022sivnuclear} or Raman-based protocols \cite{debroux2021quantum, goldman2020optical}. The large hyperfine coupling is advantageous to realise fast two-qubit gates without the typical infidelities associated with indirect driving of nearby transitions \cite{stas2022sivnuclear}. The protected nature of the nuclear spin should enable a second-long coherence time \cite{stas2022sivnuclear}. Finally, addressing nearby $^{13}$C through dynamical decoupling gates is a natural way to extend this register \cite{bradley2019tenqubit} to realise fault tolerant nodes \cite{wehner2018quantuminternet}.

Second, as a source for photonics cluster states, the detection of 5-photon events in a 24-hour period showcases the feasibility of proof-of-principle demonstrations, in particular when combined with a nuclear memory \cite{michaels2021multidimensional}. Large-scale cluster-state generation, as required for efficient and scalable quantum computing \cite{raussendorf2001clusterstates}, would require near-unity overall detection efficiency and thus integration into a nanocavity \cite{knall2022efficientsinglephotons} or open microcavity \cite{tomm2021brightsource}, but the current device is already sufficient for first demonstrations. 

Third, as a gate for photonic logic, the single-photon nonlinearity presented in this work is a quantum resource on its own. A higher reflection contrast would enable reflection-based spin-readout \cite{stas2022sivnuclear} and high-contrast spin-gated photonic switching. Its combination with coherent Raman control would enable a single-photon transistor \cite{chang2007single}. Deterministic photon-photon logic gates in combination with large scale integration into photonic integrated circuits could serve to reduce the exponential overhead of linear optical quantum computing protocols \cite{knill2001scheme, kok2007linear}.

\section*{Methods}

\subsection*{Waveguide microchiplet fabrication}

An electronic-grade diamond (Element 6, [N] $< 5$ ppb) was first chemically polished via $\text{ArCl}_2$ Reactive Ion Etching (RIE) to smooth the surface to $<1$ nm roughness, then implanted with $^{117}\text{Sn}^{++}$ ions twice. The first implantation (Cutting Edge Ions) used an implantation energy of 200 keV and a dosage of $5 \times 10^{10}$ ions/$\text{cm}^2$ whereas the second implant occured at a dosage of $1 \times 10^{11}$ ions/$\text{cm}^2$ and 350 keV implantation energy ($86$ nm implant depth; $17$ nm straggle \cite{ziegler1985stopping}). The second implantation was needed as SnV photoluminscence was not observed after the initial implatantation and subsequent high-temperature anneal. Following the second implantation, the diamond was annealed at 1200 ${}^{\circ}$C for 12 hours at $\sim$\,$10^{-7}$ mbar followed by a boiling tri-acid clean (nitric, sulfuric, and perchloric acids mixed in 1:1:1 ratio and heated to 345 ${}^{\circ}$C). Quantum microchiplets \cite{wan2020large}, each containing 8 waveguide channels, were fabricated in bulk diamond through quasi-isotropic etching \cite{mouradian2017rectangular, wan2018twodimensionalnanocavity} and subsequent high temperature annealing, again at 1200 ${}^{\circ}$C for 12 hours followed further by a boiling tri-acid clean. Finally, the diamond was then submerged in piranha solution (sulfuric acid mixed with hydrogen peroxide in a 3:1 ratio) to improve surface termination \cite{sangtawesin2019diamondsurfacenoise}. Chiplets were then pick and placed onto the edge of a fused silica substrate to enable fibre-based optical access \cite{martinez2022photonic} (SI I).

\subsection*{Tapered fibre fabrication}

A standard single-mode optical fibre (Thorlabs SM600) is first mechanically stripped and cleaved. Conically tapered fibres are then fabricated through dynamic meniscus etching in hydrofluoric acid (HF) \cite{haber2004shape, burek2017fiber} using a 80 ml solution of 40\% HF and 10 ml of ortho-xylene. A set of six fibres is lowered into the solution and progressively retracted at 55 $\upmu$m/min over the course of 95 minutes, yielding a conical half-angle of $1.5(8)^{\circ}$. A fibre is then coated with a thin layer of a photo-polymerised optical adhesive (Norland Optical Adhesive 86H) and adhered to the adiabatically tapered port of a waveguide \cite{wasserman2022cryogenic}. The final device is then mounted inside of the cryostat employed in this work such that the diamond end of the fibre is free-floating, mechanically held in place through the fibre input-output port  (SI I).

\subsection*{Laser control setup}

We employ a tunable resonant source at 619 nm (M2 SolsTiS + EMM) for all of our experiments. The off-resonant source at 532 nm is only employed for measurements of photoluminescence and is not required during on-resonance measurements to initialise the charge state of the emitter \cite{gorlitz2022coherence}. The resonant light is directed through an acousto-optic modulator  (AOM, Gooch and Housego 3080-15) and a fibre-based electro-optic modulator (EOM, Jenoptik AM635). The EOM is used for all optical pulsing and is controlled by a 25 Gsamples/s arbitrary waveform generator (Tektronix AWG70002A) through a microwave amplifier (Minicircuits ZX60-83LN12+) and a bias-tee (Minicircuits ZFBT-6G+). The resonant laser is detuned by 3.5 GHz from the middle point of the C1 and C2 transitions to minimise the laser excitation caused by leakage laser light, and near-resonant light is generated through sinusoidal modulation of the EOM with a frequency swept around 3.5 GHz. In all resonant measurements the EOM is stabilised to its interferometric minimum through a lock-in amplifier and PID loop (Red Pitaya, STEMlab 125-10). A pulse generator (Swabian Instruments Pulse Streamer 8/2) is employed as a master clock to coordinate all instruments (SI III).

\subsection*{Spectral drift compensation}

The $^{117}$SnV centres investigated in this work show a slow ($\sim$\,10 minute time-scale) spectral diffusion over a $\sim$\,100 MHz spectral range. We actively correct for this drift by conducting a 5 s CORE scan every 60 s to compensate the drifting zero-detuning ($\delta = 0$) position of the $^{117}$SnV centres. In Fig. 4 we instead recenter the drift in the data analysis by overlapping consecutive 1 minute scans based on Lorentzian fits. We further note that in panel (c) of this figure, in contrast to panels (a) and (b), the qubit is not prepared into either the $\ket{0}$ or $\ket{1}$ states. This is due to the difficulty in consistently initialising the spin while scanning the excitation power over three orders of magnitude ($10^{-2} < s < 10^{1}$). As such the sum of the two reflection-dips is taken as the nonlinearity contrast.

\begin{acknowledgments}

We would like to thank Melanie Tribble for fruitful discussions. We acknowledge support from the ERC Advanced Grant PEDESTAL (884745), the EU Quantum Flagship 2D-SIPC. J.A.M. acknowledges support from the Winton Programme and EPSRC DTP, R.A.P. from the General Sir John Monash Foundation, K.C.C. from the MITRE program, C.P.M. from the EPSRC DTP, A.M.S. from EPSRC/NQIT, and M.T. from the Army Research Laboratory ENIAC Distinguished Postdoctoral Fellowship. D.E. acknowledges further support by the MITRE Quantum Moonshot Program.

${}^{*}$ These authors contributed equally to this work.

${}^{\dagger}$ Correspondence should be addressed to:  \href{mailto:englund@mit.edu}{englund@mit.edu}, \href{mailto:ma424@cam.ac.uk}{ma424@cam.ac.uk}.

\end{acknowledgments}


\begin{thebibliography}{75}%
    \makeatletter
    \providecommand \@ifxundefined [1]{%
     \@ifx{#1\undefined}
    }%
    \providecommand \@ifnum [1]{%
     \ifnum #1\expandafter \@firstoftwo
     \else \expandafter \@secondoftwo
     \fi
    }%
    \providecommand \@ifx [1]{%
     \ifx #1\expandafter \@firstoftwo
     \else \expandafter \@secondoftwo
     \fi
    }%
    \providecommand \natexlab [1]{#1}%
    \providecommand \enquote  [1]{``#1''}%
    \providecommand \bibnamefont  [1]{#1}%
    \providecommand \bibfnamefont [1]{#1}%
    \providecommand \citenamefont [1]{#1}%
    \providecommand \href@noop [0]{\@secondoftwo}%
    \providecommand \href [0]{\begingroup \@sanitize@url \@href}%
    \providecommand \@href[1]{\@@startlink{#1}\@@href}%
    \providecommand \@@href[1]{\endgroup#1\@@endlink}%
    \providecommand \@sanitize@url [0]{\catcode `\\12\catcode `\$12\catcode
      `\&12\catcode `\#12\catcode `\^12\catcode `\_12\catcode `\%12\relax}%
    \providecommand \@@startlink[1]{}%
    \providecommand \@@endlink[0]{}%
    \providecommand \url  [0]{\begingroup\@sanitize@url \@url }%
    \providecommand \@url [1]{\endgroup\@href {#1}{\urlprefix }}%
    \providecommand \urlprefix  [0]{URL }%
    \providecommand \Eprint [0]{\href }%
    \providecommand \doibase [0]{https://doi.org/}%
    \providecommand \selectlanguage [0]{\@gobble}%
    \providecommand \bibinfo  [0]{\@secondoftwo}%
    \providecommand \bibfield  [0]{\@secondoftwo}%
    \providecommand \translation [1]{[#1]}%
    \providecommand \BibitemOpen [0]{}%
    \providecommand \bibitemStop [0]{}%
    \providecommand \bibitemNoStop [0]{.\EOS\space}%
    \providecommand \EOS [0]{\spacefactor3000\relax}%
    \providecommand \BibitemShut  [1]{\csname bibitem#1\endcsname}%
    \let\auto@bib@innerbib\@empty
    \bibitem [{\citenamefont {Kimble}(2008)}]{kimble2008quantum}%
      \BibitemOpen
      \bibfield  {author} {\bibinfo {author} {\bibfnamefont {H.~J.}\ \bibnamefont
      {Kimble}},\ }\bibfield  {title} {\bibinfo {title} {The quantum internet},\
      }\href {https://doi.org/10.1038/nature07127} {\bibfield  {journal} {\bibinfo
      {journal} {Nature}\ }\textbf {\bibinfo {volume} {453}},\ \bibinfo {pages}
      {1023} (\bibinfo {year} {2008})}\BibitemShut {NoStop}%
    \bibitem [{\citenamefont {Wehner}\ \emph {et~al.}(2018)\citenamefont {Wehner},
      \citenamefont {Elkouss},\ and\ \citenamefont
      {Hanson}}]{wehner2018quantuminternet}%
      \BibitemOpen
      \bibfield  {author} {\bibinfo {author} {\bibfnamefont {S.}~\bibnamefont
      {Wehner}}, \bibinfo {author} {\bibfnamefont {D.}~\bibnamefont {Elkouss}},\
      and\ \bibinfo {author} {\bibfnamefont {R.}~\bibnamefont {Hanson}},\
      }\bibfield  {title} {\bibinfo {title} {Quantum internet: A vision for the
      road ahead},\ }\href {https://doi.org/10.1126/science.aam9288} {\bibfield
      {journal} {\bibinfo  {journal} {Science}\ }\textbf {\bibinfo {volume}
      {362}},\ \bibinfo {pages} {eaam9288} (\bibinfo {year} {2018})}\BibitemShut
      {NoStop}%
    \bibitem [{\citenamefont {Tomm}\ \emph {et~al.}(2021)\citenamefont {Tomm},
      \citenamefont {Javadi}, \citenamefont {Antoniadis}, \citenamefont {Najer},
      \citenamefont {L{\"o}bl}, \citenamefont {Korsch}, \citenamefont {Schott},
      \citenamefont {Valentin}, \citenamefont {Wieck}, \citenamefont {Ludwig} \emph
      {et~al.}}]{tomm2021brightsource}%
      \BibitemOpen
      \bibfield  {author} {\bibinfo {author} {\bibfnamefont {N.}~\bibnamefont
      {Tomm}}, \bibinfo {author} {\bibfnamefont {A.}~\bibnamefont {Javadi}},
      \bibinfo {author} {\bibfnamefont {N.~O.}\ \bibnamefont {Antoniadis}},
      \bibinfo {author} {\bibfnamefont {D.}~\bibnamefont {Najer}}, \bibinfo
      {author} {\bibfnamefont {M.~C.}\ \bibnamefont {L{\"o}bl}}, \bibinfo {author}
      {\bibfnamefont {A.~R.}\ \bibnamefont {Korsch}}, \bibinfo {author}
      {\bibfnamefont {R.}~\bibnamefont {Schott}}, \bibinfo {author} {\bibfnamefont
      {S.~R.}\ \bibnamefont {Valentin}}, \bibinfo {author} {\bibfnamefont {A.~D.}\
      \bibnamefont {Wieck}}, \bibinfo {author} {\bibfnamefont {A.}~\bibnamefont
      {Ludwig}}, \emph {et~al.},\ }\bibfield  {title} {\bibinfo {title} {A bright
      and fast source of coherent single photons},\ }\href
      {https://doi.org/10.1038/s41565-020-00831-x} {\bibfield  {journal} {\bibinfo
      {journal} {Nature Nanotechnology}\ }\textbf {\bibinfo {volume} {16}},\
      \bibinfo {pages} {399} (\bibinfo {year} {2021})}\BibitemShut {NoStop}%
    \bibitem [{\citenamefont {Uppu}\ \emph {et~al.}(2020)\citenamefont {Uppu},
      \citenamefont {Pedersen}, \citenamefont {Wang}, \citenamefont {Olesen},
      \citenamefont {Papon}, \citenamefont {Zhou}, \citenamefont {Midolo},
      \citenamefont {Scholz}, \citenamefont {Wieck}, \citenamefont {Ludwig} \emph
      {et~al.}}]{uppu2020scalable}%
      \BibitemOpen
      \bibfield  {author} {\bibinfo {author} {\bibfnamefont {R.}~\bibnamefont
      {Uppu}}, \bibinfo {author} {\bibfnamefont {F.~T.}\ \bibnamefont {Pedersen}},
      \bibinfo {author} {\bibfnamefont {Y.}~\bibnamefont {Wang}}, \bibinfo {author}
      {\bibfnamefont {C.~T.}\ \bibnamefont {Olesen}}, \bibinfo {author}
      {\bibfnamefont {C.}~\bibnamefont {Papon}}, \bibinfo {author} {\bibfnamefont
      {X.}~\bibnamefont {Zhou}}, \bibinfo {author} {\bibfnamefont {L.}~\bibnamefont
      {Midolo}}, \bibinfo {author} {\bibfnamefont {S.}~\bibnamefont {Scholz}},
      \bibinfo {author} {\bibfnamefont {A.~D.}\ \bibnamefont {Wieck}}, \bibinfo
      {author} {\bibfnamefont {A.}~\bibnamefont {Ludwig}}, \emph {et~al.},\
      }\bibfield  {title} {\bibinfo {title} {Scalable integrated single-photon
      source},\ }\bibfield  {journal} {\bibinfo  {journal} {Science Advances}\
      }\textbf {\bibinfo {volume} {6}},\ \href
      {https://doi.org/10.1126/sciadv.abc8268} {10.1126/sciadv.abc8268} (\bibinfo
      {year} {2020})\BibitemShut {NoStop}%
    \bibitem [{\citenamefont {Papon}\ \emph {et~al.}(2022)\citenamefont {Papon},
      \citenamefont {Wang}, \citenamefont {Uppu}, \citenamefont {Scholz},
      \citenamefont {Wieck}, \citenamefont {Ludwig}, \citenamefont {Lodahl},\ and\
      \citenamefont {Midolo}}]{papon2022independent}%
      \BibitemOpen
      \bibfield  {author} {\bibinfo {author} {\bibfnamefont {C.}~\bibnamefont
      {Papon}}, \bibinfo {author} {\bibfnamefont {Y.}~\bibnamefont {Wang}},
      \bibinfo {author} {\bibfnamefont {R.}~\bibnamefont {Uppu}}, \bibinfo {author}
      {\bibfnamefont {S.}~\bibnamefont {Scholz}}, \bibinfo {author} {\bibfnamefont
      {A.~D.}\ \bibnamefont {Wieck}}, \bibinfo {author} {\bibfnamefont
      {A.}~\bibnamefont {Ludwig}}, \bibinfo {author} {\bibfnamefont
      {P.}~\bibnamefont {Lodahl}},\ and\ \bibinfo {author} {\bibfnamefont
      {L.}~\bibnamefont {Midolo}},\ }\bibfield  {title} {\bibinfo {title}
      {Independent operation of two waveguide-integrated single-photon sources},\
      }\href@noop {} {\bibfield  {journal} {\bibinfo  {journal} {arXiv preprint
      arXiv:2210.09826}\ } (\bibinfo {year} {2022})}\BibitemShut {NoStop}%
    \bibitem [{\citenamefont {Javadi}\ \emph {et~al.}(2018)\citenamefont {Javadi},
      \citenamefont {Ding}, \citenamefont {Appel}, \citenamefont {Mahmoodian},
      \citenamefont {L{\"o}bl}, \citenamefont {S{\"o}llner}, \citenamefont
      {Schott}, \citenamefont {Papon}, \citenamefont {Pregnolato}, \citenamefont
      {Stobbe} \emph {et~al.}}]{javadi2018spin}%
      \BibitemOpen
      \bibfield  {author} {\bibinfo {author} {\bibfnamefont {A.}~\bibnamefont
      {Javadi}}, \bibinfo {author} {\bibfnamefont {D.}~\bibnamefont {Ding}},
      \bibinfo {author} {\bibfnamefont {M.~H.}\ \bibnamefont {Appel}}, \bibinfo
      {author} {\bibfnamefont {S.}~\bibnamefont {Mahmoodian}}, \bibinfo {author}
      {\bibfnamefont {M.~C.}\ \bibnamefont {L{\"o}bl}}, \bibinfo {author}
      {\bibfnamefont {I.}~\bibnamefont {S{\"o}llner}}, \bibinfo {author}
      {\bibfnamefont {R.}~\bibnamefont {Schott}}, \bibinfo {author} {\bibfnamefont
      {C.}~\bibnamefont {Papon}}, \bibinfo {author} {\bibfnamefont
      {T.}~\bibnamefont {Pregnolato}}, \bibinfo {author} {\bibfnamefont
      {S.}~\bibnamefont {Stobbe}}, \emph {et~al.},\ }\bibfield  {title} {\bibinfo
      {title} {Spin--photon interface and spin-controlled photon switching in a
      nanobeam waveguide},\ }\href {https://doi.org/10.1038/s41565-018-0091-5}
      {\bibfield  {journal} {\bibinfo  {journal} {Nature Nanotechnology}\ }\textbf
      {\bibinfo {volume} {13}},\ \bibinfo {pages} {398} (\bibinfo {year}
      {2018})}\BibitemShut {NoStop}%
    \bibitem [{\citenamefont {Appel}\ \emph {et~al.}(2021)\citenamefont {Appel},
      \citenamefont {Tiranov}, \citenamefont {Javadi}, \citenamefont {L{\"o}bl},
      \citenamefont {Wang}, \citenamefont {Scholz}, \citenamefont {Wieck},
      \citenamefont {Ludwig}, \citenamefont {Warburton},\ and\ \citenamefont
      {Lodahl}}]{appel2021spinphoton}%
      \BibitemOpen
      \bibfield  {author} {\bibinfo {author} {\bibfnamefont {M.~H.}\ \bibnamefont
      {Appel}}, \bibinfo {author} {\bibfnamefont {A.}~\bibnamefont {Tiranov}},
      \bibinfo {author} {\bibfnamefont {A.}~\bibnamefont {Javadi}}, \bibinfo
      {author} {\bibfnamefont {M.~C.}\ \bibnamefont {L{\"o}bl}}, \bibinfo {author}
      {\bibfnamefont {Y.}~\bibnamefont {Wang}}, \bibinfo {author} {\bibfnamefont
      {S.}~\bibnamefont {Scholz}}, \bibinfo {author} {\bibfnamefont {A.~D.}\
      \bibnamefont {Wieck}}, \bibinfo {author} {\bibfnamefont {A.}~\bibnamefont
      {Ludwig}}, \bibinfo {author} {\bibfnamefont {R.~J.}\ \bibnamefont
      {Warburton}},\ and\ \bibinfo {author} {\bibfnamefont {P.}~\bibnamefont
      {Lodahl}},\ }\bibfield  {title} {\bibinfo {title} {Coherent spin-photon
      interface with waveguide induced cycling transitions},\ }\href
      {https://doi.org/10.1103/physrevlett.126.013602} {\bibfield  {journal}
      {\bibinfo  {journal} {Physical Review Letters}\ }\textbf {\bibinfo {volume}
      {126}},\ \bibinfo {pages} {013602} (\bibinfo {year} {2021})}\BibitemShut
      {NoStop}%
    \bibitem [{\citenamefont {Delteil}\ \emph {et~al.}(2016)\citenamefont
      {Delteil}, \citenamefont {Sun}, \citenamefont {Gao}, \citenamefont {Togan},
      \citenamefont {Faelt},\ and\ \citenamefont
      {Imamo{\u{g}}lu}}]{delteil2016generation}%
      \BibitemOpen
      \bibfield  {author} {\bibinfo {author} {\bibfnamefont {A.}~\bibnamefont
      {Delteil}}, \bibinfo {author} {\bibfnamefont {Z.}~\bibnamefont {Sun}},
      \bibinfo {author} {\bibfnamefont {W.-b.}\ \bibnamefont {Gao}}, \bibinfo
      {author} {\bibfnamefont {E.}~\bibnamefont {Togan}}, \bibinfo {author}
      {\bibfnamefont {S.}~\bibnamefont {Faelt}},\ and\ \bibinfo {author}
      {\bibfnamefont {A.}~\bibnamefont {Imamo{\u{g}}lu}},\ }\bibfield  {title}
      {\bibinfo {title} {Generation of heralded entanglement between distant hole
      spins},\ }\href {https://doi.org/10.1038/nphys3605} {\bibfield  {journal}
      {\bibinfo  {journal} {Nature Physics}\ }\textbf {\bibinfo {volume} {12}},\
      \bibinfo {pages} {218} (\bibinfo {year} {2016})}\BibitemShut {NoStop}%
    \bibitem [{\citenamefont {Stockill}\ \emph {et~al.}(2017)\citenamefont
      {Stockill}, \citenamefont {Stanley}, \citenamefont {Huthmacher},
      \citenamefont {Clarke}, \citenamefont {Hugues}, \citenamefont {Miller},
      \citenamefont {Matthiesen}, \citenamefont {Le~Gall},\ and\ \citenamefont
      {Atat{\"u}re}}]{stockill2017phase}%
      \BibitemOpen
      \bibfield  {author} {\bibinfo {author} {\bibfnamefont {R.}~\bibnamefont
      {Stockill}}, \bibinfo {author} {\bibfnamefont {M.}~\bibnamefont {Stanley}},
      \bibinfo {author} {\bibfnamefont {L.}~\bibnamefont {Huthmacher}}, \bibinfo
      {author} {\bibfnamefont {E.}~\bibnamefont {Clarke}}, \bibinfo {author}
      {\bibfnamefont {M.}~\bibnamefont {Hugues}}, \bibinfo {author} {\bibfnamefont
      {A.}~\bibnamefont {Miller}}, \bibinfo {author} {\bibfnamefont
      {C.}~\bibnamefont {Matthiesen}}, \bibinfo {author} {\bibfnamefont
      {C.}~\bibnamefont {Le~Gall}},\ and\ \bibinfo {author} {\bibfnamefont
      {M.}~\bibnamefont {Atat{\"u}re}},\ }\bibfield  {title} {\bibinfo {title}
      {Phase-tuned entangled state generation between distant spin qubits},\ }\href
      {https://doi.org/10.1103/physrevlett.119.010503} {\bibfield  {journal}
      {\bibinfo  {journal} {Physical review letters}\ }\textbf {\bibinfo {volume}
      {119}},\ \bibinfo {pages} {010503} (\bibinfo {year} {2017})}\BibitemShut
      {NoStop}%
    \bibitem [{\citenamefont {Istrati}\ \emph
      {et~al.}(2020{\natexlab{a}})\citenamefont {Istrati}, \citenamefont {Pilnyak},
      \citenamefont {Loredo}, \citenamefont {Ant{\'o}n}, \citenamefont {Somaschi},
      \citenamefont {Hilaire}, \citenamefont {Ollivier}, \citenamefont {Esmann},
      \citenamefont {Cohen}, \citenamefont {Vidro} \emph
      {et~al.}}]{istrati2020clusterstates}%
      \BibitemOpen
      \bibfield  {author} {\bibinfo {author} {\bibfnamefont {D.}~\bibnamefont
      {Istrati}}, \bibinfo {author} {\bibfnamefont {Y.}~\bibnamefont {Pilnyak}},
      \bibinfo {author} {\bibfnamefont {J.}~\bibnamefont {Loredo}}, \bibinfo
      {author} {\bibfnamefont {C.}~\bibnamefont {Ant{\'o}n}}, \bibinfo {author}
      {\bibfnamefont {N.}~\bibnamefont {Somaschi}}, \bibinfo {author}
      {\bibfnamefont {P.}~\bibnamefont {Hilaire}}, \bibinfo {author} {\bibfnamefont
      {H.}~\bibnamefont {Ollivier}}, \bibinfo {author} {\bibfnamefont
      {M.}~\bibnamefont {Esmann}}, \bibinfo {author} {\bibfnamefont
      {L.}~\bibnamefont {Cohen}}, \bibinfo {author} {\bibfnamefont
      {L.}~\bibnamefont {Vidro}}, \emph {et~al.},\ }\bibfield  {title} {\bibinfo
      {title} {Sequential generation of linear cluster states from a single photon
      emitter},\ }\href {https://doi.org/10.1364/quantum.2020.qtu8a.10} {\bibfield
      {journal} {\bibinfo  {journal} {Nature Communications}\ }\textbf {\bibinfo
      {volume} {11}},\ \bibinfo {pages} {5501} (\bibinfo {year}
      {2020}{\natexlab{a}})}\BibitemShut {NoStop}%
    \bibitem [{\citenamefont {Cogan}\ \emph {et~al.}(2023)\citenamefont {Cogan},
      \citenamefont {Su}, \citenamefont {Kenneth},\ and\ \citenamefont
      {Gershoni}}]{cogan2023clusterstate}%
      \BibitemOpen
      \bibfield  {author} {\bibinfo {author} {\bibfnamefont {D.}~\bibnamefont
      {Cogan}}, \bibinfo {author} {\bibfnamefont {Z.-E.}\ \bibnamefont {Su}},
      \bibinfo {author} {\bibfnamefont {O.}~\bibnamefont {Kenneth}},\ and\ \bibinfo
      {author} {\bibfnamefont {D.}~\bibnamefont {Gershoni}},\ }\bibfield  {title}
      {\bibinfo {title} {Deterministic generation of indistinguishable photons in a
      cluster state},\ }\href {https://doi.org/10.1038/s41566-022-01152-2}
      {\bibfield  {journal} {\bibinfo  {journal} {Nature Photonics}\ ,\ \bibinfo
      {pages} {1}} (\bibinfo {year} {2023})}\BibitemShut {NoStop}%
    \bibitem [{\citenamefont {Zaporski}\ \emph {et~al.}(2023)\citenamefont
      {Zaporski}, \citenamefont {Shofer}, \citenamefont {Bodey}, \citenamefont
      {Manna}, \citenamefont {Gillard}, \citenamefont {Appel}, \citenamefont
      {Schimpf}, \citenamefont {Covre~da Silva}, \citenamefont {Jarman},
      \citenamefont {Delamare} \emph {et~al.}}]{zaporski2023idealrefocusing}%
      \BibitemOpen
      \bibfield  {author} {\bibinfo {author} {\bibfnamefont {L.}~\bibnamefont
      {Zaporski}}, \bibinfo {author} {\bibfnamefont {N.}~\bibnamefont {Shofer}},
      \bibinfo {author} {\bibfnamefont {J.~H.}\ \bibnamefont {Bodey}}, \bibinfo
      {author} {\bibfnamefont {S.}~\bibnamefont {Manna}}, \bibinfo {author}
      {\bibfnamefont {G.}~\bibnamefont {Gillard}}, \bibinfo {author} {\bibfnamefont
      {M.~H.}\ \bibnamefont {Appel}}, \bibinfo {author} {\bibfnamefont
      {C.}~\bibnamefont {Schimpf}}, \bibinfo {author} {\bibfnamefont {S.~F.}\
      \bibnamefont {Covre~da Silva}}, \bibinfo {author} {\bibfnamefont
      {J.}~\bibnamefont {Jarman}}, \bibinfo {author} {\bibfnamefont
      {G.}~\bibnamefont {Delamare}}, \emph {et~al.},\ }\bibfield  {title} {\bibinfo
      {title} {Ideal refocusing of an optically active spin qubit under strong
      hyperfine interactions},\ }\href {https://doi.org/10.1038/s41565-022-01282-2}
      {\bibfield  {journal} {\bibinfo  {journal} {Nature Nanotechnology}\ ,\
      \bibinfo {pages} {1}} (\bibinfo {year} {2023})}\BibitemShut {NoStop}%
    \bibitem [{\citenamefont {Denning}\ \emph {et~al.}(2019)\citenamefont
      {Denning}, \citenamefont {Gangloff}, \citenamefont {Atat{\"u}re},
      \citenamefont {M{\o}rk},\ and\ \citenamefont {Le~Gall}}]{denning2019magnon}%
      \BibitemOpen
      \bibfield  {author} {\bibinfo {author} {\bibfnamefont {E.~V.}\ \bibnamefont
      {Denning}}, \bibinfo {author} {\bibfnamefont {D.~A.}\ \bibnamefont
      {Gangloff}}, \bibinfo {author} {\bibfnamefont {M.}~\bibnamefont
      {Atat{\"u}re}}, \bibinfo {author} {\bibfnamefont {J.}~\bibnamefont
      {M{\o}rk}},\ and\ \bibinfo {author} {\bibfnamefont {C.}~\bibnamefont
      {Le~Gall}},\ }\bibfield  {title} {\bibinfo {title} {Collective quantum memory
      activated by a driven central spin},\ }\href
      {https://doi.org/10.1103/physrevlett.123.140502} {\bibfield  {journal}
      {\bibinfo  {journal} {Physical Review Letters}\ }\textbf {\bibinfo {volume}
      {123}},\ \bibinfo {pages} {140502} (\bibinfo {year} {2019})}\BibitemShut
      {NoStop}%
    \bibitem [{\citenamefont {Bar-Gill}\ \emph {et~al.}(2013)\citenamefont
      {Bar-Gill}, \citenamefont {Pham}, \citenamefont {Jarmola}, \citenamefont
      {Budker},\ and\ \citenamefont {Walsworth}}]{bar2013solid}%
      \BibitemOpen
      \bibfield  {author} {\bibinfo {author} {\bibfnamefont {N.}~\bibnamefont
      {Bar-Gill}}, \bibinfo {author} {\bibfnamefont {L.~M.}\ \bibnamefont {Pham}},
      \bibinfo {author} {\bibfnamefont {A.}~\bibnamefont {Jarmola}}, \bibinfo
      {author} {\bibfnamefont {D.}~\bibnamefont {Budker}},\ and\ \bibinfo {author}
      {\bibfnamefont {R.~L.}\ \bibnamefont {Walsworth}},\ }\bibfield  {title}
      {\bibinfo {title} {Solid-state electronic spin coherence time approaching one
      second},\ }\href {https://doi.org/10.1038/ncomms2771} {\bibfield  {journal}
      {\bibinfo  {journal} {Nature Communications}\ }\textbf {\bibinfo {volume}
      {4}},\ \bibinfo {pages} {1743} (\bibinfo {year} {2013})}\BibitemShut
      {NoStop}%
    \bibitem [{\citenamefont {Balasubramanian}\ \emph {et~al.}(2009)\citenamefont
      {Balasubramanian}, \citenamefont {Neumann}, \citenamefont {Twitchen},
      \citenamefont {Markham}, \citenamefont {Kolesov}, \citenamefont {Mizuochi},
      \citenamefont {Isoya}, \citenamefont {Achard}, \citenamefont {Beck},
      \citenamefont {Tissler} \emph {et~al.}}]{balasubramanian2009ultralong}%
      \BibitemOpen
      \bibfield  {author} {\bibinfo {author} {\bibfnamefont {G.}~\bibnamefont
      {Balasubramanian}}, \bibinfo {author} {\bibfnamefont {P.}~\bibnamefont
      {Neumann}}, \bibinfo {author} {\bibfnamefont {D.}~\bibnamefont {Twitchen}},
      \bibinfo {author} {\bibfnamefont {M.}~\bibnamefont {Markham}}, \bibinfo
      {author} {\bibfnamefont {R.}~\bibnamefont {Kolesov}}, \bibinfo {author}
      {\bibfnamefont {N.}~\bibnamefont {Mizuochi}}, \bibinfo {author}
      {\bibfnamefont {J.}~\bibnamefont {Isoya}}, \bibinfo {author} {\bibfnamefont
      {J.}~\bibnamefont {Achard}}, \bibinfo {author} {\bibfnamefont
      {J.}~\bibnamefont {Beck}}, \bibinfo {author} {\bibfnamefont {J.}~\bibnamefont
      {Tissler}}, \emph {et~al.},\ }\bibfield  {title} {\bibinfo {title} {Ultralong
      spin coherence time in isotopically engineered diamond},\ }\href
      {https://doi.org/10.1038/nmat2420} {\bibfield  {journal} {\bibinfo  {journal}
      {Nature materials}\ }\textbf {\bibinfo {volume} {8}},\ \bibinfo {pages} {383}
      (\bibinfo {year} {2009})}\BibitemShut {NoStop}%
    \bibitem [{\citenamefont {Bradley}\ \emph {et~al.}(2019)\citenamefont
      {Bradley}, \citenamefont {Randall}, \citenamefont {Abobeih}, \citenamefont
      {Berrevoets}, \citenamefont {Degen}, \citenamefont {Bakker}, \citenamefont
      {Markham}, \citenamefont {Twitchen},\ and\ \citenamefont
      {Taminiau}}]{bradley2019tenqubit}%
      \BibitemOpen
      \bibfield  {author} {\bibinfo {author} {\bibfnamefont {C.~E.}\ \bibnamefont
      {Bradley}}, \bibinfo {author} {\bibfnamefont {J.}~\bibnamefont {Randall}},
      \bibinfo {author} {\bibfnamefont {M.~H.}\ \bibnamefont {Abobeih}}, \bibinfo
      {author} {\bibfnamefont {R.}~\bibnamefont {Berrevoets}}, \bibinfo {author}
      {\bibfnamefont {M.}~\bibnamefont {Degen}}, \bibinfo {author} {\bibfnamefont
      {M.~A.}\ \bibnamefont {Bakker}}, \bibinfo {author} {\bibfnamefont
      {M.}~\bibnamefont {Markham}}, \bibinfo {author} {\bibfnamefont
      {D.}~\bibnamefont {Twitchen}},\ and\ \bibinfo {author} {\bibfnamefont
      {T.~H.}\ \bibnamefont {Taminiau}},\ }\bibfield  {title} {\bibinfo {title} {A
      ten-qubit solid-state spin register with quantum memory up to one minute},\
      }\href {https://doi.org/10.1103/physrevx.9.031045} {\bibfield  {journal}
      {\bibinfo  {journal} {Physical Review X}\ }\textbf {\bibinfo {volume} {9}},\
      \bibinfo {pages} {031045} (\bibinfo {year} {2019})}\BibitemShut {NoStop}%
    \bibitem [{\citenamefont {Johnson}\ \emph {et~al.}(2015)\citenamefont
      {Johnson}, \citenamefont {Dolan}, \citenamefont {Grange}, \citenamefont
      {Trichet}, \citenamefont {Hornecker}, \citenamefont {Chen}, \citenamefont
      {Weng}, \citenamefont {Hughes}, \citenamefont {Watt}, \citenamefont
      {Auff{\`e}ves} \emph {et~al.}}]{johnson2015tunable}%
      \BibitemOpen
      \bibfield  {author} {\bibinfo {author} {\bibfnamefont {S.}~\bibnamefont
      {Johnson}}, \bibinfo {author} {\bibfnamefont {P.}~\bibnamefont {Dolan}},
      \bibinfo {author} {\bibfnamefont {T.}~\bibnamefont {Grange}}, \bibinfo
      {author} {\bibfnamefont {A.}~\bibnamefont {Trichet}}, \bibinfo {author}
      {\bibfnamefont {G.}~\bibnamefont {Hornecker}}, \bibinfo {author}
      {\bibfnamefont {Y.-C.}\ \bibnamefont {Chen}}, \bibinfo {author}
      {\bibfnamefont {L.}~\bibnamefont {Weng}}, \bibinfo {author} {\bibfnamefont
      {G.}~\bibnamefont {Hughes}}, \bibinfo {author} {\bibfnamefont
      {A.}~\bibnamefont {Watt}}, \bibinfo {author} {\bibfnamefont {A.}~\bibnamefont
      {Auff{\`e}ves}}, \emph {et~al.},\ }\bibfield  {title} {\bibinfo {title}
      {Tunable cavity coupling of the zero phonon line of a nitrogen-vacancy defect
      in diamond},\ }\href {https://doi.org/10.1088/1367-2630/17/12/122003}
      {\bibfield  {journal} {\bibinfo  {journal} {New Journal of Physics}\ }\textbf
      {\bibinfo {volume} {17}},\ \bibinfo {pages} {122003} (\bibinfo {year}
      {2015})}\BibitemShut {NoStop}%
    \bibitem [{\citenamefont {Pompili}\ \emph {et~al.}(2021)\citenamefont
      {Pompili}, \citenamefont {Hermans}, \citenamefont {Baier}, \citenamefont
      {Beukers}, \citenamefont {Humphreys}, \citenamefont {Schouten}, \citenamefont
      {Vermeulen}, \citenamefont {Tiggelman}, \citenamefont {dos Santos~Martins},
      \citenamefont {Dirkse} \emph {et~al.}}]{pompili2021realization}%
      \BibitemOpen
      \bibfield  {author} {\bibinfo {author} {\bibfnamefont {M.}~\bibnamefont
      {Pompili}}, \bibinfo {author} {\bibfnamefont {S.~L.}\ \bibnamefont
      {Hermans}}, \bibinfo {author} {\bibfnamefont {S.}~\bibnamefont {Baier}},
      \bibinfo {author} {\bibfnamefont {H.~K.}\ \bibnamefont {Beukers}}, \bibinfo
      {author} {\bibfnamefont {P.~C.}\ \bibnamefont {Humphreys}}, \bibinfo {author}
      {\bibfnamefont {R.~N.}\ \bibnamefont {Schouten}}, \bibinfo {author}
      {\bibfnamefont {R.~F.}\ \bibnamefont {Vermeulen}}, \bibinfo {author}
      {\bibfnamefont {M.~J.}\ \bibnamefont {Tiggelman}}, \bibinfo {author}
      {\bibfnamefont {L.}~\bibnamefont {dos Santos~Martins}}, \bibinfo {author}
      {\bibfnamefont {B.}~\bibnamefont {Dirkse}}, \emph {et~al.},\ }\bibfield
      {title} {\bibinfo {title} {Realization of a multinode quantum network of
      remote solid-state qubits},\ }\href {https://doi.org/10.1364/qim.2021.m2a.2}
      {\bibfield  {journal} {\bibinfo  {journal} {Science}\ }\textbf {\bibinfo
      {volume} {372}},\ \bibinfo {pages} {259} (\bibinfo {year}
      {2021})}\BibitemShut {NoStop}%
    \bibitem [{\citenamefont {Hermans}\ \emph {et~al.}(2022)\citenamefont
      {Hermans}, \citenamefont {Pompili}, \citenamefont {Beukers}, \citenamefont
      {Baier}, \citenamefont {Borregaard},\ and\ \citenamefont
      {Hanson}}]{hermans2022qubit}%
      \BibitemOpen
      \bibfield  {author} {\bibinfo {author} {\bibfnamefont {S.}~\bibnamefont
      {Hermans}}, \bibinfo {author} {\bibfnamefont {M.}~\bibnamefont {Pompili}},
      \bibinfo {author} {\bibfnamefont {H.}~\bibnamefont {Beukers}}, \bibinfo
      {author} {\bibfnamefont {S.}~\bibnamefont {Baier}}, \bibinfo {author}
      {\bibfnamefont {J.}~\bibnamefont {Borregaard}},\ and\ \bibinfo {author}
      {\bibfnamefont {R.}~\bibnamefont {Hanson}},\ }\bibfield  {title} {\bibinfo
      {title} {Qubit teleportation between non-neighbouring nodes in a quantum
      network},\ }\href {https://doi.org/10.1038/s41586-022-04697-y} {\bibfield
      {journal} {\bibinfo  {journal} {Nature}\ }\textbf {\bibinfo {volume} {605}},\
      \bibinfo {pages} {663} (\bibinfo {year} {2022})}\BibitemShut {NoStop}%
    \bibitem [{\citenamefont {Knall}\ \emph {et~al.}(2022)\citenamefont {Knall},
      \citenamefont {Knaut}, \citenamefont {Bekenstein}, \citenamefont {Assumpcao},
      \citenamefont {Stroganov}, \citenamefont {Gong}, \citenamefont {Huan},
      \citenamefont {Stas}, \citenamefont {Machielse}, \citenamefont {Chalupnik}
      \emph {et~al.}}]{knall2022efficientsinglephotons}%
      \BibitemOpen
      \bibfield  {author} {\bibinfo {author} {\bibfnamefont {E.~N.}\ \bibnamefont
      {Knall}}, \bibinfo {author} {\bibfnamefont {C.~M.}\ \bibnamefont {Knaut}},
      \bibinfo {author} {\bibfnamefont {R.}~\bibnamefont {Bekenstein}}, \bibinfo
      {author} {\bibfnamefont {D.~R.}\ \bibnamefont {Assumpcao}}, \bibinfo {author}
      {\bibfnamefont {P.~L.}\ \bibnamefont {Stroganov}}, \bibinfo {author}
      {\bibfnamefont {W.}~\bibnamefont {Gong}}, \bibinfo {author} {\bibfnamefont
      {Y.~Q.}\ \bibnamefont {Huan}}, \bibinfo {author} {\bibfnamefont {P.-J.}\
      \bibnamefont {Stas}}, \bibinfo {author} {\bibfnamefont {B.}~\bibnamefont
      {Machielse}}, \bibinfo {author} {\bibfnamefont {M.}~\bibnamefont
      {Chalupnik}}, \emph {et~al.},\ }\bibfield  {title} {\bibinfo {title}
      {Efficient source of shaped single photons based on an integrated diamond
      nanophotonic system},\ }\href
      {https://doi.org/10.1103/physrevlett.129.053603} {\bibfield  {journal}
      {\bibinfo  {journal} {Physical Review Letters}\ }\textbf {\bibinfo {volume}
      {129}},\ \bibinfo {pages} {053603} (\bibinfo {year} {2022})}\BibitemShut
      {NoStop}%
    \bibitem [{\citenamefont {Bhaskar}\ \emph {et~al.}(2017)\citenamefont
      {Bhaskar}, \citenamefont {Sukachev}, \citenamefont {Sipahigil}, \citenamefont
      {Evans}, \citenamefont {Burek}, \citenamefont {Nguyen}, \citenamefont
      {Rogers}, \citenamefont {Siyushev}, \citenamefont {Metsch}, \citenamefont
      {Park} \emph {et~al.}}]{bhaskar2017quantum}%
      \BibitemOpen
      \bibfield  {author} {\bibinfo {author} {\bibfnamefont {M.~K.}\ \bibnamefont
      {Bhaskar}}, \bibinfo {author} {\bibfnamefont {D.~D.}\ \bibnamefont
      {Sukachev}}, \bibinfo {author} {\bibfnamefont {A.}~\bibnamefont {Sipahigil}},
      \bibinfo {author} {\bibfnamefont {R.~E.}\ \bibnamefont {Evans}}, \bibinfo
      {author} {\bibfnamefont {M.~J.}\ \bibnamefont {Burek}}, \bibinfo {author}
      {\bibfnamefont {C.~T.}\ \bibnamefont {Nguyen}}, \bibinfo {author}
      {\bibfnamefont {L.~J.}\ \bibnamefont {Rogers}}, \bibinfo {author}
      {\bibfnamefont {P.}~\bibnamefont {Siyushev}}, \bibinfo {author}
      {\bibfnamefont {M.~H.}\ \bibnamefont {Metsch}}, \bibinfo {author}
      {\bibfnamefont {H.}~\bibnamefont {Park}}, \emph {et~al.},\ }\bibfield
      {title} {\bibinfo {title} {Quantum nonlinear optics with a germanium-vacancy
      color center in a nanoscale diamond waveguide},\ }\href
      {https://doi.org/10.1103/physrevlett.118.223603} {\bibfield  {journal}
      {\bibinfo  {journal} {Physical Review Letters}\ }\textbf {\bibinfo {volume}
      {118}},\ \bibinfo {pages} {223603} (\bibinfo {year} {2017})}\BibitemShut
      {NoStop}%
    \bibitem [{\citenamefont {Mart{\'\i}nez}\ \emph {et~al.}(2022)\citenamefont
      {Mart{\'\i}nez}, \citenamefont {Parker}, \citenamefont {Chen}, \citenamefont
      {Purser}, \citenamefont {Li}, \citenamefont {Michaels}, \citenamefont
      {Stramma}, \citenamefont {Debroux}, \citenamefont {Harris}, \citenamefont
      {Appel} \emph {et~al.}}]{martinez2022photonic}%
      \BibitemOpen
      \bibfield  {author} {\bibinfo {author} {\bibfnamefont {J.~A.}\ \bibnamefont
      {Mart{\'\i}nez}}, \bibinfo {author} {\bibfnamefont {R.~A.}\ \bibnamefont
      {Parker}}, \bibinfo {author} {\bibfnamefont {K.~C.}\ \bibnamefont {Chen}},
      \bibinfo {author} {\bibfnamefont {C.~M.}\ \bibnamefont {Purser}}, \bibinfo
      {author} {\bibfnamefont {L.}~\bibnamefont {Li}}, \bibinfo {author}
      {\bibfnamefont {C.~P.}\ \bibnamefont {Michaels}}, \bibinfo {author}
      {\bibfnamefont {A.~M.}\ \bibnamefont {Stramma}}, \bibinfo {author}
      {\bibfnamefont {R.}~\bibnamefont {Debroux}}, \bibinfo {author} {\bibfnamefont
      {I.~B.}\ \bibnamefont {Harris}}, \bibinfo {author} {\bibfnamefont {M.~H.}\
      \bibnamefont {Appel}}, \emph {et~al.},\ }\bibfield  {title} {\bibinfo {title}
      {Photonic indistinguishability of the tin-vacancy center in nanostructured
      diamond},\ }\href {https://doi.org/10.1103/physrevlett.129.173603} {\bibfield
       {journal} {\bibinfo  {journal} {Physical Review Letters}\ }\textbf {\bibinfo
      {volume} {129}},\ \bibinfo {pages} {173603} (\bibinfo {year}
      {2022})}\BibitemShut {NoStop}%
    \bibitem [{\citenamefont {Sukachev}\ \emph {et~al.}(2017)\citenamefont
      {Sukachev}, \citenamefont {Sipahigil}, \citenamefont {Nguyen}, \citenamefont
      {Bhaskar}, \citenamefont {Evans}, \citenamefont {Jelezko},\ and\
      \citenamefont {Lukin}}]{sukachev2017silicon}%
      \BibitemOpen
      \bibfield  {author} {\bibinfo {author} {\bibfnamefont {D.~D.}\ \bibnamefont
      {Sukachev}}, \bibinfo {author} {\bibfnamefont {A.}~\bibnamefont {Sipahigil}},
      \bibinfo {author} {\bibfnamefont {C.~T.}\ \bibnamefont {Nguyen}}, \bibinfo
      {author} {\bibfnamefont {M.~K.}\ \bibnamefont {Bhaskar}}, \bibinfo {author}
      {\bibfnamefont {R.~E.}\ \bibnamefont {Evans}}, \bibinfo {author}
      {\bibfnamefont {F.}~\bibnamefont {Jelezko}},\ and\ \bibinfo {author}
      {\bibfnamefont {M.~D.}\ \bibnamefont {Lukin}},\ }\bibfield  {title} {\bibinfo
      {title} {Silicon-vacancy spin qubit in diamond: a quantum memory exceeding 10
      ms with single-shot state readout},\ }\href
      {https://doi.org/10.1103/physrevlett.119.223602} {\bibfield  {journal}
      {\bibinfo  {journal} {Physical Review Letters}\ }\textbf {\bibinfo {volume}
      {119}},\ \bibinfo {pages} {223602} (\bibinfo {year} {2017})}\BibitemShut
      {NoStop}%
    \bibitem [{\citenamefont {Debroux}\ \emph {et~al.}(2021)\citenamefont
      {Debroux}, \citenamefont {Michaels}, \citenamefont {Purser}, \citenamefont
      {Wan}, \citenamefont {Trusheim}, \citenamefont {Mart{\'\i}nez}, \citenamefont
      {Parker}, \citenamefont {Stramma}, \citenamefont {Chen}, \citenamefont
      {de~Santis} \emph {et~al.}}]{debroux2021quantum}%
      \BibitemOpen
      \bibfield  {author} {\bibinfo {author} {\bibfnamefont {R.}~\bibnamefont
      {Debroux}}, \bibinfo {author} {\bibfnamefont {C.~P.}\ \bibnamefont
      {Michaels}}, \bibinfo {author} {\bibfnamefont {C.~M.}\ \bibnamefont
      {Purser}}, \bibinfo {author} {\bibfnamefont {N.}~\bibnamefont {Wan}},
      \bibinfo {author} {\bibfnamefont {M.~E.}\ \bibnamefont {Trusheim}}, \bibinfo
      {author} {\bibfnamefont {J.~A.}\ \bibnamefont {Mart{\'\i}nez}}, \bibinfo
      {author} {\bibfnamefont {R.~A.}\ \bibnamefont {Parker}}, \bibinfo {author}
      {\bibfnamefont {A.~M.}\ \bibnamefont {Stramma}}, \bibinfo {author}
      {\bibfnamefont {K.~C.}\ \bibnamefont {Chen}}, \bibinfo {author}
      {\bibfnamefont {L.}~\bibnamefont {de~Santis}}, \emph {et~al.},\ }\bibfield
      {title} {\bibinfo {title} {Quantum control of the tin-vacancy spin qubit in
      diamond},\ }\href {https://doi.org/10.1364/qim.2021.f2a.2} {\bibfield
      {journal} {\bibinfo  {journal} {Physical Review X}\ }\textbf {\bibinfo
      {volume} {11}},\ \bibinfo {pages} {041041} (\bibinfo {year}
      {2021})}\BibitemShut {NoStop}%
    \bibitem [{\citenamefont {Nguyen}\ \emph {et~al.}(2019)\citenamefont {Nguyen},
      \citenamefont {Sukachev}, \citenamefont {Bhaskar}, \citenamefont {Machielse},
      \citenamefont {Levonian}, \citenamefont {Knall}, \citenamefont {Stroganov},
      \citenamefont {Chia}, \citenamefont {Burek}, \citenamefont {Riedinger} \emph
      {et~al.}}]{nguyen2019integrated}%
      \BibitemOpen
      \bibfield  {author} {\bibinfo {author} {\bibfnamefont {C.}~\bibnamefont
      {Nguyen}}, \bibinfo {author} {\bibfnamefont {D.}~\bibnamefont {Sukachev}},
      \bibinfo {author} {\bibfnamefont {M.}~\bibnamefont {Bhaskar}}, \bibinfo
      {author} {\bibfnamefont {B.}~\bibnamefont {Machielse}}, \bibinfo {author}
      {\bibfnamefont {D.}~\bibnamefont {Levonian}}, \bibinfo {author}
      {\bibfnamefont {E.}~\bibnamefont {Knall}}, \bibinfo {author} {\bibfnamefont
      {P.}~\bibnamefont {Stroganov}}, \bibinfo {author} {\bibfnamefont
      {C.}~\bibnamefont {Chia}}, \bibinfo {author} {\bibfnamefont {M.}~\bibnamefont
      {Burek}}, \bibinfo {author} {\bibfnamefont {R.}~\bibnamefont {Riedinger}},
      \emph {et~al.},\ }\bibfield  {title} {\bibinfo {title} {An integrated
      nanophotonic quantum register based on silicon-vacancy spins in diamond},\
      }\href {https://doi.org/10.1103/physrevb.100.165428} {\bibfield  {journal}
      {\bibinfo  {journal} {Physical Review B}\ }\textbf {\bibinfo {volume}
      {100}},\ \bibinfo {pages} {165428} (\bibinfo {year} {2019})}\BibitemShut
      {NoStop}%
    \bibitem [{\citenamefont {Stas}\ \emph {et~al.}(2022)\citenamefont {Stas},
      \citenamefont {Huan}, \citenamefont {Machielse}, \citenamefont {Knall},
      \citenamefont {Suleymanzade}, \citenamefont {Pingault}, \citenamefont
      {Sutula}, \citenamefont {Ding}, \citenamefont {Knaut}, \citenamefont
      {Assumpcao} \emph {et~al.}}]{stas2022sivnuclear}%
      \BibitemOpen
      \bibfield  {author} {\bibinfo {author} {\bibfnamefont {P.-J.}\ \bibnamefont
      {Stas}}, \bibinfo {author} {\bibfnamefont {Y.~Q.}\ \bibnamefont {Huan}},
      \bibinfo {author} {\bibfnamefont {B.}~\bibnamefont {Machielse}}, \bibinfo
      {author} {\bibfnamefont {E.~N.}\ \bibnamefont {Knall}}, \bibinfo {author}
      {\bibfnamefont {A.}~\bibnamefont {Suleymanzade}}, \bibinfo {author}
      {\bibfnamefont {B.}~\bibnamefont {Pingault}}, \bibinfo {author}
      {\bibfnamefont {M.}~\bibnamefont {Sutula}}, \bibinfo {author} {\bibfnamefont
      {S.~W.}\ \bibnamefont {Ding}}, \bibinfo {author} {\bibfnamefont {C.~M.}\
      \bibnamefont {Knaut}}, \bibinfo {author} {\bibfnamefont {D.~R.}\ \bibnamefont
      {Assumpcao}}, \emph {et~al.},\ }\bibfield  {title} {\bibinfo {title} {Robust
      multi-qubit quantum network node with integrated error detection},\ }\href
      {https://doi.org/10.26226/m.6275705866d5dcf63a311413} {\bibfield  {journal}
      {\bibinfo  {journal} {Science}\ }\textbf {\bibinfo {volume} {378}},\ \bibinfo
      {pages} {557} (\bibinfo {year} {2022})}\BibitemShut {NoStop}%
    \bibitem [{\citenamefont {Rugar}\ \emph
      {et~al.}(2020{\natexlab{a}})\citenamefont {Rugar}, \citenamefont {Dory},
      \citenamefont {Aghaeimeibodi}, \citenamefont {Lu}, \citenamefont {Sun},
      \citenamefont {Mishra}, \citenamefont {Shen}, \citenamefont {Melosh},\ and\
      \citenamefont {Vuckovic}}]{rugar2020narrow}%
      \BibitemOpen
      \bibfield  {author} {\bibinfo {author} {\bibfnamefont {A.~E.}\ \bibnamefont
      {Rugar}}, \bibinfo {author} {\bibfnamefont {C.}~\bibnamefont {Dory}},
      \bibinfo {author} {\bibfnamefont {S.}~\bibnamefont {Aghaeimeibodi}}, \bibinfo
      {author} {\bibfnamefont {H.}~\bibnamefont {Lu}}, \bibinfo {author}
      {\bibfnamefont {S.}~\bibnamefont {Sun}}, \bibinfo {author} {\bibfnamefont
      {S.~D.}\ \bibnamefont {Mishra}}, \bibinfo {author} {\bibfnamefont {Z.-X.}\
      \bibnamefont {Shen}}, \bibinfo {author} {\bibfnamefont {N.~A.}\ \bibnamefont
      {Melosh}},\ and\ \bibinfo {author} {\bibfnamefont {J.}~\bibnamefont
      {Vuckovic}},\ }\bibfield  {title} {\bibinfo {title} {Narrow-linewidth
      tin-vacancy centers in a diamond waveguide},\ }\href
      {https://doi.org/10.1021/acsphotonics.0c00833} {\bibfield  {journal}
      {\bibinfo  {journal} {ACS Photonics}\ }\textbf {\bibinfo {volume} {7}},\
      \bibinfo {pages} {2356} (\bibinfo {year} {2020}{\natexlab{a}})}\BibitemShut
      {NoStop}%
    \bibitem [{\citenamefont {Bradac}\ \emph {et~al.}(2019)\citenamefont {Bradac},
      \citenamefont {Gao}, \citenamefont {Forneris}, \citenamefont {Trusheim},\
      and\ \citenamefont {Aharonovich}}]{bradac2019quantum}%
      \BibitemOpen
      \bibfield  {author} {\bibinfo {author} {\bibfnamefont {C.}~\bibnamefont
      {Bradac}}, \bibinfo {author} {\bibfnamefont {W.}~\bibnamefont {Gao}},
      \bibinfo {author} {\bibfnamefont {J.}~\bibnamefont {Forneris}}, \bibinfo
      {author} {\bibfnamefont {M.~E.}\ \bibnamefont {Trusheim}},\ and\ \bibinfo
      {author} {\bibfnamefont {I.}~\bibnamefont {Aharonovich}},\ }\bibfield
      {title} {\bibinfo {title} {Quantum nanophotonics with group iv defects in
      diamond},\ }\href {https://doi.org/10.1038/s41467-019-13332-w} {\bibfield
      {journal} {\bibinfo  {journal} {Nature Communications}\ }\textbf {\bibinfo
      {volume} {10}},\ \bibinfo {pages} {5625} (\bibinfo {year}
      {2019})}\BibitemShut {NoStop}%
    \bibitem [{\citenamefont {Koch}\ \emph {et~al.}(2022)\citenamefont {Koch},
      \citenamefont {Hoese}, \citenamefont {Bharadwaj}, \citenamefont {Lang},
      \citenamefont {Hadden}, \citenamefont {Ramponi}, \citenamefont {Jelezko},
      \citenamefont {Eaton},\ and\ \citenamefont {Kubanek}}]{koch2022super}%
      \BibitemOpen
      \bibfield  {author} {\bibinfo {author} {\bibfnamefont {M.~K.}\ \bibnamefont
      {Koch}}, \bibinfo {author} {\bibfnamefont {M.}~\bibnamefont {Hoese}},
      \bibinfo {author} {\bibfnamefont {V.}~\bibnamefont {Bharadwaj}}, \bibinfo
      {author} {\bibfnamefont {J.}~\bibnamefont {Lang}}, \bibinfo {author}
      {\bibfnamefont {J.~P.}\ \bibnamefont {Hadden}}, \bibinfo {author}
      {\bibfnamefont {R.}~\bibnamefont {Ramponi}}, \bibinfo {author} {\bibfnamefont
      {F.}~\bibnamefont {Jelezko}}, \bibinfo {author} {\bibfnamefont {S.~M.}\
      \bibnamefont {Eaton}},\ and\ \bibinfo {author} {\bibfnamefont
      {A.}~\bibnamefont {Kubanek}},\ }\bibfield  {title} {\bibinfo {title}
      {Super-poissonian light statistics from individual silicon vacancy centers
      coupled to a laser-written diamond waveguide},\ }\href
      {https://doi.org/10.1021/acsphotonics.2c00774} {\bibfield  {journal}
      {\bibinfo  {journal} {ACS Photonics}\ }\textbf {\bibinfo {volume} {9}},\
      \bibinfo {pages} {3366} (\bibinfo {year} {2022})}\BibitemShut {NoStop}%
    \bibitem [{\citenamefont {Evans}\ \emph {et~al.}(2018)\citenamefont {Evans},
      \citenamefont {Bhaskar}, \citenamefont {Sukachev}, \citenamefont {Nguyen},
      \citenamefont {Sipahigil}, \citenamefont {Burek}, \citenamefont {Machielse},
      \citenamefont {Zhang}, \citenamefont {Zibrov}, \citenamefont {Bielejec} \emph
      {et~al.}}]{evans2018photon}%
      \BibitemOpen
      \bibfield  {author} {\bibinfo {author} {\bibfnamefont {R.~E.}\ \bibnamefont
      {Evans}}, \bibinfo {author} {\bibfnamefont {M.~K.}\ \bibnamefont {Bhaskar}},
      \bibinfo {author} {\bibfnamefont {D.~D.}\ \bibnamefont {Sukachev}}, \bibinfo
      {author} {\bibfnamefont {C.~T.}\ \bibnamefont {Nguyen}}, \bibinfo {author}
      {\bibfnamefont {A.}~\bibnamefont {Sipahigil}}, \bibinfo {author}
      {\bibfnamefont {M.~J.}\ \bibnamefont {Burek}}, \bibinfo {author}
      {\bibfnamefont {B.}~\bibnamefont {Machielse}}, \bibinfo {author}
      {\bibfnamefont {G.~H.}\ \bibnamefont {Zhang}}, \bibinfo {author}
      {\bibfnamefont {A.~S.}\ \bibnamefont {Zibrov}}, \bibinfo {author}
      {\bibfnamefont {E.}~\bibnamefont {Bielejec}}, \emph {et~al.},\ }\bibfield
      {title} {\bibinfo {title} {Photon-mediated interactions between quantum
      emitters in a diamond nanocavity},\ }\href
      {https://doi.org/10.1126/science.aau4691} {\bibfield  {journal} {\bibinfo
      {journal} {Science}\ }\textbf {\bibinfo {volume} {362}},\ \bibinfo {pages}
      {662} (\bibinfo {year} {2018})}\BibitemShut {NoStop}%
    \bibitem [{\citenamefont {Sipahigil}\ \emph {et~al.}(2016)\citenamefont
      {Sipahigil}, \citenamefont {Evans}, \citenamefont {Sukachev}, \citenamefont
      {Burek}, \citenamefont {Borregaard}, \citenamefont {Bhaskar}, \citenamefont
      {Nguyen}, \citenamefont {Pacheco}, \citenamefont {Atikian}, \citenamefont
      {Meuwly} \emph {et~al.}}]{sipahigil2016integrated}%
      \BibitemOpen
      \bibfield  {author} {\bibinfo {author} {\bibfnamefont {A.}~\bibnamefont
      {Sipahigil}}, \bibinfo {author} {\bibfnamefont {R.~E.}\ \bibnamefont
      {Evans}}, \bibinfo {author} {\bibfnamefont {D.~D.}\ \bibnamefont {Sukachev}},
      \bibinfo {author} {\bibfnamefont {M.~J.}\ \bibnamefont {Burek}}, \bibinfo
      {author} {\bibfnamefont {J.}~\bibnamefont {Borregaard}}, \bibinfo {author}
      {\bibfnamefont {M.~K.}\ \bibnamefont {Bhaskar}}, \bibinfo {author}
      {\bibfnamefont {C.~T.}\ \bibnamefont {Nguyen}}, \bibinfo {author}
      {\bibfnamefont {J.~L.}\ \bibnamefont {Pacheco}}, \bibinfo {author}
      {\bibfnamefont {H.~A.}\ \bibnamefont {Atikian}}, \bibinfo {author}
      {\bibfnamefont {C.}~\bibnamefont {Meuwly}}, \emph {et~al.},\ }\bibfield
      {title} {\bibinfo {title} {An integrated diamond nanophotonics platform for
      quantum-optical networks},\ }\href {https://doi.org/10.1126/science.aah6875}
      {\bibfield  {journal} {\bibinfo  {journal} {Science}\ }\textbf {\bibinfo
      {volume} {354}},\ \bibinfo {pages} {847} (\bibinfo {year}
      {2016})}\BibitemShut {NoStop}%
    \bibitem [{\citenamefont {Iwasaki}\ \emph {et~al.}(2017)\citenamefont
      {Iwasaki}, \citenamefont {Miyamoto}, \citenamefont {Taniguchi}, \citenamefont
      {Siyushev}, \citenamefont {Metsch}, \citenamefont {Jelezko},\ and\
      \citenamefont {Hatano}}]{iwasaki2017tin}%
      \BibitemOpen
      \bibfield  {author} {\bibinfo {author} {\bibfnamefont {T.}~\bibnamefont
      {Iwasaki}}, \bibinfo {author} {\bibfnamefont {Y.}~\bibnamefont {Miyamoto}},
      \bibinfo {author} {\bibfnamefont {T.}~\bibnamefont {Taniguchi}}, \bibinfo
      {author} {\bibfnamefont {P.}~\bibnamefont {Siyushev}}, \bibinfo {author}
      {\bibfnamefont {M.~H.}\ \bibnamefont {Metsch}}, \bibinfo {author}
      {\bibfnamefont {F.}~\bibnamefont {Jelezko}},\ and\ \bibinfo {author}
      {\bibfnamefont {M.}~\bibnamefont {Hatano}},\ }\bibfield  {title} {\bibinfo
      {title} {Tin-vacancy quantum emitters in diamond},\ }\href
      {https://doi.org/10.1103/physrevlett.119.253601} {\bibfield  {journal}
      {\bibinfo  {journal} {Physical Review Letters}\ }\textbf {\bibinfo {volume}
      {119}},\ \bibinfo {pages} {253601} (\bibinfo {year} {2017})}\BibitemShut
      {NoStop}%
    \bibitem [{\citenamefont {Trusheim}\ \emph {et~al.}(2020)\citenamefont
      {Trusheim}, \citenamefont {Pingault}, \citenamefont {Wan}, \citenamefont
      {G{\"u}ndo{\u{g}}an}, \citenamefont {De~Santis}, \citenamefont {Debroux},
      \citenamefont {Gangloff}, \citenamefont {Purser}, \citenamefont {Chen},
      \citenamefont {Walsh} \emph {et~al.}}]{trusheim2020transformlimited}%
      \BibitemOpen
      \bibfield  {author} {\bibinfo {author} {\bibfnamefont {M.~E.}\ \bibnamefont
      {Trusheim}}, \bibinfo {author} {\bibfnamefont {B.}~\bibnamefont {Pingault}},
      \bibinfo {author} {\bibfnamefont {N.~H.}\ \bibnamefont {Wan}}, \bibinfo
      {author} {\bibfnamefont {M.}~\bibnamefont {G{\"u}ndo{\u{g}}an}}, \bibinfo
      {author} {\bibfnamefont {L.}~\bibnamefont {De~Santis}}, \bibinfo {author}
      {\bibfnamefont {R.}~\bibnamefont {Debroux}}, \bibinfo {author} {\bibfnamefont
      {D.}~\bibnamefont {Gangloff}}, \bibinfo {author} {\bibfnamefont
      {C.}~\bibnamefont {Purser}}, \bibinfo {author} {\bibfnamefont {K.~C.}\
      \bibnamefont {Chen}}, \bibinfo {author} {\bibfnamefont {M.}~\bibnamefont
      {Walsh}}, \emph {et~al.},\ }\bibfield  {title} {\bibinfo {title}
      {Transform-limited photons from a coherent tin-vacancy spin in diamond},\
      }\href {https://doi.org/10.1103/physrevlett.124.023602} {\bibfield  {journal}
      {\bibinfo  {journal} {Physical Review Letters}\ }\textbf {\bibinfo {volume}
      {124}},\ \bibinfo {pages} {023602} (\bibinfo {year} {2020})}\BibitemShut
      {NoStop}%
    \bibitem [{\citenamefont {Le~Jeannic}\ \emph {et~al.}(2022)\citenamefont
      {Le~Jeannic}, \citenamefont {Tiranov}, \citenamefont {Carolan}, \citenamefont
      {Ramos}, \citenamefont {Wang}, \citenamefont {Appel}, \citenamefont {Scholz},
      \citenamefont {Wieck}, \citenamefont {Ludwig}, \citenamefont {Rotenberg}
      \emph {et~al.}}]{le2022dynamical}%
      \BibitemOpen
      \bibfield  {author} {\bibinfo {author} {\bibfnamefont {H.}~\bibnamefont
      {Le~Jeannic}}, \bibinfo {author} {\bibfnamefont {A.}~\bibnamefont {Tiranov}},
      \bibinfo {author} {\bibfnamefont {J.}~\bibnamefont {Carolan}}, \bibinfo
      {author} {\bibfnamefont {T.}~\bibnamefont {Ramos}}, \bibinfo {author}
      {\bibfnamefont {Y.}~\bibnamefont {Wang}}, \bibinfo {author} {\bibfnamefont
      {M.~H.}\ \bibnamefont {Appel}}, \bibinfo {author} {\bibfnamefont
      {S.}~\bibnamefont {Scholz}}, \bibinfo {author} {\bibfnamefont {A.~D.}\
      \bibnamefont {Wieck}}, \bibinfo {author} {\bibfnamefont {A.}~\bibnamefont
      {Ludwig}}, \bibinfo {author} {\bibfnamefont {N.}~\bibnamefont {Rotenberg}},
      \emph {et~al.},\ }\bibfield  {title} {\bibinfo {title} {Dynamical
      photon--photon interaction mediated by a quantum emitter},\ }\href
      {https://doi.org/10.21203/rs.3.rs-1167448/v1} {\bibfield  {journal} {\bibinfo
       {journal} {Nature Physics}\ }\textbf {\bibinfo {volume} {18}},\ \bibinfo
      {pages} {1191} (\bibinfo {year} {2022})}\BibitemShut {NoStop}%
    \bibitem [{\citenamefont {Wan}\ \emph {et~al.}(2020)\citenamefont {Wan},
      \citenamefont {Lu}, \citenamefont {Chen}, \citenamefont {Walsh},
      \citenamefont {Trusheim}, \citenamefont {De~Santis}, \citenamefont {Bersin},
      \citenamefont {Harris}, \citenamefont {Mouradian}, \citenamefont {Christen}
      \emph {et~al.}}]{wan2020large}%
      \BibitemOpen
      \bibfield  {author} {\bibinfo {author} {\bibfnamefont {N.~H.}\ \bibnamefont
      {Wan}}, \bibinfo {author} {\bibfnamefont {T.-J.}\ \bibnamefont {Lu}},
      \bibinfo {author} {\bibfnamefont {K.~C.}\ \bibnamefont {Chen}}, \bibinfo
      {author} {\bibfnamefont {M.~P.}\ \bibnamefont {Walsh}}, \bibinfo {author}
      {\bibfnamefont {M.~E.}\ \bibnamefont {Trusheim}}, \bibinfo {author}
      {\bibfnamefont {L.}~\bibnamefont {De~Santis}}, \bibinfo {author}
      {\bibfnamefont {E.~A.}\ \bibnamefont {Bersin}}, \bibinfo {author}
      {\bibfnamefont {I.~B.}\ \bibnamefont {Harris}}, \bibinfo {author}
      {\bibfnamefont {S.~L.}\ \bibnamefont {Mouradian}}, \bibinfo {author}
      {\bibfnamefont {I.~R.}\ \bibnamefont {Christen}}, \emph {et~al.},\ }\bibfield
       {title} {\bibinfo {title} {Large-scale integration of artificial atoms in
      hybrid photonic circuits},\ }\href
      {https://doi.org/10.1038/s41586-020-2441-3} {\bibfield  {journal} {\bibinfo
      {journal} {Nature}\ }\textbf {\bibinfo {volume} {583}},\ \bibinfo {pages}
      {226} (\bibinfo {year} {2020})}\BibitemShut {NoStop}%
    \bibitem [{\citenamefont {Douglas}\ \emph {et~al.}(2015)\citenamefont
      {Douglas}, \citenamefont {Habibian}, \citenamefont {Hung}, \citenamefont
      {Gorshkov}, \citenamefont {Kimble},\ and\ \citenamefont
      {Chang}}]{douglas2015quantum}%
      \BibitemOpen
      \bibfield  {author} {\bibinfo {author} {\bibfnamefont {J.~S.}\ \bibnamefont
      {Douglas}}, \bibinfo {author} {\bibfnamefont {H.}~\bibnamefont {Habibian}},
      \bibinfo {author} {\bibfnamefont {C.-L.}\ \bibnamefont {Hung}}, \bibinfo
      {author} {\bibfnamefont {A.~V.}\ \bibnamefont {Gorshkov}}, \bibinfo {author}
      {\bibfnamefont {H.~J.}\ \bibnamefont {Kimble}},\ and\ \bibinfo {author}
      {\bibfnamefont {D.~E.}\ \bibnamefont {Chang}},\ }\bibfield  {title} {\bibinfo
      {title} {Quantum many-body models with cold atoms coupled to photonic
      crystals},\ }\href {https://doi.org/10.1038/nphoton.2015.57} {\bibfield
      {journal} {\bibinfo  {journal} {Nature Photonics}\ }\textbf {\bibinfo
      {volume} {9}},\ \bibinfo {pages} {326} (\bibinfo {year} {2015})}\BibitemShut
      {NoStop}%
    \bibitem [{\citenamefont {Kim}\ and\ \citenamefont
      {Noh}(2020)}]{kim2020analytical}%
      \BibitemOpen
      \bibfield  {author} {\bibinfo {author} {\bibfnamefont {J.}~\bibnamefont
      {Kim}}\ and\ \bibinfo {author} {\bibfnamefont {H.-R.}\ \bibnamefont {Noh}},\
      }\bibfield  {title} {\bibinfo {title} {Analytical solutions of the
      second-order correlation function in resonance fluorescence for two-level
      systems},\ }\href {https://doi.org/10.3938/jkps.76.463} {\bibfield  {journal}
      {\bibinfo  {journal} {Journal of the Korean Physical Society}\ }\textbf
      {\bibinfo {volume} {76}},\ \bibinfo {pages} {463} (\bibinfo {year}
      {2020})}\BibitemShut {NoStop}%
    \bibitem [{\citenamefont {Michaels}\ \emph {et~al.}(2021)\citenamefont
      {Michaels}, \citenamefont {Mart{\'\i}nez}, \citenamefont {Debroux},
      \citenamefont {Parker}, \citenamefont {Stramma}, \citenamefont {Huber},
      \citenamefont {Purser}, \citenamefont {Atat{\"u}re},\ and\ \citenamefont
      {Gangloff}}]{michaels2021multidimensional}%
      \BibitemOpen
      \bibfield  {author} {\bibinfo {author} {\bibfnamefont {C.~P.}\ \bibnamefont
      {Michaels}}, \bibinfo {author} {\bibfnamefont {J.~A.}\ \bibnamefont
      {Mart{\'\i}nez}}, \bibinfo {author} {\bibfnamefont {R.}~\bibnamefont
      {Debroux}}, \bibinfo {author} {\bibfnamefont {R.~A.}\ \bibnamefont {Parker}},
      \bibinfo {author} {\bibfnamefont {A.~M.}\ \bibnamefont {Stramma}}, \bibinfo
      {author} {\bibfnamefont {L.~I.}\ \bibnamefont {Huber}}, \bibinfo {author}
      {\bibfnamefont {C.~M.}\ \bibnamefont {Purser}}, \bibinfo {author}
      {\bibfnamefont {M.}~\bibnamefont {Atat{\"u}re}},\ and\ \bibinfo {author}
      {\bibfnamefont {D.~A.}\ \bibnamefont {Gangloff}},\ }\bibfield  {title}
      {\bibinfo {title} {Multidimensional cluster states using a single spin-photon
      interface coupled strongly to an intrinsic nuclear register},\ }\href
      {https://doi.org/10.22331/q-2021-10-19-565} {\bibfield  {journal} {\bibinfo
      {journal} {Quantum}\ }\textbf {\bibinfo {volume} {5}},\ \bibinfo {pages}
      {565} (\bibinfo {year} {2021})}\BibitemShut {NoStop}%
    \bibitem [{\citenamefont {Mouradian}\ \emph {et~al.}(2017)\citenamefont
      {Mouradian}, \citenamefont {Wan}, \citenamefont {Schr{\"o}der},\ and\
      \citenamefont {Englund}}]{mouradian2017rectangular}%
      \BibitemOpen
      \bibfield  {author} {\bibinfo {author} {\bibfnamefont {S.}~\bibnamefont
      {Mouradian}}, \bibinfo {author} {\bibfnamefont {N.~H.}\ \bibnamefont {Wan}},
      \bibinfo {author} {\bibfnamefont {T.}~\bibnamefont {Schr{\"o}der}},\ and\
      \bibinfo {author} {\bibfnamefont {D.}~\bibnamefont {Englund}},\ }\bibfield
      {title} {\bibinfo {title} {Rectangular photonic crystal nanobeam cavities in
      bulk diamond},\ }\href {https://doi.org/10.1063/1.4992118} {\bibfield
      {journal} {\bibinfo  {journal} {Applied Physics Letters}\ }\textbf {\bibinfo
      {volume} {111}},\ \bibinfo {pages} {021103} (\bibinfo {year}
      {2017})}\BibitemShut {NoStop}%
    \bibitem [{\citenamefont {Hepp}\ \emph {et~al.}(2014)\citenamefont {Hepp},
      \citenamefont {M{\"u}ller}, \citenamefont {Waselowski}, \citenamefont
      {Becker}, \citenamefont {Pingault}, \citenamefont {Sternschulte},
      \citenamefont {Steinm{\"u}ller-Nethl}, \citenamefont {Gali}, \citenamefont
      {Maze}, \citenamefont {Atat{\"u}re} \emph {et~al.}}]{hepp2014electronic}%
      \BibitemOpen
      \bibfield  {author} {\bibinfo {author} {\bibfnamefont {C.}~\bibnamefont
      {Hepp}}, \bibinfo {author} {\bibfnamefont {T.}~\bibnamefont {M{\"u}ller}},
      \bibinfo {author} {\bibfnamefont {V.}~\bibnamefont {Waselowski}}, \bibinfo
      {author} {\bibfnamefont {J.~N.}\ \bibnamefont {Becker}}, \bibinfo {author}
      {\bibfnamefont {B.}~\bibnamefont {Pingault}}, \bibinfo {author}
      {\bibfnamefont {H.}~\bibnamefont {Sternschulte}}, \bibinfo {author}
      {\bibfnamefont {D.}~\bibnamefont {Steinm{\"u}ller-Nethl}}, \bibinfo {author}
      {\bibfnamefont {A.}~\bibnamefont {Gali}}, \bibinfo {author} {\bibfnamefont
      {J.~R.}\ \bibnamefont {Maze}}, \bibinfo {author} {\bibfnamefont
      {M.}~\bibnamefont {Atat{\"u}re}}, \emph {et~al.},\ }\bibfield  {title}
      {\bibinfo {title} {Electronic structure of the silicon vacancy color center
      in diamond},\ }\href {https://doi.org/10.1103/physrevlett.112.036405}
      {\bibfield  {journal} {\bibinfo  {journal} {Physical Review Letters}\
      }\textbf {\bibinfo {volume} {112}},\ \bibinfo {pages} {036405} (\bibinfo
      {year} {2014})}\BibitemShut {NoStop}%
    \bibitem [{\citenamefont {McConnell}\ and\ \citenamefont
      {Robertson}(1958)}]{mcconnell1958isotropic}%
      \BibitemOpen
      \bibfield  {author} {\bibinfo {author} {\bibfnamefont {H.~M.}\ \bibnamefont
      {McConnell}}\ and\ \bibinfo {author} {\bibfnamefont {R.~E.}\ \bibnamefont
      {Robertson}},\ }\bibfield  {title} {\bibinfo {title} {Isotropic nuclear
      resonance shifts},\ }\href {https://doi.org/10.1063/1.1744723} {\bibfield
      {journal} {\bibinfo  {journal} {The Journal of Chemical Physics}\ }\textbf
      {\bibinfo {volume} {29}},\ \bibinfo {pages} {1361} (\bibinfo {year}
      {1958})}\BibitemShut {NoStop}%
    \bibitem [{\citenamefont {Laaksonen}\ and\ \citenamefont
      {Wasylishen}(1995)}]{laaksonen1995absolute}%
      \BibitemOpen
      \bibfield  {author} {\bibinfo {author} {\bibfnamefont {A.}~\bibnamefont
      {Laaksonen}}\ and\ \bibinfo {author} {\bibfnamefont {R.~E.}\ \bibnamefont
      {Wasylishen}},\ }\bibfield  {title} {\bibinfo {title} {An absolute chemical
      shielding scale for tin from nmr relaxation studies and molecular dynamics
      simulations},\ }\href {https://doi.org/10.1021/ja00106a044} {\bibfield
      {journal} {\bibinfo  {journal} {Journal of the American Chemical Society}\
      }\textbf {\bibinfo {volume} {117}},\ \bibinfo {pages} {392} (\bibinfo {year}
      {1995})}\BibitemShut {NoStop}%
    \bibitem [{\citenamefont {Lassigne}\ \emph {et~al.}(1981)\citenamefont
      {Lassigne}, \citenamefont {Wells}, \citenamefont {Farrugia},\ and\
      \citenamefont {James}}]{lassigne1981heteronuclear}%
      \BibitemOpen
      \bibfield  {author} {\bibinfo {author} {\bibfnamefont {C.~R.}\ \bibnamefont
      {Lassigne}}, \bibinfo {author} {\bibfnamefont {E.}~\bibnamefont {Wells}},
      \bibinfo {author} {\bibfnamefont {L.~J.}\ \bibnamefont {Farrugia}},\ and\
      \bibinfo {author} {\bibfnamefont {B.~R.}\ \bibnamefont {James}},\ }\bibfield
      {title} {\bibinfo {title} {Heteronuclear second-order coupling effects in
      high-resolution tin nmr spectra},\ }\href
      {https://doi.org/10.1016/0022-2364(81)90062-7} {\bibfield  {journal}
      {\bibinfo  {journal} {Journal of Magnetic Resonance}\ }\textbf {\bibinfo
      {volume} {43}},\ \bibinfo {pages} {488} (\bibinfo {year} {1981})}\BibitemShut
      {NoStop}%
    \bibitem [{\citenamefont {Defo}\ \emph {et~al.}(2021)\citenamefont {Defo},
      \citenamefont {Kaxiras},\ and\ \citenamefont
      {Richardson}}]{defo2021carbon13snv}%
      \BibitemOpen
      \bibfield  {author} {\bibinfo {author} {\bibfnamefont {R.~K.}\ \bibnamefont
      {Defo}}, \bibinfo {author} {\bibfnamefont {E.}~\bibnamefont {Kaxiras}},\ and\
      \bibinfo {author} {\bibfnamefont {S.~L.}\ \bibnamefont {Richardson}},\
      }\bibfield  {title} {\bibinfo {title} {Calculating the hyperfine tensors for
      group-iv impurity-vacancy centers in diamond using hybrid density functional
      theory},\ }\href {https://doi.org/10.1103/physrevb.104.075158} {\bibfield
      {journal} {\bibinfo  {journal} {Physical Review B}\ }\textbf {\bibinfo
      {volume} {104}},\ \bibinfo {pages} {075158} (\bibinfo {year}
      {2021})}\BibitemShut {NoStop}%
    \bibitem [{\citenamefont {Waldherr}\ \emph {et~al.}(2014)\citenamefont
      {Waldherr}, \citenamefont {Wang}, \citenamefont {Zaiser}, \citenamefont
      {Jamali}, \citenamefont {Schulte-Herbr{\"u}ggen}, \citenamefont {Abe},
      \citenamefont {Ohshima}, \citenamefont {Isoya}, \citenamefont {Du},
      \citenamefont {Neumann} \emph {et~al.}}]{waldherr2014quantum}%
      \BibitemOpen
      \bibfield  {author} {\bibinfo {author} {\bibfnamefont {G.}~\bibnamefont
      {Waldherr}}, \bibinfo {author} {\bibfnamefont {Y.}~\bibnamefont {Wang}},
      \bibinfo {author} {\bibfnamefont {S.}~\bibnamefont {Zaiser}}, \bibinfo
      {author} {\bibfnamefont {M.}~\bibnamefont {Jamali}}, \bibinfo {author}
      {\bibfnamefont {T.}~\bibnamefont {Schulte-Herbr{\"u}ggen}}, \bibinfo {author}
      {\bibfnamefont {H.}~\bibnamefont {Abe}}, \bibinfo {author} {\bibfnamefont
      {T.}~\bibnamefont {Ohshima}}, \bibinfo {author} {\bibfnamefont
      {J.}~\bibnamefont {Isoya}}, \bibinfo {author} {\bibfnamefont
      {J.}~\bibnamefont {Du}}, \bibinfo {author} {\bibfnamefont {P.}~\bibnamefont
      {Neumann}}, \emph {et~al.},\ }\bibfield  {title} {\bibinfo {title} {Quantum
      error correction in a solid-state hybrid spin register},\ }\href
      {https://doi.org/10.1038/nature12919} {\bibfield  {journal} {\bibinfo
      {journal} {Nature}\ }\textbf {\bibinfo {volume} {506}},\ \bibinfo {pages}
      {204} (\bibinfo {year} {2014})}\BibitemShut {NoStop}%
    \bibitem [{\citenamefont {Scheuer}\ \emph {et~al.}(2016)\citenamefont
      {Scheuer}, \citenamefont {Schwartz}, \citenamefont {Chen}, \citenamefont
      {Schulze-S{\"u}nninghausen}, \citenamefont {Carl}, \citenamefont {H{\"o}fer},
      \citenamefont {Retzker}, \citenamefont {Sumiya}, \citenamefont {Isoya},
      \citenamefont {Luy} \emph {et~al.}}]{scheuer2016optically}%
      \BibitemOpen
      \bibfield  {author} {\bibinfo {author} {\bibfnamefont {J.}~\bibnamefont
      {Scheuer}}, \bibinfo {author} {\bibfnamefont {I.}~\bibnamefont {Schwartz}},
      \bibinfo {author} {\bibfnamefont {Q.}~\bibnamefont {Chen}}, \bibinfo {author}
      {\bibfnamefont {D.}~\bibnamefont {Schulze-S{\"u}nninghausen}}, \bibinfo
      {author} {\bibfnamefont {P.}~\bibnamefont {Carl}}, \bibinfo {author}
      {\bibfnamefont {P.}~\bibnamefont {H{\"o}fer}}, \bibinfo {author}
      {\bibfnamefont {A.}~\bibnamefont {Retzker}}, \bibinfo {author} {\bibfnamefont
      {H.}~\bibnamefont {Sumiya}}, \bibinfo {author} {\bibfnamefont
      {J.}~\bibnamefont {Isoya}}, \bibinfo {author} {\bibfnamefont
      {B.}~\bibnamefont {Luy}}, \emph {et~al.},\ }\bibfield  {title} {\bibinfo
      {title} {Optically induced dynamic nuclear spin polarisation in diamond},\
      }\href {https://doi.org/10.1088/1367-2630/18/1/013040} {\bibfield  {journal}
      {\bibinfo  {journal} {New Journal of Physics}\ }\textbf {\bibinfo {volume}
      {18}},\ \bibinfo {pages} {013040} (\bibinfo {year} {2016})}\BibitemShut
      {NoStop}%
    \bibitem [{\citenamefont {Wolfowicz}\ \emph {et~al.}(2020)\citenamefont
      {Wolfowicz}, \citenamefont {Anderson}, \citenamefont {Diler}, \citenamefont
      {Poluektov}, \citenamefont {Heremans},\ and\ \citenamefont
      {Awschalom}}]{wolfowicz2020vanadium}%
      \BibitemOpen
      \bibfield  {author} {\bibinfo {author} {\bibfnamefont {G.}~\bibnamefont
      {Wolfowicz}}, \bibinfo {author} {\bibfnamefont {C.~P.}\ \bibnamefont
      {Anderson}}, \bibinfo {author} {\bibfnamefont {B.}~\bibnamefont {Diler}},
      \bibinfo {author} {\bibfnamefont {O.~G.}\ \bibnamefont {Poluektov}}, \bibinfo
      {author} {\bibfnamefont {F.~J.}\ \bibnamefont {Heremans}},\ and\ \bibinfo
      {author} {\bibfnamefont {D.~D.}\ \bibnamefont {Awschalom}},\ }\bibfield
      {title} {\bibinfo {title} {Vanadium spin qubits as telecom quantum emitters
      in silicon carbide},\ }\href {https://doi.org/10.1126/sciadv.aaz1192}
      {\bibfield  {journal} {\bibinfo  {journal} {Science Advances}\ }\textbf
      {\bibinfo {volume} {6}},\ \bibinfo {pages} {eaaz1192} (\bibinfo {year}
      {2020})}\BibitemShut {NoStop}%
    \bibitem [{\citenamefont {Hedges}\ \emph {et~al.}(2010)\citenamefont {Hedges},
      \citenamefont {Longdell}, \citenamefont {Li},\ and\ \citenamefont
      {Sellars}}]{hedges2010efficient}%
      \BibitemOpen
      \bibfield  {author} {\bibinfo {author} {\bibfnamefont {M.~P.}\ \bibnamefont
      {Hedges}}, \bibinfo {author} {\bibfnamefont {J.~J.}\ \bibnamefont
      {Longdell}}, \bibinfo {author} {\bibfnamefont {Y.}~\bibnamefont {Li}},\ and\
      \bibinfo {author} {\bibfnamefont {M.~J.}\ \bibnamefont {Sellars}},\
      }\bibfield  {title} {\bibinfo {title} {Efficient quantum memory for light},\
      }\href {https://doi.org/10.1038/nature09081} {\bibfield  {journal} {\bibinfo
      {journal} {Nature}\ }\textbf {\bibinfo {volume} {465}},\ \bibinfo {pages}
      {1052} (\bibinfo {year} {2010})}\BibitemShut {NoStop}%
    \bibitem [{\citenamefont {Liu}\ \emph {et~al.}(2021)\citenamefont {Liu},
      \citenamefont {Hu}, \citenamefont {Li}, \citenamefont {Li}, \citenamefont
      {Li}, \citenamefont {Liang}, \citenamefont {Zhou}, \citenamefont {Li},\ and\
      \citenamefont {Guo}}]{liu2021heralded}%
      \BibitemOpen
      \bibfield  {author} {\bibinfo {author} {\bibfnamefont {X.}~\bibnamefont
      {Liu}}, \bibinfo {author} {\bibfnamefont {J.}~\bibnamefont {Hu}}, \bibinfo
      {author} {\bibfnamefont {Z.-F.}\ \bibnamefont {Li}}, \bibinfo {author}
      {\bibfnamefont {X.}~\bibnamefont {Li}}, \bibinfo {author} {\bibfnamefont
      {P.-Y.}\ \bibnamefont {Li}}, \bibinfo {author} {\bibfnamefont {P.-J.}\
      \bibnamefont {Liang}}, \bibinfo {author} {\bibfnamefont {Z.-Q.}\ \bibnamefont
      {Zhou}}, \bibinfo {author} {\bibfnamefont {C.-F.}\ \bibnamefont {Li}},\ and\
      \bibinfo {author} {\bibfnamefont {G.-C.}\ \bibnamefont {Guo}},\ }\bibfield
      {title} {\bibinfo {title} {Heralded entanglement distribution between two
      absorptive quantum memories},\ }\href
      {https://doi.org/10.1038/s41586-021-03505-3} {\bibfield  {journal} {\bibinfo
      {journal} {Nature}\ }\textbf {\bibinfo {volume} {594}},\ \bibinfo {pages}
      {41} (\bibinfo {year} {2021})}\BibitemShut {NoStop}%
    \bibitem [{\citenamefont {Saglamyurek}\ \emph {et~al.}(2011)\citenamefont
      {Saglamyurek}, \citenamefont {Sinclair}, \citenamefont {Jin}, \citenamefont
      {Slater}, \citenamefont {Oblak}, \citenamefont {Bussieres}, \citenamefont
      {George}, \citenamefont {Ricken}, \citenamefont {Sohler},\ and\ \citenamefont
      {Tittel}}]{saglamyurek2011broadband}%
      \BibitemOpen
      \bibfield  {author} {\bibinfo {author} {\bibfnamefont {E.}~\bibnamefont
      {Saglamyurek}}, \bibinfo {author} {\bibfnamefont {N.}~\bibnamefont
      {Sinclair}}, \bibinfo {author} {\bibfnamefont {J.}~\bibnamefont {Jin}},
      \bibinfo {author} {\bibfnamefont {J.~A.}\ \bibnamefont {Slater}}, \bibinfo
      {author} {\bibfnamefont {D.}~\bibnamefont {Oblak}}, \bibinfo {author}
      {\bibfnamefont {F.}~\bibnamefont {Bussieres}}, \bibinfo {author}
      {\bibfnamefont {M.}~\bibnamefont {George}}, \bibinfo {author} {\bibfnamefont
      {R.}~\bibnamefont {Ricken}}, \bibinfo {author} {\bibfnamefont
      {W.}~\bibnamefont {Sohler}},\ and\ \bibinfo {author} {\bibfnamefont
      {W.}~\bibnamefont {Tittel}},\ }\bibfield  {title} {\bibinfo {title}
      {Broadband waveguide quantum memory for entangled photons},\ }\href
      {https://doi.org/10.1364/icqi.2011.qtug2} {\bibfield  {journal} {\bibinfo
      {journal} {Nature}\ }\textbf {\bibinfo {volume} {469}},\ \bibinfo {pages}
      {512} (\bibinfo {year} {2011})}\BibitemShut {NoStop}%
    \bibitem [{\citenamefont {Zhong}\ \emph {et~al.}(2017)\citenamefont {Zhong},
      \citenamefont {Kindem}, \citenamefont {Bartholomew}, \citenamefont {Rochman},
      \citenamefont {Craiciu}, \citenamefont {Miyazono}, \citenamefont
      {Bettinelli}, \citenamefont {Cavalli}, \citenamefont {Verma}, \citenamefont
      {Nam} \emph {et~al.}}]{zhong2017nanophotonic}%
      \BibitemOpen
      \bibfield  {author} {\bibinfo {author} {\bibfnamefont {T.}~\bibnamefont
      {Zhong}}, \bibinfo {author} {\bibfnamefont {J.~M.}\ \bibnamefont {Kindem}},
      \bibinfo {author} {\bibfnamefont {J.~G.}\ \bibnamefont {Bartholomew}},
      \bibinfo {author} {\bibfnamefont {J.}~\bibnamefont {Rochman}}, \bibinfo
      {author} {\bibfnamefont {I.}~\bibnamefont {Craiciu}}, \bibinfo {author}
      {\bibfnamefont {E.}~\bibnamefont {Miyazono}}, \bibinfo {author}
      {\bibfnamefont {M.}~\bibnamefont {Bettinelli}}, \bibinfo {author}
      {\bibfnamefont {E.}~\bibnamefont {Cavalli}}, \bibinfo {author} {\bibfnamefont
      {V.}~\bibnamefont {Verma}}, \bibinfo {author} {\bibfnamefont {S.~W.}\
      \bibnamefont {Nam}}, \emph {et~al.},\ }\bibfield  {title} {\bibinfo {title}
      {Nanophotonic rare-earth quantum memory with optically controlled
      retrieval},\ }\href {https://doi.org/10.1126/science.aan5959} {\bibfield
      {journal} {\bibinfo  {journal} {Science}\ }\textbf {\bibinfo {volume}
      {357}},\ \bibinfo {pages} {1392} (\bibinfo {year} {2017})}\BibitemShut
      {NoStop}%
    \bibitem [{\citenamefont {Merkulov}\ \emph {et~al.}(2002)\citenamefont
      {Merkulov}, \citenamefont {Efros},\ and\ \citenamefont
      {Rosen}}]{merkulov2002electron}%
      \BibitemOpen
      \bibfield  {author} {\bibinfo {author} {\bibfnamefont {I.}~\bibnamefont
      {Merkulov}}, \bibinfo {author} {\bibfnamefont {A.~L.}\ \bibnamefont
      {Efros}},\ and\ \bibinfo {author} {\bibfnamefont {M.}~\bibnamefont {Rosen}},\
      }\bibfield  {title} {\bibinfo {title} {Electron spin relaxation by nuclei in
      semiconductor quantum dots},\ }\href
      {https://doi.org/10.1103/physrevb.65.205309} {\bibfield  {journal} {\bibinfo
      {journal} {Physical Review B}\ }\textbf {\bibinfo {volume} {65}},\ \bibinfo
      {pages} {205309} (\bibinfo {year} {2002})}\BibitemShut {NoStop}%
    \bibitem [{\citenamefont {Gillard}\ \emph {et~al.}(2021)\citenamefont
      {Gillard}, \citenamefont {Griffiths}, \citenamefont {Ragunathan},
      \citenamefont {Ulhaq}, \citenamefont {McEwan}, \citenamefont {Clarke},\ and\
      \citenamefont {Chekhovich}}]{gillard2021fundamental}%
      \BibitemOpen
      \bibfield  {author} {\bibinfo {author} {\bibfnamefont {G.}~\bibnamefont
      {Gillard}}, \bibinfo {author} {\bibfnamefont {I.~M.}\ \bibnamefont
      {Griffiths}}, \bibinfo {author} {\bibfnamefont {G.}~\bibnamefont
      {Ragunathan}}, \bibinfo {author} {\bibfnamefont {A.}~\bibnamefont {Ulhaq}},
      \bibinfo {author} {\bibfnamefont {C.}~\bibnamefont {McEwan}}, \bibinfo
      {author} {\bibfnamefont {E.}~\bibnamefont {Clarke}},\ and\ \bibinfo {author}
      {\bibfnamefont {E.~A.}\ \bibnamefont {Chekhovich}},\ }\bibfield  {title}
      {\bibinfo {title} {Fundamental limits of electron and nuclear spin qubit
      lifetimes in an isolated self-assembled quantum dot},\ }\href
      {https://doi.org/10.1038/s41534-021-00378-2} {\bibfield  {journal} {\bibinfo
      {journal} {npj Quantum Information}\ }\textbf {\bibinfo {volume} {7}},\
      \bibinfo {pages} {43} (\bibinfo {year} {2021})}\BibitemShut {NoStop}%
    \bibitem [{\citenamefont {Seo}\ \emph {et~al.}(2016)\citenamefont {Seo},
      \citenamefont {Falk}, \citenamefont {Klimov}, \citenamefont {Miao},
      \citenamefont {Galli},\ and\ \citenamefont {Awschalom}}]{seo2016quantum}%
      \BibitemOpen
      \bibfield  {author} {\bibinfo {author} {\bibfnamefont {H.}~\bibnamefont
      {Seo}}, \bibinfo {author} {\bibfnamefont {A.~L.}\ \bibnamefont {Falk}},
      \bibinfo {author} {\bibfnamefont {P.~V.}\ \bibnamefont {Klimov}}, \bibinfo
      {author} {\bibfnamefont {K.~C.}\ \bibnamefont {Miao}}, \bibinfo {author}
      {\bibfnamefont {G.}~\bibnamefont {Galli}},\ and\ \bibinfo {author}
      {\bibfnamefont {D.~D.}\ \bibnamefont {Awschalom}},\ }\bibfield  {title}
      {\bibinfo {title} {Quantum decoherence dynamics of divacancy spins in silicon
      carbide},\ }\href {https://doi.org/10.1038/ncomms12935} {\bibfield  {journal}
      {\bibinfo  {journal} {Nature Communications}\ }\textbf {\bibinfo {volume}
      {7}},\ \bibinfo {pages} {12935} (\bibinfo {year} {2016})}\BibitemShut
      {NoStop}%
    \bibitem [{\citenamefont {Radulaski}\ \emph {et~al.}(2017)\citenamefont
      {Radulaski}, \citenamefont {Widmann}, \citenamefont {Niethammer},
      \citenamefont {Zhang}, \citenamefont {Lee}, \citenamefont {Rendler},
      \citenamefont {Lagoudakis}, \citenamefont {Son}, \citenamefont {Janzen},
      \citenamefont {Ohshima} \emph {et~al.}}]{radulaski2017scalable}%
      \BibitemOpen
      \bibfield  {author} {\bibinfo {author} {\bibfnamefont {M.}~\bibnamefont
      {Radulaski}}, \bibinfo {author} {\bibfnamefont {M.}~\bibnamefont {Widmann}},
      \bibinfo {author} {\bibfnamefont {M.}~\bibnamefont {Niethammer}}, \bibinfo
      {author} {\bibfnamefont {J.~L.}\ \bibnamefont {Zhang}}, \bibinfo {author}
      {\bibfnamefont {S.-Y.}\ \bibnamefont {Lee}}, \bibinfo {author} {\bibfnamefont
      {T.}~\bibnamefont {Rendler}}, \bibinfo {author} {\bibfnamefont {K.~G.}\
      \bibnamefont {Lagoudakis}}, \bibinfo {author} {\bibfnamefont {N.~T.}\
      \bibnamefont {Son}}, \bibinfo {author} {\bibfnamefont {E.}~\bibnamefont
      {Janzen}}, \bibinfo {author} {\bibfnamefont {T.}~\bibnamefont {Ohshima}},
      \emph {et~al.},\ }\bibfield  {title} {\bibinfo {title} {Scalable quantum
      photonics with single color centers in silicon carbide},\ }\href
      {https://doi.org/10.1021/acs.nanolett.6b05102} {\bibfield  {journal}
      {\bibinfo  {journal} {Nano Letters}\ }\textbf {\bibinfo {volume} {17}},\
      \bibinfo {pages} {1782} (\bibinfo {year} {2017})}\BibitemShut {NoStop}%
    \bibitem [{\citenamefont {G{\"o}rlitz}\ \emph {et~al.}(2022)\citenamefont
      {G{\"o}rlitz}, \citenamefont {Herrmann}, \citenamefont {Fuchs}, \citenamefont
      {Iwasaki}, \citenamefont {Taniguchi}, \citenamefont {Rogalla}, \citenamefont
      {Hardeman}, \citenamefont {Colard}, \citenamefont {Markham}, \citenamefont
      {Hatano} \emph {et~al.}}]{gorlitz2022coherence}%
      \BibitemOpen
      \bibfield  {author} {\bibinfo {author} {\bibfnamefont {J.}~\bibnamefont
      {G{\"o}rlitz}}, \bibinfo {author} {\bibfnamefont {D.}~\bibnamefont
      {Herrmann}}, \bibinfo {author} {\bibfnamefont {P.}~\bibnamefont {Fuchs}},
      \bibinfo {author} {\bibfnamefont {T.}~\bibnamefont {Iwasaki}}, \bibinfo
      {author} {\bibfnamefont {T.}~\bibnamefont {Taniguchi}}, \bibinfo {author}
      {\bibfnamefont {D.}~\bibnamefont {Rogalla}}, \bibinfo {author} {\bibfnamefont
      {D.}~\bibnamefont {Hardeman}}, \bibinfo {author} {\bibfnamefont {P.-O.}\
      \bibnamefont {Colard}}, \bibinfo {author} {\bibfnamefont {M.}~\bibnamefont
      {Markham}}, \bibinfo {author} {\bibfnamefont {M.}~\bibnamefont {Hatano}},
      \emph {et~al.},\ }\bibfield  {title} {\bibinfo {title} {Coherence of a charge
      stabilised tin-vacancy spin in diamond},\ }\href
      {https://doi.org/10.1038/s41534-022-00552-0} {\bibfield  {journal} {\bibinfo
      {journal} {npj Quantum Information}\ }\textbf {\bibinfo {volume} {8}},\
      \bibinfo {pages} {45} (\bibinfo {year} {2022})}\BibitemShut {NoStop}%
    \bibitem [{\citenamefont {Istrati}\ \emph
      {et~al.}(2020{\natexlab{b}})\citenamefont {Istrati}, \citenamefont {Pilnyak},
      \citenamefont {Loredo}, \citenamefont {Ant{\'o}n}, \citenamefont {Somaschi},
      \citenamefont {Hilaire}, \citenamefont {Ollivier}, \citenamefont {Esmann},
      \citenamefont {Cohen}, \citenamefont {Vidro} \emph
      {et~al.}}]{istrati2020sequential}%
      \BibitemOpen
      \bibfield  {author} {\bibinfo {author} {\bibfnamefont {D.}~\bibnamefont
      {Istrati}}, \bibinfo {author} {\bibfnamefont {Y.}~\bibnamefont {Pilnyak}},
      \bibinfo {author} {\bibfnamefont {J.}~\bibnamefont {Loredo}}, \bibinfo
      {author} {\bibfnamefont {C.}~\bibnamefont {Ant{\'o}n}}, \bibinfo {author}
      {\bibfnamefont {N.}~\bibnamefont {Somaschi}}, \bibinfo {author}
      {\bibfnamefont {P.}~\bibnamefont {Hilaire}}, \bibinfo {author} {\bibfnamefont
      {H.}~\bibnamefont {Ollivier}}, \bibinfo {author} {\bibfnamefont
      {M.}~\bibnamefont {Esmann}}, \bibinfo {author} {\bibfnamefont
      {L.}~\bibnamefont {Cohen}}, \bibinfo {author} {\bibfnamefont
      {L.}~\bibnamefont {Vidro}}, \emph {et~al.},\ }\bibfield  {title} {\bibinfo
      {title} {Sequential generation of linear cluster states from a single photon
      emitter},\ }\href {https://doi.org/10.1364/quantum.2020.qtu8a.10} {\bibfield
      {journal} {\bibinfo  {journal} {Nature Communications}\ }\textbf {\bibinfo
      {volume} {11}},\ \bibinfo {pages} {5501} (\bibinfo {year}
      {2020}{\natexlab{b}})}\BibitemShut {NoStop}%
    \bibitem [{\citenamefont {Grangier}\ \emph {et~al.}(2000)\citenamefont
      {Grangier}, \citenamefont {Reymond},\ and\ \citenamefont
      {Schlosser}}]{grangier2000implementations}%
      \BibitemOpen
      \bibfield  {author} {\bibinfo {author} {\bibfnamefont {P.}~\bibnamefont
      {Grangier}}, \bibinfo {author} {\bibfnamefont {G.}~\bibnamefont {Reymond}},\
      and\ \bibinfo {author} {\bibfnamefont {N.}~\bibnamefont {Schlosser}},\
      }\bibfield  {title} {\bibinfo {title} {Implementations of quantum computing
      using cavity quantum electrodynamics schemes},\ }\href
      {https://doi.org/10.1002/3527603182.ch7} {\bibfield  {journal} {\bibinfo
      {journal} {Scalable Quantum Computers: Paving the Way to Realization}\ ,\
      \bibinfo {pages} {89}} (\bibinfo {year} {2000})}\BibitemShut {NoStop}%
    \bibitem [{\citenamefont {Bhaskar}\ \emph {et~al.}(2020)\citenamefont
      {Bhaskar}, \citenamefont {Riedinger}, \citenamefont {Machielse},
      \citenamefont {Levonian}, \citenamefont {Nguyen}, \citenamefont {Knall},
      \citenamefont {Park}, \citenamefont {Englund}, \citenamefont {Lon{\v{c}}ar},
      \citenamefont {Sukachev} \emph {et~al.}}]{bhaskar2020experimental}%
      \BibitemOpen
      \bibfield  {author} {\bibinfo {author} {\bibfnamefont {M.~K.}\ \bibnamefont
      {Bhaskar}}, \bibinfo {author} {\bibfnamefont {R.}~\bibnamefont {Riedinger}},
      \bibinfo {author} {\bibfnamefont {B.}~\bibnamefont {Machielse}}, \bibinfo
      {author} {\bibfnamefont {D.~S.}\ \bibnamefont {Levonian}}, \bibinfo {author}
      {\bibfnamefont {C.~T.}\ \bibnamefont {Nguyen}}, \bibinfo {author}
      {\bibfnamefont {E.~N.}\ \bibnamefont {Knall}}, \bibinfo {author}
      {\bibfnamefont {H.}~\bibnamefont {Park}}, \bibinfo {author} {\bibfnamefont
      {D.}~\bibnamefont {Englund}}, \bibinfo {author} {\bibfnamefont
      {M.}~\bibnamefont {Lon{\v{c}}ar}}, \bibinfo {author} {\bibfnamefont {D.~D.}\
      \bibnamefont {Sukachev}}, \emph {et~al.},\ }\bibfield  {title} {\bibinfo
      {title} {Experimental demonstration of memory-enhanced quantum
      communication},\ }\href {https://doi.org/10.1038/s41586-020-2103-5}
      {\bibfield  {journal} {\bibinfo  {journal} {Nature}\ }\textbf {\bibinfo
      {volume} {580}},\ \bibinfo {pages} {60} (\bibinfo {year} {2020})}\BibitemShut
      {NoStop}%
    \bibitem [{\citenamefont {Rugar}\ \emph
      {et~al.}(2020{\natexlab{b}})\citenamefont {Rugar}, \citenamefont {Lu},
      \citenamefont {Dory}, \citenamefont {Sun}, \citenamefont {McQuade},
      \citenamefont {Shen}, \citenamefont {Melosh},\ and\ \citenamefont
      {Vuckovic}}]{rugar2020generation}%
      \BibitemOpen
      \bibfield  {author} {\bibinfo {author} {\bibfnamefont {A.~E.}\ \bibnamefont
      {Rugar}}, \bibinfo {author} {\bibfnamefont {H.}~\bibnamefont {Lu}}, \bibinfo
      {author} {\bibfnamefont {C.}~\bibnamefont {Dory}}, \bibinfo {author}
      {\bibfnamefont {S.}~\bibnamefont {Sun}}, \bibinfo {author} {\bibfnamefont
      {P.~J.}\ \bibnamefont {McQuade}}, \bibinfo {author} {\bibfnamefont {Z.-X.}\
      \bibnamefont {Shen}}, \bibinfo {author} {\bibfnamefont {N.~A.}\ \bibnamefont
      {Melosh}},\ and\ \bibinfo {author} {\bibfnamefont {J.}~\bibnamefont
      {Vuckovic}},\ }\bibfield  {title} {\bibinfo {title} {Generation of
      tin-vacancy centers in diamond via shallow ion implantation and subsequent
      diamond overgrowth},\ }\href {https://doi.org/10.1021/acs.nanolett.9b04495}
      {\bibfield  {journal} {\bibinfo  {journal} {Nano Letters}\ }\textbf {\bibinfo
      {volume} {20}},\ \bibinfo {pages} {1614} (\bibinfo {year}
      {2020}{\natexlab{b}})}\BibitemShut {NoStop}%
    \bibitem [{\citenamefont {Rugar}\ \emph {et~al.}(2021)\citenamefont {Rugar},
      \citenamefont {Aghaeimeibodi}, \citenamefont {Riedel}, \citenamefont {Dory},
      \citenamefont {Lu}, \citenamefont {McQuade}, \citenamefont {Shen},
      \citenamefont {Melosh},\ and\ \citenamefont
      {Vu{\v{c}}kovi{\'c}}}]{rugar2021quantum}%
      \BibitemOpen
      \bibfield  {author} {\bibinfo {author} {\bibfnamefont {A.~E.}\ \bibnamefont
      {Rugar}}, \bibinfo {author} {\bibfnamefont {S.}~\bibnamefont
      {Aghaeimeibodi}}, \bibinfo {author} {\bibfnamefont {D.}~\bibnamefont
      {Riedel}}, \bibinfo {author} {\bibfnamefont {C.}~\bibnamefont {Dory}},
      \bibinfo {author} {\bibfnamefont {H.}~\bibnamefont {Lu}}, \bibinfo {author}
      {\bibfnamefont {P.~J.}\ \bibnamefont {McQuade}}, \bibinfo {author}
      {\bibfnamefont {Z.-X.}\ \bibnamefont {Shen}}, \bibinfo {author}
      {\bibfnamefont {N.~A.}\ \bibnamefont {Melosh}},\ and\ \bibinfo {author}
      {\bibfnamefont {J.}~\bibnamefont {Vu{\v{c}}kovi{\'c}}},\ }\bibfield  {title}
      {\bibinfo {title} {Quantum photonic interface for tin-vacancy centers in
      diamond},\ }\href {https://doi.org/10.1103/physrevx.11.031021} {\bibfield
      {journal} {\bibinfo  {journal} {Physical Review X}\ }\textbf {\bibinfo
      {volume} {11}},\ \bibinfo {pages} {031021} (\bibinfo {year}
      {2021})}\BibitemShut {NoStop}%
    \bibitem [{\citenamefont {Thiering}\ and\ \citenamefont
      {Gali}(2018)}]{thiering2018abinitio}%
      \BibitemOpen
      \bibfield  {author} {\bibinfo {author} {\bibfnamefont {G.}~\bibnamefont
      {Thiering}}\ and\ \bibinfo {author} {\bibfnamefont {A.}~\bibnamefont
      {Gali}},\ }\bibfield  {title} {\bibinfo {title} {Ab initio magneto-optical
      spectrum of group-iv vacancy color centers in diamond},\ }\href
      {https://doi.org/10.1103/physrevx.8.021063} {\bibfield  {journal} {\bibinfo
      {journal} {Physical Review X}\ }\textbf {\bibinfo {volume} {8}},\ \bibinfo
      {pages} {021063} (\bibinfo {year} {2018})}\BibitemShut {NoStop}%
    \bibitem [{\citenamefont {Neu}\ \emph {et~al.}(2012)\citenamefont {Neu},
      \citenamefont {Agio},\ and\ \citenamefont {Becher}}]{neu2012photophysics}%
      \BibitemOpen
      \bibfield  {author} {\bibinfo {author} {\bibfnamefont {E.}~\bibnamefont
      {Neu}}, \bibinfo {author} {\bibfnamefont {M.}~\bibnamefont {Agio}},\ and\
      \bibinfo {author} {\bibfnamefont {C.}~\bibnamefont {Becher}},\ }\bibfield
      {title} {\bibinfo {title} {Photophysics of single silicon vacancy centers in
      diamond: implications for single photon emission},\ }\href
      {https://doi.org/10.1364/oe.20.019956} {\bibfield  {journal} {\bibinfo
      {journal} {Optics Express}\ }\textbf {\bibinfo {volume} {20}},\ \bibinfo
      {pages} {19956} (\bibinfo {year} {2012})}\BibitemShut {NoStop}%
    \bibitem [{\citenamefont {Baier}\ \emph {et~al.}(2021)\citenamefont {Baier},
      \citenamefont {Pompili}, \citenamefont {Hermans}, \citenamefont {Beukers},
      \citenamefont {Humphreys}, \citenamefont {Schouten}, \citenamefont
      {Vermeulen}, \citenamefont {Tiggelman}, \citenamefont {dos Santos~Martins},
      \citenamefont {Dirkse} \emph {et~al.}}]{baier2021realization}%
      \BibitemOpen
      \bibfield  {author} {\bibinfo {author} {\bibfnamefont {S.}~\bibnamefont
      {Baier}}, \bibinfo {author} {\bibfnamefont {M.}~\bibnamefont {Pompili}},
      \bibinfo {author} {\bibfnamefont {S.~L.}\ \bibnamefont {Hermans}}, \bibinfo
      {author} {\bibfnamefont {H.~K.}\ \bibnamefont {Beukers}}, \bibinfo {author}
      {\bibfnamefont {P.~C.}\ \bibnamefont {Humphreys}}, \bibinfo {author}
      {\bibfnamefont {R.~N.}\ \bibnamefont {Schouten}}, \bibinfo {author}
      {\bibfnamefont {R.~F.}\ \bibnamefont {Vermeulen}}, \bibinfo {author}
      {\bibfnamefont {M.~J.}\ \bibnamefont {Tiggelman}}, \bibinfo {author}
      {\bibfnamefont {L.}~\bibnamefont {dos Santos~Martins}}, \bibinfo {author}
      {\bibfnamefont {B.}~\bibnamefont {Dirkse}}, \emph {et~al.},\ }\bibfield
      {title} {\bibinfo {title} {Realization of a multi-node quantum network of
      remote solid-state qubits},\ }in\ \href
      {https://doi.org/10.1126/science.abg1919} {\emph {\bibinfo {booktitle}
      {Quantum Information and Measurement}}}\ (\bibinfo {organization} {Optica
      Publishing Group},\ \bibinfo {year} {2021})\ pp.\ \bibinfo {pages}
      {M2A--2}\BibitemShut {NoStop}%
    \bibitem [{\citenamefont {Goldman}\ \emph {et~al.}(2020)\citenamefont
      {Goldman}, \citenamefont {Patti}, \citenamefont {Levonian}, \citenamefont
      {Yelin},\ and\ \citenamefont {Lukin}}]{goldman2020optical}%
      \BibitemOpen
      \bibfield  {author} {\bibinfo {author} {\bibfnamefont {M.}~\bibnamefont
      {Goldman}}, \bibinfo {author} {\bibfnamefont {T.}~\bibnamefont {Patti}},
      \bibinfo {author} {\bibfnamefont {D.}~\bibnamefont {Levonian}}, \bibinfo
      {author} {\bibfnamefont {S.}~\bibnamefont {Yelin}},\ and\ \bibinfo {author}
      {\bibfnamefont {M.}~\bibnamefont {Lukin}},\ }\bibfield  {title} {\bibinfo
      {title} {Optical control of a single nuclear spin in the solid state},\
      }\href {https://doi.org/10.1103/physrevlett.124.153203} {\bibfield  {journal}
      {\bibinfo  {journal} {Physical Review Letters}\ }\textbf {\bibinfo {volume}
      {124}},\ \bibinfo {pages} {153203} (\bibinfo {year} {2020})}\BibitemShut
      {NoStop}%
    \bibitem [{\citenamefont {Raussendorf}\ and\ \citenamefont
      {Briegel}(2001)}]{raussendorf2001clusterstates}%
      \BibitemOpen
      \bibfield  {author} {\bibinfo {author} {\bibfnamefont {R.}~\bibnamefont
      {Raussendorf}}\ and\ \bibinfo {author} {\bibfnamefont {H.~J.}\ \bibnamefont
      {Briegel}},\ }\bibfield  {title} {\bibinfo {title} {A one-way quantum
      computer},\ }\href {https://doi.org/10.1103/physrevlett.86.5188} {\bibfield
      {journal} {\bibinfo  {journal} {Physical Review Letters}\ }\textbf {\bibinfo
      {volume} {86}},\ \bibinfo {pages} {5188} (\bibinfo {year}
      {2001})}\BibitemShut {NoStop}%
    \bibitem [{\citenamefont {Chang}\ \emph {et~al.}(2007)\citenamefont {Chang},
      \citenamefont {S{\o}rensen}, \citenamefont {Demler},\ and\ \citenamefont
      {Lukin}}]{chang2007single}%
      \BibitemOpen
      \bibfield  {author} {\bibinfo {author} {\bibfnamefont {D.~E.}\ \bibnamefont
      {Chang}}, \bibinfo {author} {\bibfnamefont {A.~S.}\ \bibnamefont
      {S{\o}rensen}}, \bibinfo {author} {\bibfnamefont {E.~A.}\ \bibnamefont
      {Demler}},\ and\ \bibinfo {author} {\bibfnamefont {M.~D.}\ \bibnamefont
      {Lukin}},\ }\bibfield  {title} {\bibinfo {title} {A single-photon transistor
      using nanoscale surface plasmons},\ }\href {https://doi.org/10.1038/nphys708}
      {\bibfield  {journal} {\bibinfo  {journal} {Nature Physics}\ }\textbf
      {\bibinfo {volume} {3}},\ \bibinfo {pages} {807} (\bibinfo {year}
      {2007})}\BibitemShut {NoStop}%
    \bibitem [{\citenamefont {Knill}\ \emph {et~al.}(2001)\citenamefont {Knill},
      \citenamefont {Laflamme},\ and\ \citenamefont {Milburn}}]{knill2001scheme}%
      \BibitemOpen
      \bibfield  {author} {\bibinfo {author} {\bibfnamefont {E.}~\bibnamefont
      {Knill}}, \bibinfo {author} {\bibfnamefont {R.}~\bibnamefont {Laflamme}},\
      and\ \bibinfo {author} {\bibfnamefont {G.~J.}\ \bibnamefont {Milburn}},\
      }\bibfield  {title} {\bibinfo {title} {A scheme for efficient quantum
      computation with linear optics},\ }\href {https://doi.org/10.1038/35051009}
      {\bibfield  {journal} {\bibinfo  {journal} {nature}\ }\textbf {\bibinfo
      {volume} {409}},\ \bibinfo {pages} {46} (\bibinfo {year} {2001})}\BibitemShut
      {NoStop}%
    \bibitem [{\citenamefont {Kok}\ \emph {et~al.}(2007)\citenamefont {Kok},
      \citenamefont {Munro}, \citenamefont {Nemoto}, \citenamefont {Ralph},
      \citenamefont {Dowling},\ and\ \citenamefont {Milburn}}]{kok2007linear}%
      \BibitemOpen
      \bibfield  {author} {\bibinfo {author} {\bibfnamefont {P.}~\bibnamefont
      {Kok}}, \bibinfo {author} {\bibfnamefont {W.~J.}\ \bibnamefont {Munro}},
      \bibinfo {author} {\bibfnamefont {K.}~\bibnamefont {Nemoto}}, \bibinfo
      {author} {\bibfnamefont {T.~C.}\ \bibnamefont {Ralph}}, \bibinfo {author}
      {\bibfnamefont {J.~P.}\ \bibnamefont {Dowling}},\ and\ \bibinfo {author}
      {\bibfnamefont {G.~J.}\ \bibnamefont {Milburn}},\ }\bibfield  {title}
      {\bibinfo {title} {Linear optical quantum computing with photonic qubits},\
      }\href {https://doi.org/10.1142/9789812774491_0030} {\bibfield  {journal}
      {\bibinfo  {journal} {Reviews of modern physics}\ }\textbf {\bibinfo {volume}
      {79}},\ \bibinfo {pages} {135} (\bibinfo {year} {2007})}\BibitemShut
      {NoStop}%
    \bibitem [{\citenamefont {Ziegler}\ and\ \citenamefont
      {Biersack}(1985)}]{ziegler1985stopping}%
      \BibitemOpen
      \bibfield  {author} {\bibinfo {author} {\bibfnamefont {J.~F.}\ \bibnamefont
      {Ziegler}}\ and\ \bibinfo {author} {\bibfnamefont {J.~P.}\ \bibnamefont
      {Biersack}},\ }\bibfield  {title} {\bibinfo {title} {The stopping and range
      of ions in matter},\ }in\ \href {https://doi.org/10.1007/978-1-4615-8103-1_3}
      {\emph {\bibinfo {booktitle} {Treatise on heavy-ion science}}}\ (\bibinfo
      {publisher} {Springer},\ \bibinfo {year} {1985})\ pp.\ \bibinfo {pages}
      {93--129}\BibitemShut {NoStop}%
    \bibitem [{\citenamefont {Wan}\ \emph {et~al.}(2018)\citenamefont {Wan},
      \citenamefont {Mouradian},\ and\ \citenamefont
      {Englund}}]{wan2018twodimensionalnanocavity}%
      \BibitemOpen
      \bibfield  {author} {\bibinfo {author} {\bibfnamefont {N.~H.}\ \bibnamefont
      {Wan}}, \bibinfo {author} {\bibfnamefont {S.}~\bibnamefont {Mouradian}},\
      and\ \bibinfo {author} {\bibfnamefont {D.}~\bibnamefont {Englund}},\
      }\bibfield  {title} {\bibinfo {title} {Two-dimensional photonic crystal slab
      nanocavities on bulk single-crystal diamond},\ }\href
      {https://doi.org/10.1063/1.5021349} {\bibfield  {journal} {\bibinfo
      {journal} {Applied Physics Letters}\ }\textbf {\bibinfo {volume} {112}},\
      \bibinfo {pages} {141102} (\bibinfo {year} {2018})}\BibitemShut {NoStop}%
    \bibitem [{\citenamefont {Sangtawesin}\ \emph {et~al.}(2019)\citenamefont
      {Sangtawesin}, \citenamefont {Dwyer}, \citenamefont {Srinivasan},
      \citenamefont {Allred}, \citenamefont {Rodgers}, \citenamefont {De~Greve},
      \citenamefont {Stacey}, \citenamefont {Dontschuk}, \citenamefont
      {O’Donnell}, \citenamefont {Hu} \emph
      {et~al.}}]{sangtawesin2019diamondsurfacenoise}%
      \BibitemOpen
      \bibfield  {author} {\bibinfo {author} {\bibfnamefont {S.}~\bibnamefont
      {Sangtawesin}}, \bibinfo {author} {\bibfnamefont {B.~L.}\ \bibnamefont
      {Dwyer}}, \bibinfo {author} {\bibfnamefont {S.}~\bibnamefont {Srinivasan}},
      \bibinfo {author} {\bibfnamefont {J.~J.}\ \bibnamefont {Allred}}, \bibinfo
      {author} {\bibfnamefont {L.~V.}\ \bibnamefont {Rodgers}}, \bibinfo {author}
      {\bibfnamefont {K.}~\bibnamefont {De~Greve}}, \bibinfo {author}
      {\bibfnamefont {A.}~\bibnamefont {Stacey}}, \bibinfo {author} {\bibfnamefont
      {N.}~\bibnamefont {Dontschuk}}, \bibinfo {author} {\bibfnamefont {K.~M.}\
      \bibnamefont {O’Donnell}}, \bibinfo {author} {\bibfnamefont
      {D.}~\bibnamefont {Hu}}, \emph {et~al.},\ }\bibfield  {title} {\bibinfo
      {title} {Origins of diamond surface noise probed by correlating single-spin
      measurements with surface spectroscopy},\ }\href
      {https://doi.org/10.1103/physrevx.9.031052} {\bibfield  {journal} {\bibinfo
      {journal} {Physical Review X}\ }\textbf {\bibinfo {volume} {9}},\ \bibinfo
      {pages} {031052} (\bibinfo {year} {2019})}\BibitemShut {NoStop}%
    \bibitem [{\citenamefont {Haber}\ \emph {et~al.}(2004)\citenamefont {Haber},
      \citenamefont {Schaller}, \citenamefont {Johnson},\ and\ \citenamefont
      {Saykally}}]{haber2004shape}%
      \BibitemOpen
      \bibfield  {author} {\bibinfo {author} {\bibfnamefont {L.}~\bibnamefont
      {Haber}}, \bibinfo {author} {\bibfnamefont {R.}~\bibnamefont {Schaller}},
      \bibinfo {author} {\bibfnamefont {J.}~\bibnamefont {Johnson}},\ and\ \bibinfo
      {author} {\bibfnamefont {R.}~\bibnamefont {Saykally}},\ }\bibfield  {title}
      {\bibinfo {title} {Shape control of near-field probes using dynamic meniscus
      etching},\ }\href {https://doi.org/10.1111/j.0022-2720.2004.01308.x}
      {\bibfield  {journal} {\bibinfo  {journal} {Journal of Microscopy}\ }\textbf
      {\bibinfo {volume} {214}},\ \bibinfo {pages} {27} (\bibinfo {year}
      {2004})}\BibitemShut {NoStop}%
    \bibitem [{\citenamefont {Burek}\ \emph {et~al.}(2017)\citenamefont {Burek},
      \citenamefont {Meuwly}, \citenamefont {Evans}, \citenamefont {Bhaskar},
      \citenamefont {Sipahigil}, \citenamefont {Meesala}, \citenamefont
      {Machielse}, \citenamefont {Sukachev}, \citenamefont {Nguyen}, \citenamefont
      {Pacheco} \emph {et~al.}}]{burek2017fiber}%
      \BibitemOpen
      \bibfield  {author} {\bibinfo {author} {\bibfnamefont {M.~J.}\ \bibnamefont
      {Burek}}, \bibinfo {author} {\bibfnamefont {C.}~\bibnamefont {Meuwly}},
      \bibinfo {author} {\bibfnamefont {R.~E.}\ \bibnamefont {Evans}}, \bibinfo
      {author} {\bibfnamefont {M.~K.}\ \bibnamefont {Bhaskar}}, \bibinfo {author}
      {\bibfnamefont {A.}~\bibnamefont {Sipahigil}}, \bibinfo {author}
      {\bibfnamefont {S.}~\bibnamefont {Meesala}}, \bibinfo {author} {\bibfnamefont
      {B.}~\bibnamefont {Machielse}}, \bibinfo {author} {\bibfnamefont {D.~D.}\
      \bibnamefont {Sukachev}}, \bibinfo {author} {\bibfnamefont {C.~T.}\
      \bibnamefont {Nguyen}}, \bibinfo {author} {\bibfnamefont {J.~L.}\
      \bibnamefont {Pacheco}}, \emph {et~al.},\ }\bibfield  {title} {\bibinfo
      {title} {Fiber-coupled diamond quantum nanophotonic interface},\ }\href
      {https://doi.org/10.1103/physrevapplied.8.024026} {\bibfield  {journal}
      {\bibinfo  {journal} {Physical Review Applied}\ }\textbf {\bibinfo {volume}
      {8}},\ \bibinfo {pages} {024026} (\bibinfo {year} {2017})}\BibitemShut
      {NoStop}%
    \bibitem [{\citenamefont {Wasserman}\ \emph {et~al.}(2022)\citenamefont
      {Wasserman}, \citenamefont {Harrison}, \citenamefont {Harris}, \citenamefont
      {Sawadsky}, \citenamefont {Sfendla}, \citenamefont {Bowen},\ and\
      \citenamefont {Baker}}]{wasserman2022cryogenic}%
      \BibitemOpen
      \bibfield  {author} {\bibinfo {author} {\bibfnamefont {W.}~\bibnamefont
      {Wasserman}}, \bibinfo {author} {\bibfnamefont {R.}~\bibnamefont {Harrison}},
      \bibinfo {author} {\bibfnamefont {G.}~\bibnamefont {Harris}}, \bibinfo
      {author} {\bibfnamefont {A.}~\bibnamefont {Sawadsky}}, \bibinfo {author}
      {\bibfnamefont {Y.}~\bibnamefont {Sfendla}}, \bibinfo {author} {\bibfnamefont
      {W.}~\bibnamefont {Bowen}},\ and\ \bibinfo {author} {\bibfnamefont
      {C.}~\bibnamefont {Baker}},\ }\bibfield  {title} {\bibinfo {title} {Cryogenic
      and hermetically sealed packaging of photonic chips for optomechanics},\
      }\href {https://doi.org/10.1364/oe.463752} {\bibfield  {journal} {\bibinfo
      {journal} {Optics Express}\ }\textbf {\bibinfo {volume} {30}},\ \bibinfo
      {pages} {30822} (\bibinfo {year} {2022})}\BibitemShut {NoStop}%
    \end{thebibliography}
\end{document}


\preprint{APS/123-QED}

\title{Supplemental information for: \\ 
A diamond nanophotonic interface with an optically accessible \\ deterministic electronuclear spin register}

\affiliation{\textsuperscript{1}Cavendish Laboratory, University of Cambridge, JJ Thomson Avenue, Cambridge CB3 0HE, United Kingdom}
\affiliation{\textsuperscript{2}Department of Electrical Engineering and Computer Science, Massachusetts Institute of Technology, Cambridge, MA 02139, USA}
\affiliation{\textsuperscript{3}Cambridge Graphene Centre, University of Cambridge, Cambridge CB3 0FA, United Kingdom
\\ \ \\
\textsuperscript{*}\,These authors contributed equally to this work. \\
\textsuperscript{$\dagger$}\,Correspondence should be addressed to: englund@mit.edu, dag50@cam.ac.uk, and ma424@cam.ac.uk. \\}

\maketitle
\tableofcontents

\pagebreak

\section{Tapered fibre fabrication}

\begin{figure*}[h]
    \centering
    \includegraphics[width=0.75\textwidth]{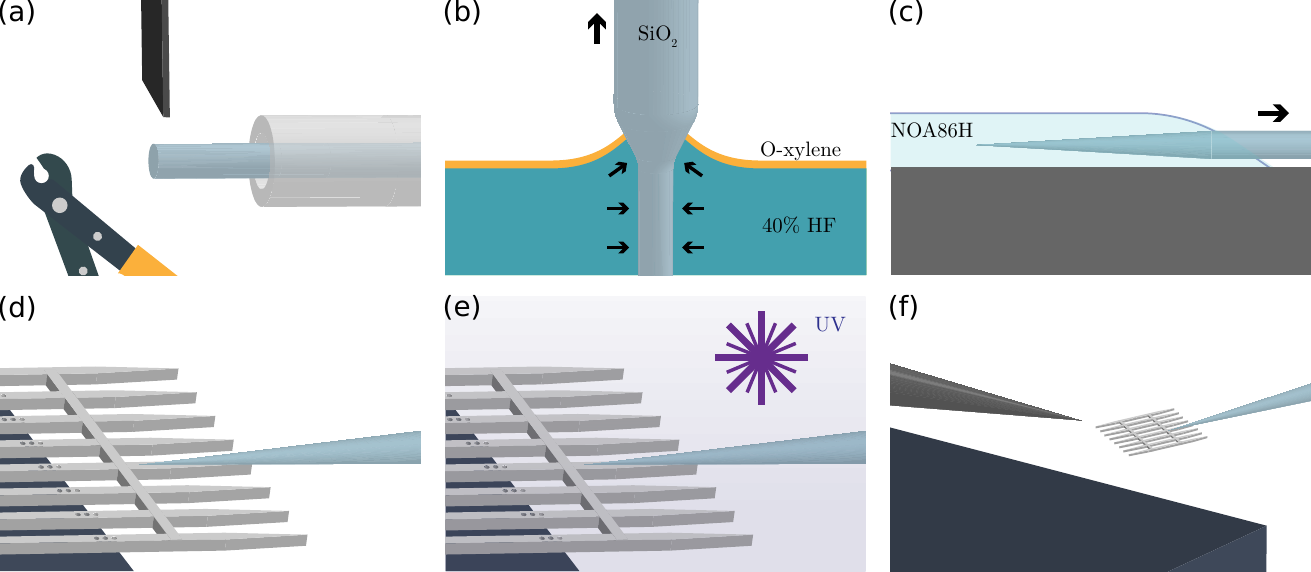}
    \caption{
        Schematic description of the photonic packaging process. An optical fibre is (a) stripped and cleaved, (b) inserted and then gradually pulled from a 40\% HF solution to form a taper, (c) inserted and removed from an optical adhesive, (d) contacted onto a diamond quantum microchiplet, (e) cured through UV irradiation and, (f) removed from the substrate.
    }
    \label{fig:process}
\end{figure*}

\begin{figure*}[h]
    \centering
    \includegraphics[width=1.0\textwidth]{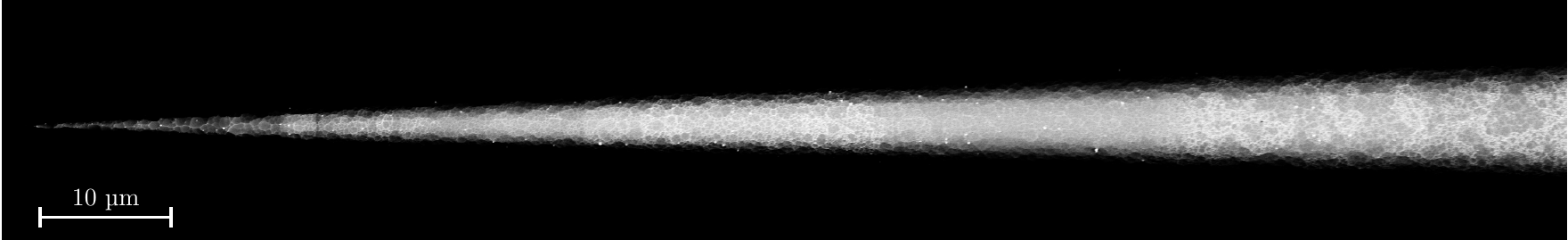}
    \caption{
        Adiabatically tapered fibre imaged through Scanning Electron Microscopy (SEM).
    }
    \label{fig:sem}
\end{figure*}

\begin{figure*}[h]
    \centering
    \includegraphics[width=0.6\textwidth]{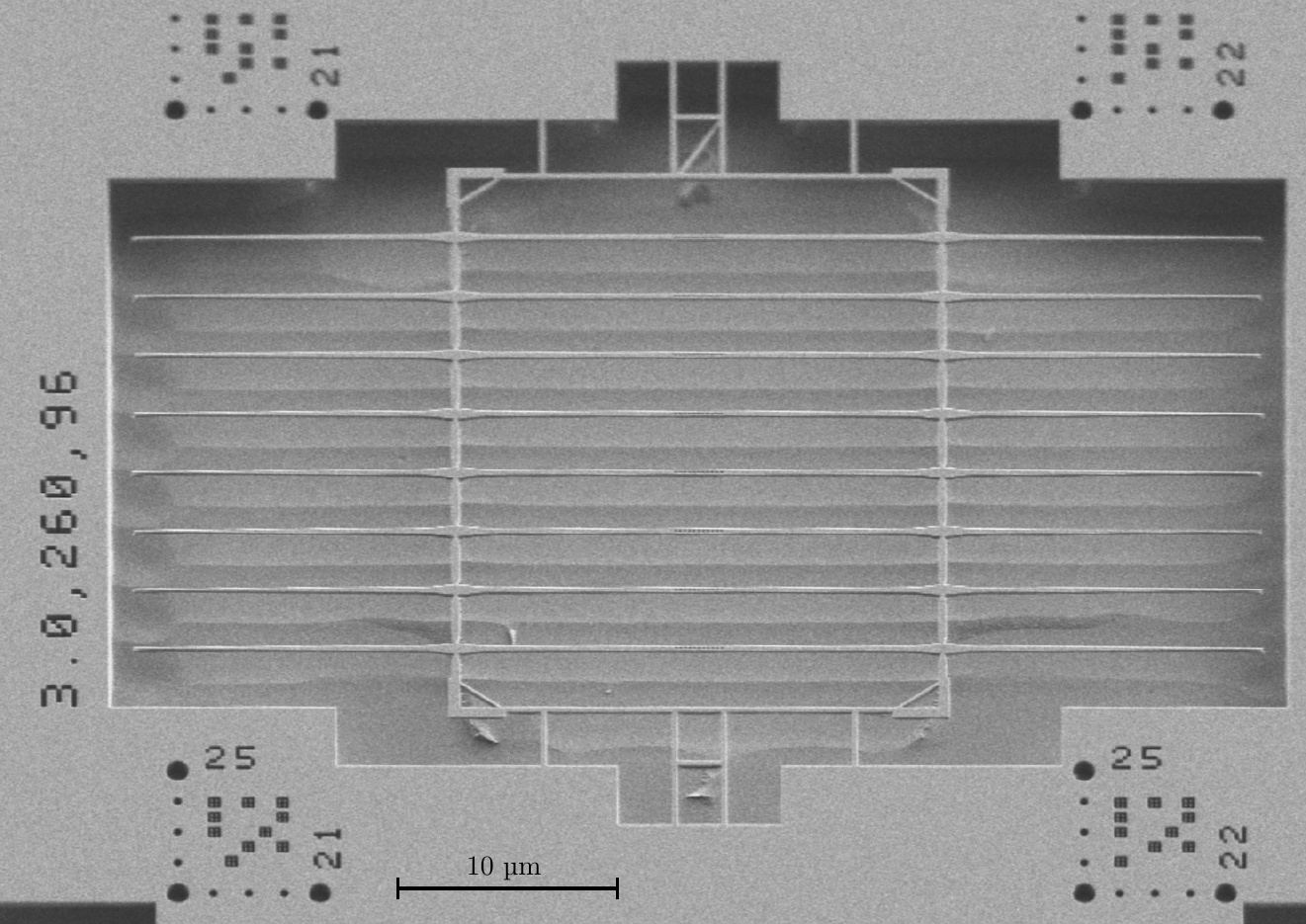}
    \caption{
        Scanning Electron Microscopy (SEM) image of a characteristic chiplet device.
    }
    \label{fig:chiplet}
\end{figure*}

The packaging process is illustrated in Fig. \ref{fig:process}. A standard single-mode optical fibre (Thorlabs SM600) is first mechanically stripped and cleaved. Adiabatically tapered fibres are then fabricated through dynamic meniscus etching in hydrofluoric acid (HF) \cite{haber2004shape, burek2017fiber}. We start with a solution of 80 ml 40\% HF and introduce 10 ml of ortho-xylene. Due to their immiscibility, the two liquids separate whereby the lower density ortho-xylene forms a layer at the top of the solution, preventing HF evaporation. A set of fibres is then lowered into the solution and progressively retracted over the course of 95 minutes. We observe an etching rate of $1.5(6)$ $\upmu$m/min and thus obtain fibres with a half-angle $1.5(8)^{\circ}$ with a pulling rate of 55 $\upmu$m/min. A characteristic fibre obtained through this process is shown in Fig. \ref{fig:sem}. A fibre is then coated with a thin layer of a photo-polymerised optical adhesive (Norland Optical Adhesive 86H) \cite{wasserman2022cryogenic} by pulling the tapered end of the fibre through a droplet of adhesive. Upon removing the fibre, we visually inspect under an optical microscope that no droplets have formed near the tip of the fibre, indicating that only a thin layer of the adhesive is present. The fibre is then contacted onto the adiabatically tapered port of a diamond waveguide and aligned in a custom-built micromanipulator setup by monitoring the reflection efficiency through the device at 655 nm. A characteristic chiplet device is shown in Fig. \ref{fig:chiplet}. The adhesive is then polymerised through UV irradiation at 405 nm with a total irradiation energy density of over 10 Joules cm$^{-2}$. Finally, a tungsten probe is used to remove the chiplet from the substrate edge, forming a packaged device. To obtain optimal adhesion, we anneal the adhesive at 80$^{\circ}\text{C}$ for 3 hours, followed by 50$^{\circ}\text{C}$ for 12 hours.

\section{Analysis of the overall detection efficiency}

\subsection{Simulations of waveguide-fibre coupling}

\begin{figure*}[h]
    \centering
    \includegraphics[width=1.0\textwidth]{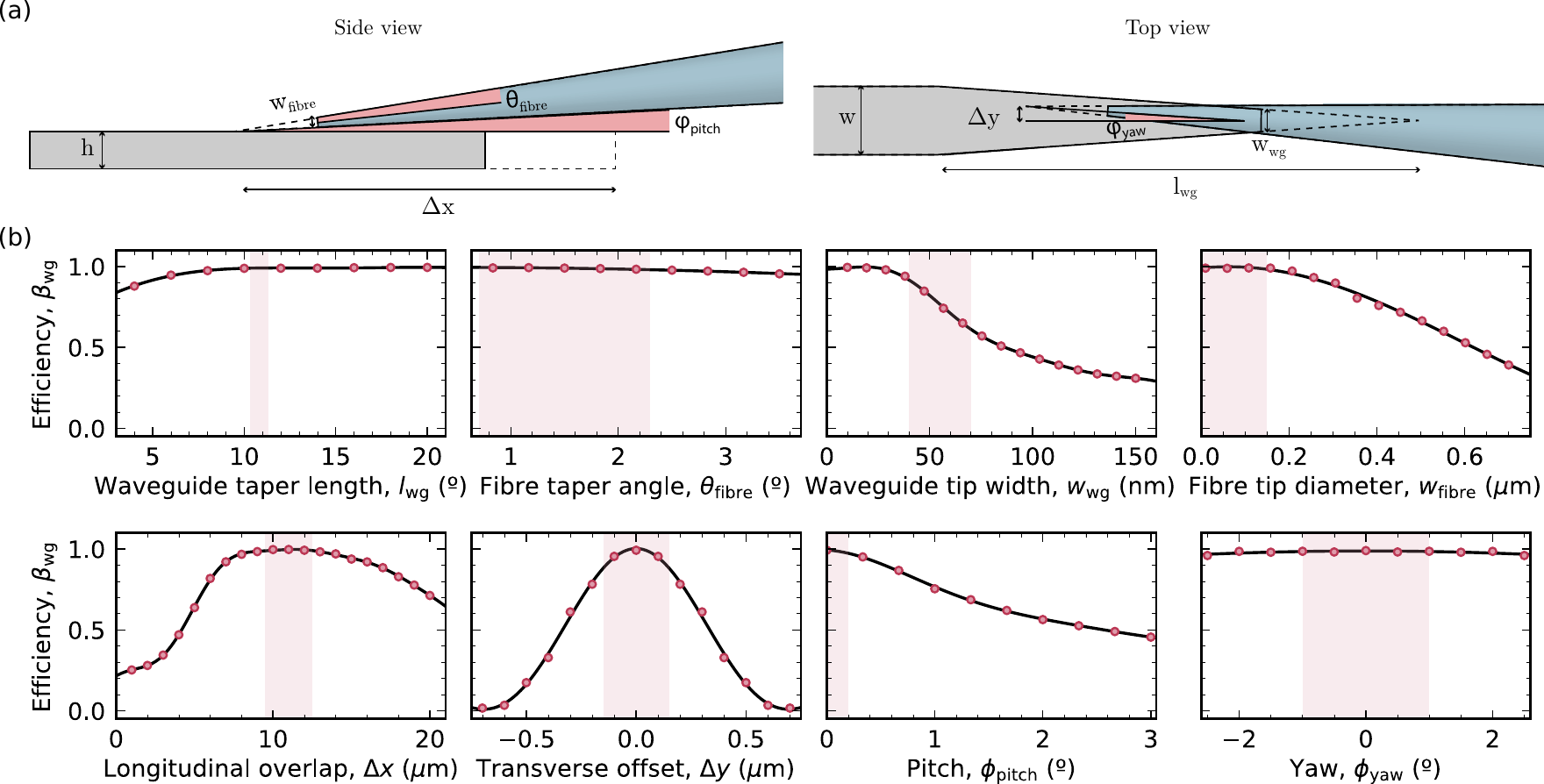}
    \caption{ 
        (a) Diagramatic representation of an imperfect diamond waveguide to fibre alignment. Length dimensions are indicated with arrows whereas angular dimensions are represented with shaded red regions. (b) Coupling efficiency as a function of deviations for each of the parameters in the model. Shaded regions correspond to experimentally realistic ranges for each of the parameters. Solid black lines correspond to smooth fits, based on Gaussian process regression, solely as a guide to the eye.
        }
    \label{fig:simulation}
\end{figure*}

In this section we numerically evaluate the fibre to waveguide coupling using Finite-Difference Time-Domain simulations including several realistic alignment and fabrication imperfections. In these simulations the fundamental TE mode of the fibre is excited, and its coupling to the fundamental TE mode of the waveguide is evaluated. Figure \ref{fig:simulation}(a) shows a diagramatic representation of the fibre to waveguide alignment. This is parameterised by the waveguide height ($h$) and width ($w$), the waveguide taper length ($l\text{wg}$) and fibre tapering angle ($\theta_\text{fibre}$), the pitch ($\phi_\text{pitch}$) and yaw misalignment ($\phi_\text{yaw}$), the finite diamond tip width ($w_\text{wg}$) and fibre tip diameter ($w_\text{fibre}$), and the longitudinal overlap ($\Delta x$) and transverse misalignment ($\Delta y$). In the ideal case $h = 200$ nm, $w = 270$ nm, $\l_\text{wg} = 11.3$ $\upmu$m, $\theta_\text{fibre} = 1.5^\circ$, $\phi_\text{pitch} = 0^\circ$, $\phi_\text{yaw} = 0^\circ$, $w_\text{wg} = 0$, $w_\text{fibre} = 0$, $\Delta y = 0$ and $\Delta y = 12~\upmu$m. For this ideal case, the waveguide to fibre coupling efficiency is higher than 99\%.

Figure \ref{fig:simulation}(b) shows how this coupling efficiency varies as fabrication (top row) and alignment (bottom row) imperfections are introduced by perturbing these parameters. The deviation of one parameter can often be compensated by another coupled parameter.  For instance, a slight yaw misalignment ($\phi_\text{yaw}$) can be compensated through a small transverse deviation ($\Delta y$), or a finite waveguide fibre tip diameter ($w_\text{fibre}$) can be compensated by increasing the longitudinal overlap ($\Delta x$). In these graphs, the coupling efficiency is thus computed at each point by first optimising the other coupled parameters. As described in SI II-B, we experimentally optimise the alignment by monitoring the reflection before curing the device. As such, it is generally realistic to assume that the maximal point for a given constraint would be found.

Our simulations show that, due to the adiabaticity of the platform, the coupling efficiency is remarkably insensitive to many imperfections. The strong Van der Waals adhesion between the diamond and fibre further ensures that many of the alignment parameters, such as $\Delta y$ or $\phi_\text{pitch}$ are likely very close to the ideal case. The numerical analysis suggests that the limiting factor in the coupling efficiency is the finite width of the diamond tip. This is currently limited by fabrication imperfections in the diamond etching process. From these simulations we estimate that the coupling efficiency should be in the range of 60\% - 90\%, in reasonable agreement with our experimental reflection measurements.

\subsection{Reflection measurements}

We measure the round-trip optical intensity transmission efficiency by injecting a known amount of optical power at 620 nm (quasi-resonant light) into the input port and measuring the reflected intensity back-propagated through the 90:10 beamsplitter, as shown in Fig. 1(b) of the main text. Correcting for the beamsplitter transmission:reflection ratio of 88(1)\%:12(1)\% at 620 nm, the 0.4 dB excess insertion loss of the beamsplitter and the maximum simulated Bragg reflector efficiency of 95\%, we derive a total roundtrip transmission efficiency of $27(3)\%$, or 52(6)\% for a single forward or backward pass. This total transmission efficiency is the product of various losses. To obtain the value for the fibre to waveguide coupling efficiency we correct for the 0.04 dB loss per optical fibre splice, 4\% scattering losses from scattering off of the diamond support structure orthogonal to the waveguide's optical axis (as estimated from FDTD), and 12 dB/km Rayleigh scattering losses from our optical fibres (Thorlabs SM600). Correcting for these factors we extract a fibre-waveguide adiabatic mode transfer efficiency of  $>57(6)\%$. We highlight that this corresponds to a lower-bound on the adiabatic mode transfer efficiency as the simulated efficiency of the Bragg reflector and the inclusion of no scattering losses due to surface roughness assume an idealised waveguide without fabrication imperfections.

\subsection{Breakdown of the overall detection efficiency}

\begin{table}[h]
\centering
\begin{tblr}{
  cells = {c},
  hline{1,10} = {-}{0.08em},
  hline{2,9} = {-}{},
}
\textbf{Parameter}                & \textbf{Description}                    & \textbf{Efficiency}      \\
$F_\pi$               & $\pi$-pulse fidelity         & 80(1)\%         \\
$\eta_\text{QE}$      & Quantum efficiency             & 79(3)\% \cite{iwasaki2017tin}         \\
$1 - \eta_\text{DW}$ & PSB fraction                   & 43(1)\%  \cite{gorlitz2020spectroscopic}       \\
$\beta_\text{wg}$     & Waveguide beta factor          & 32.5\%            \\
$\eta_\text{fibre}$  & Waveguide to fibre coupling~   & 57(6)\%         \\
$\eta_\text{setup}$   & Setup efficiency               & 51\%            \\
$\eta_\text{det}$     & Detector efficiency            & 68\%            \\
$\bm{\eta_\text{tot}}$     & \textbf{Overall detection
  efficiency} & $\bm{\sim}$\,$\bm{1.7\%}$ 
\end{tblr}
\caption{Breakdown of the overall detection efficiency. }
\label{tab:efficiencies}
\end{table}

Table \ref{tab:efficiencies} presents a breakdown of the factors involved in independently estimating the overall photon detection efficiency. The $\pi$-pulse fidelity is characterised through pulsed excitation experiments in SI VI-B. The setup efficiency includes a factor of 79\% corresponding to the spectral fraction collected from the phonon sideband. This is a result of only collecting the 638-700 nm region, and is derived from the spectral characterisation in \cite{gorlitz2020spectroscopic}. The estimated overall detection efficiency is in reasonable agreement with the value of 1.40(5)\%, extracted from the slope of Fig. 3(c) in the main text.

\section{Extended methods}

\subsection{Setup}

\begin{figure*}[h]
    \centering
    \includegraphics[width=1.0\textwidth]{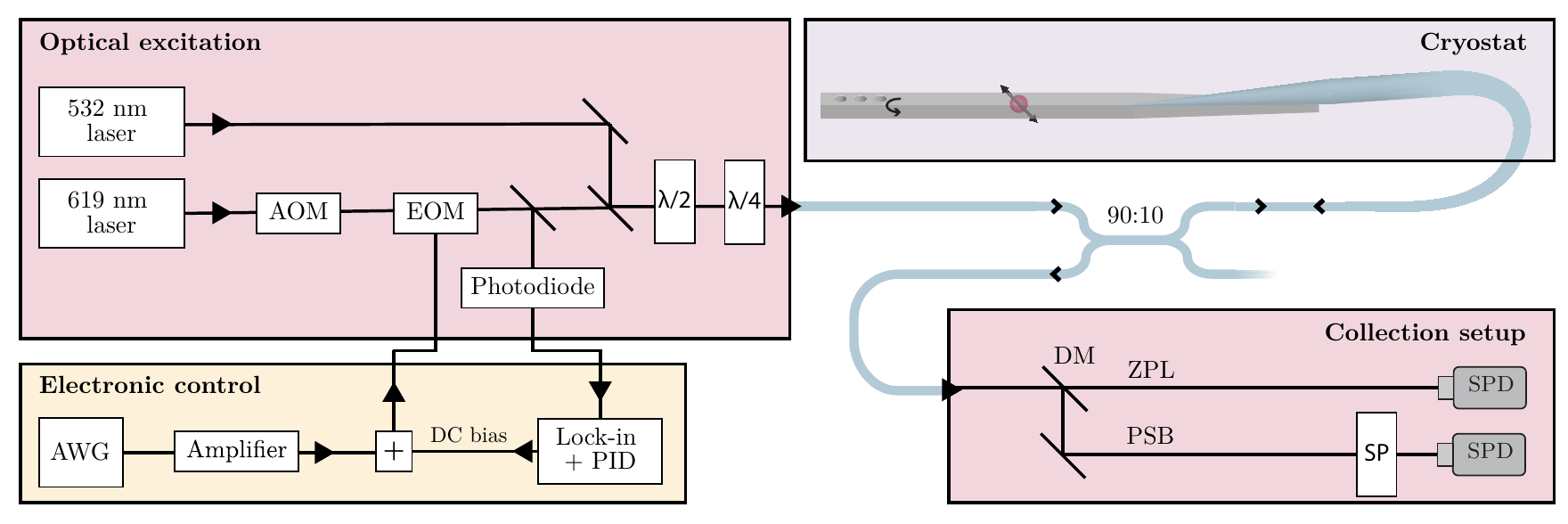}
    \caption{
        Schematic of the experimental setup. AOM: Acousto-Optic Modulator. EOM: Electro-Optic Modulator. $\lambda/2$: Half-Wave Plate. $\lambda/4$: Quarter-Wave Plate. DM: Dichroic Mirror. SP: Short-Pass filter. SPD: Single-Photon Detector. AWG: Arbitrary Waveform Generator. The $+$ sign denotes a bias-tee.
        }
    \label{fig:setup}
\end{figure*}

Figure \ref{fig:setup} shows a schematic of our experimental setup. We employ a tunable resonant source at 619 nm (M2 SolsTiS + EMM) for all of our experiments. The off-resonant source at 532 nm is only employed for measurements of photoluminescence and is not required during on-resonance measurements to initialise the charge state of the emitter \cite{gorlitz2022coherence}. The resonant light is directed through an acousto-optic modulator  (AOM, Gooch and Housego 3080-15) and a fibre-based electro-optic modulator (EOM, Jenoptik AM635). The EOM is used for all optical pulsing and is controlled by a 25 Gsamples/s arbitrary waveform generator (Tektronix AWG70002A) through a microwave amplifier (Minicircuits ZX60-83LN12+) and a bias-tee (Minicircuits ZFBT-6G+). The resonant laser is detuned by 3.5 GHz from the middle point of the C1 and C2 transitions to minimise the laser excitation caused by leakage laser light, and near-resonant light is generated through sinusoidal modulation of the EOM with a frequency swept around 3.5 GHz. In all resonant measurements the EOM is stabilised to its interferometric minimum through a lock-in amplifier and PID loop (Red Pitaya, STEMlab 125-10). A pulse generator (Swabian Instruments Pulse Streamer 8/2) is employed as a master clock to coordinate all instruments.

After modulation and polarisation control, the light is routed via a 90:10 fibre based beamsplitter (Thorlabs broadband custom SM600 90:10 beampslitter), to the tapered SM600 fibre inside the cryostat (Bluefors LD250 He dilution refrigerator at 120 mK). Connections between SM600 fibres are spliced to minimise insertion losses. Vacuum is maintained inside the cryostat by an epoxy seal (Loctite Ablestick 285, 24LV catalyst) hardened after threading the fibre through a custom feedthrough. 

The light emitted by the SnV centre is separated into phonon sideband (PSB) and zero phonon line (ZPL) components by a dichroic mirror (Thorlabs DMLP638) with a tunable long-pass filter employed for further suppression of the resonant light in the PSB path (Semrock BLP01-633R-25). A shortpass filter FESH0700 is also employed in the PSB path. All detection is performed by fibre-coupled single photon detectors  (Excelitas SPCM-AQRH-16-FC). In Fig. 4 of the main text neutral density  filters (Thorlabs NE20A and NE03A) are employed to reduce the photon flux at the ZPL single-photon detector.

\subsection{Data acquisition and analysis}

In Fig. 2(c) and Fig. 3(b) of the main text, optimal initialisation via the C1 and C2 transitions was achieved when the frequency of the optical pump was blue-detuned or red-detuned by $\sim$\,40 MHz respectively, that is when the laser frequency is closer to the centre of the two transitions. Similarly, optimal readout was achieved when the frequency of the optical probe was detuned by $\sim$\,40 MHz, in this case away from the centre of the two transitions. A possible explanation for this effect is a splitting due to hyperfine coupling in the excited state that results in split nuclear-conserving and nuclear-flipping transitions to the excited state.

All the $^{117}$SnV centres investigated in this work show a bi-stable emission behaviour between a bright and a dark state. This phenomenon has been consistently observed across multiple independent studies of SnV centres \cite{martinez2022photonic, debroux2021quantum, gorlitz2020spectroscopic} and has been postulated to be due to a charge-state instability \cite{gorlitz2022coherence}. In this work this manifests as a $\sim$\,40\% duty-cycle for the bright-state of the $^{117}$SnV. The overall detection efficiency presented in the main text enables single-shot identification of this state based on the timetrace of the fluorescence. Accordingly, the data presented in Fig. 2(c-d), Fig. 3(a-b) and Fig. 4 is post-selected for when the $^{117}$SnV centre is in the bright-state. We do not post-select Fig. 3(c), to show the true probability of detection in a 24 h period. In Fig. 4, to correct the power fluctuations in the excitation intensity as the EOM sideband is scanned over the transition, we take a reference measurement off-resonant from the emitter and use it to normalise the reflection data. Additionally, as the EOM generates positive and negative sidebands of equal intensity, but only one of them is being scanned through the emitter resonance, we correct the data for a 50\% background.

\section{Electro-nuclear Hamiltonian}

\subsection{Theoretical description}

We consider the electronuclear Hamiltonian of the ground state

\begin{equation}
H = \underbrace{\lambda L_z S_z}_{H_\text{SO}}
+ \underbrace{\gamma_S \bm{B} \cdot \bm{S}}_{H_\text{Zeeman,S}}
+ \underbrace{\gamma_I \bm{B} \cdot \bm{I}}_{H_\text{Zeeman,I}}
+ \underbrace{A_{\parallel} S_z I_z 
+ A_{\perp} (S_x I_x + S_y I_y)}_{H_\text{hf}},
\end{equation}
where $\lambda/2\pi = 850$ GHz \cite{trusheim2020transformlimited} is the spin-orbit coupling constant, $\gamma_S/2\pi = 28$ GHz/T the electron gyromagnetic ratio, $\gamma_I/2\pi = - 15.24$ MHz/T is the nuclear gyromagnetic ratio for $^{117}\text{Sn}$ and $A_\parallel$ and $A_\perp$ are the parallel and perpendicular hyperfine coupling constants respectively. The angular momentum operators can explicitely be written in the $\ket{e_{g+}}$, $\ket{e_{g-}}$ orbital basis \cite{hepp2014electronic} by $L_z = \sigma_z$, $\bm{S} = \bm{\sigma}/2$, $\bm{I} = \bm{\sigma}/2$, where $\bm{\sigma} = (\sigma_x, \sigma_y, \sigma_z)$ is a vector containing the three Pauli matrices. This includes the dominant effects of spin-orbit coupling and electronic Zeeman effect for the electron; the nuclear Zeeman effect for the nucleus; and the hyperfine interaction between the two. We do not include the effects of strain, Jahn-Teller, and the orbital Zeeman effect in this discussion. 

At zero-magnetic field and with no hyperfine interaction present, the eigenstates of the Hamiltonian are split into two branches, with energies $\pm \lambda/2$. In the lower branch, which forms the ground state, the eigenvectors are $\ket{e_+ \downarrow}$ and $\ket{e_- \uparrow}$. Notably, this implies that even at $\bm{B} = \bm{0}$, spin-orbit constrains the electron spin to be aligned to the symmetry axis of the defect, $z$. When an off-axis magnetic field is applied ($B_x$, $B_y$), it results in a competition between the magnetic and spin-orbit quantisation axes and thus only in a negligible quadratic energy shift. A magnetic field along the $z$ axis, on the contrary, results in a linear splitting at a rate $\pm \gamma_S B_z$. For this reason, the hamiltonian is often reduced for just the two ground state levels to include an anisotropic g-factor $H_\text{Zeeman,S} \rightarrow \gamma_S B_z S_z$.

The addition of a hyperfine interaction results in a similar quantisation axis competition for the nuclear spin. The lower angular branch of the ground state consists of two levels, each doubly degenerate, with energies given by,

\begin{equation}
E_{\uparrow\Uparrow, \downarrow\Downarrow} = -\frac{\lambda}{2} + \frac{A_\parallel}{4}, \quad{ E_{\uparrow\Downarrow, \downarrow\Uparrow} = -\frac{\sqrt{\lambda^2 + A_\perp^2}}{2} - \frac{A_\parallel}{4} \approx -\frac{\lambda}{2} - \frac{A_\parallel}{4}}
\end{equation}
where $E_{\uparrow\Uparrow, \downarrow\Downarrow}$ and $E_{\uparrow\Downarrow, \downarrow\Uparrow}$ refer respectively to the energy of the states with aligned and anti-aligned electron and nuclear spins. To first order, only the parallel component of the hyperfine coupling affects the eigenenergies. The strong spin-orbit coupling which resulted in a pinning of the quantisation axis of the electron now results, via the electron, in a pinning of the quantisation axis of the nucleus. When a magnetic field is applied along $B_z$, the eigenstate splitting is given by $\pm \gamma_S B_z \pm \gamma_N B_z$. The reduced Hamiltonian for the lower orbital branch of the ground state can thus be written in the form

\begin{equation}
H \approx \gamma_S^* B_z S_z + \gamma_I B_z I_z + A_\parallel S_z I_z,
\end{equation}
where $\gamma_S^*$ is now an effective gyromagnetic ratio that now includes the effects of the orbital Zeeman effect \cite{hepp2014electronic}. We note that $\gamma_I B_z \ll A_\parallel$ for all realistic magnetic fields probed in this work and thus the nuclear gyromagnetic ratio $\gamma_I$ can be neglected in most cases.

\subsection{Magnetic field splitting}

\begin{figure*}[h]
    \centering
    \includegraphics[width=0.6\textwidth]{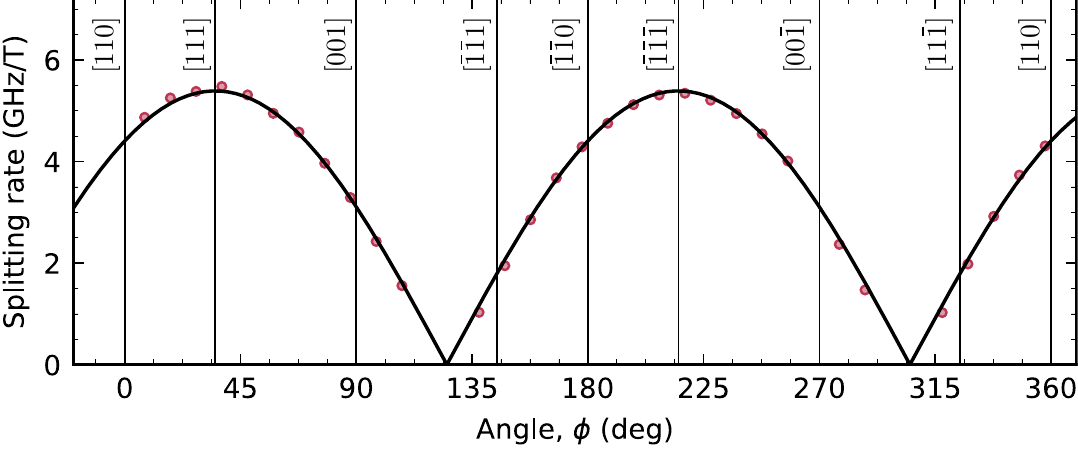}
    \caption{
        Splitting rate of the two spin conserving transitions derived from Lorentzian fits to CORE scans while the magnetic field is scanned in a plane orthogonal to the $[\overline{1}10]$ crystallographic axis at a fixed magnitude of 50 mT. The solid black line is a fit to $|A \cos(\phi +\phi_0)|$, with free fit parameters $A$ and $\phi_0$.
        }
    \label{fig:angular_ple}
\end{figure*}

Figure \ref{fig:angular_ple} shows the splitting rate of the two spin conserving transitions seen in PLE as a function of the magnetic field orientation. The magnetic field is scanned in a plane orthogonal to the $[\overline{1}10]$ crystallographic axis. A magnetic field parallel to the axis of the emitter is expected to lead to a first-order splitting, while a perpendicular field is expected to lead to no first-order splitting. We thus extract that the symmetry axis of the emitter studied in this work is the $[111]$ axis. In Fig. 2(a) of the main text the magnetic field was scanned along the [110] axis. We thus extract $B_z$, the magnetic field along the emitter axis, through a factor of $\cos(35^{\circ})$

\subsection{PLE of multiple emitters}

\begin{figure*}[h]
    \centering
    \includegraphics[width=0.8\textwidth]{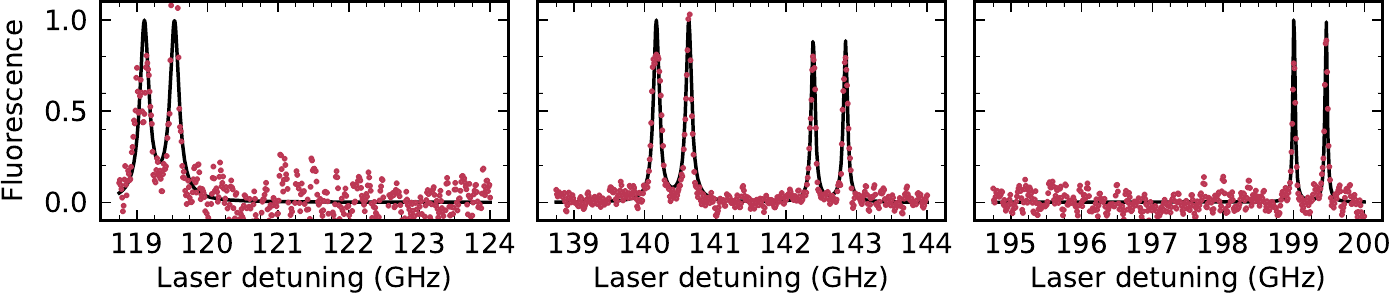}
    \caption{
        A characteristic selection of CORE scans across multiple $^{117}$SnV centres, with laser frequency measured from the relative detuning to 484.000 THz. (left) The $^{117}$SnV centre at the minimum of the inhomogeneous distribution. (middle) Two $^{117}$SnV centres near the centre of the inhomogeneous distribution. (right) The $^{117}$SnV centre at the maximum of the inhomogeneous distribution.
    }
    \label{fig:ple}
\end{figure*}

A characteristic selection of CORE scans across distinct emitters is shown in Fig. \ref{fig:ple}. The inhomogeneous strain in the device results in an inhomogeneous spectral distribution of $^{117}$SnV centres. Accordingly, we find distinct emitters located in a $\sim$\,$85$ GHz spectral range from 484.115 THz to 484.200 THz. Figure \ref{fig:ple} shows PLE scans of SnV centres sampled from the tails and the centre of the inhomogeneous distribution. In all cases, the same optically resolvable hyperfine splitting is observed.

\section{Supplementary emitter characterisation}

\subsection{Lifetime measurement}

\begin{figure*}[h]
    \centering
    \includegraphics[width=0.5\textwidth]{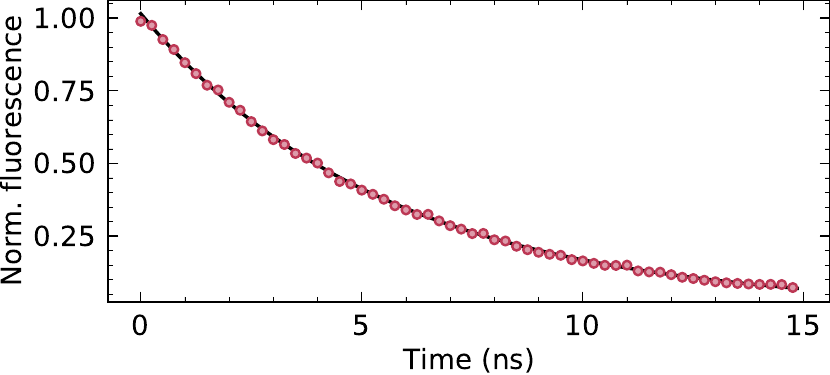}
    \caption{
        PSB fluorescence decay following an optical $\pi$-pulse. The fit is to a single exponential function with decay constant $\tau_{\text{sp}} = 5.56(3)$ ns.
    }
    \label{fig:lifetime}
\end{figure*}

Measurements of the lifetime are presented in Fig. \ref{fig:lifetime}, obtained by applying a $\pi$-pulse to the excited state and collecting the subsequent decay. This yields a single-exponential decay with lifetime $\tau_{\text{sp}} = 5.56(3)$ ns or equivalently a Fourier limit of $\gamma_0/2\pi = 28.6(1)$ MHz.

\subsection{Optical $\pi$-pulse characterisation}

\begin{figure*}[h!]
    \centering
    \includegraphics[width=0.5\textwidth]{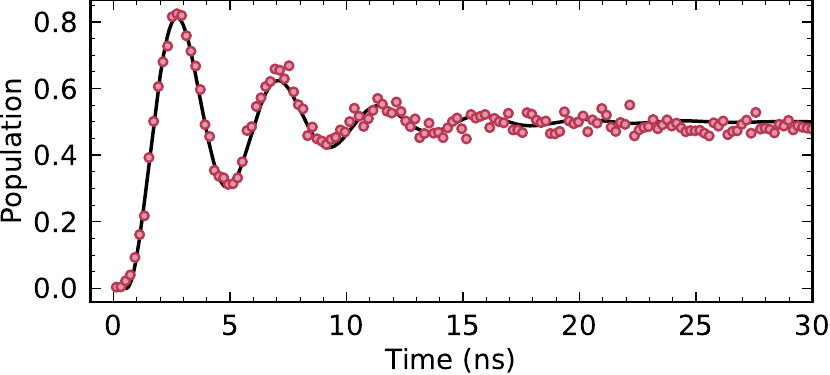}
    \caption{
        Optical Rabi oscillations between the ground and excited-state manifolds.
    }
    \label{fig:rabi}
\end{figure*}

To measure the photon detection events in Fig. 3c of the main text, the $^{117}$ SnV was initialised into the $\ket{1}$ state and subsequently excited with a train of optical $\pi$-pulses with a separation of 100 ns. Figure \ref{fig:rabi} shows Rabi oscillations of the excited state population from applying a pulse on resonance with the C1 transition at $t=0$. A $\pi$-pulse is achieved in 1.8(1) ns with 80(1)\% fidelity. The fit to the excited state population ($\rho_{\text{ee}}$) is of the form \cite{kim2020analytical}:

\begin{equation}
\rho_{\text{ee}} = \frac{1}{2}[1-\{\cos{(\Omega t)} + \frac{1}{\Omega T_1}\sin{(\Omega t)}\}e^{-t/T_1}],
\end{equation}
where $T_1 = 4.7(2)$ ns is the optical relaxation time, $\Omega = 2\pi \cdot 230(10)$ MHz the Rabi rate and $t$ the resonant drive time, where the steady-state population has been normalised to 0.5 at long drive times ($t > 30~\text{ns}$). The $\pi$-pulse fidelity is given by the maximum population achieved in the excited state

\section{Theory of optical nonlinearity}
In this section, we describe the theoretical reflection spectrum of our waveguide device and its dependence on the cooperativity. Based on the measured cooperativity and other parameters, we estimate a lower bound on the quantum efficiency.

For a cavity resonantly coupled to a two-level emitter (TLE), the reflected amplitude of light in near resonance with the TLE is given by \cite{stas2022sivnuclear, reiserer2015cavityreview}
\begin{equation}
r(\delta) = 1 - \frac{2 \kappa_\text{in}/\kappa_\text{tot}}{1 + C/(1 + 2\text{i}\delta/\gamma_{\text{h}})},
\end{equation}
where $C$ is the cooperativity of the light-matter interface, $\gamma_\text{h}$ is the homogeneous linewidth of the emitter in the absence of Purcell enhancement, and $\delta$ is the detuning of the incident light from the TLE. $\kappa_\text{tot}$ is the total decay rate of the cavity and $\kappa_\text{in}$ is the decay rate into the input port of the cavity, which in our case is towards the adiabatically tapered fibre. For compactness, we define the ratio $f=\kappa_\text{in}/\kappa_\text{tot}$ which is ideally unity. To apply the cavity formalism to a Bragg reflector yielding constructive interference, we have assumed the case of a highly broadband cavity ($\kappa_\text{tot} \gg \gamma_{\text{h}}$), which ensures near-perfect resonance between the cavity and the TLE. Additionally, we assume that the TLE is driven well below saturation.

In the experiment, we report the reflection intensity normalised to the far-detuned reflection intensity
\begin{align}
    \mathcal{R}(\delta)_{norm}=\frac{|r(\delta)|^2}{|r(\delta\gg \gamma_h)|^2}.\label{eq:RNorm}
\end{align}

Of particular interest is the normalised reflection at zero detuning
\begin{align}
    \mathcal{R}(\delta=0)_{norm}=\frac{|r(\delta=0)|^2}{|r(\delta\gg \gamma_h)|^2}=\left(\frac{1-2f/(1+C)}{1-2f}\right)^2
\end{align}
where the cooperativity is given by $C=\Gamma/\gamma_h$, where $\Gamma$ is the rate of ZPL emission via the C-transition into the cavity mode. $\Gamma$ can be related to the total emitter decay rate $\gamma_0$ via
\begin{align}
\Gamma=\underbrace{\beta_{cav}\times\eta_{DW}\times\eta_{QE}\times\eta_{orb}}_{\beta_{tot}}\times\gamma_0,\label{eq:GammaDecomposition}
\end{align}
where $\beta_{cav}$ is a photonic property representing the fraction of photons emitted into the cavity mode, $\eta_{DW}$ is the Debye-Waller factor representing the fraction of photons emitted into the ZPL, $\eta_{QE}$ is the quantum efficiency and $\eta_{orb}$ is the orbital branching ratio.

Using Eq. \ref{eq:RNorm}, we fit the results in Fig. 4 of the main text, giving a cooperativity of $C=0.027(3)$. This is the average value extracted from the fits of the two transitions in Fig. 4(b) after correcting for the finite initialisation fidelity of the state. This slightly reduced initialisation fidelity, relative to the high fidelity realised after 30 $\mu$s of pumping in Fig. 2(c) of the main text, is chosen to increase the repetition rate of the experiment. We additionally constrain $f = 0.95$ based on numerical simulations of the Braff reflectors.

Equation \ref{eq:GammaDecomposition} then allows us to estimate $\eta_{QE}$ from the values of $\gamma_h$, $\beta_{cav}$, $\eta_{DW}$, $\eta_{orb}$ and $\gamma_0$. From PLE scans we extract a homogenous linewidth of $\gamma_h=70(5)$ MHz. The measured radiative lifetime $1/\gamma_0=5.56(3)$ ns only deviates slightly from the value measured in bulk \cite{iwasaki2017tin} implying negligible Purcell enhancement, and we thus assume that the measured homogenous linewidth is representative of $\gamma_{h}$ in the absence of Purcell enhancement. We employ $\beta_\text{wg} = 32\%$, as derived from finite-difference time-domain (FDTD) simulations; $\eta_{DW} = 0.57$, from previous spectroscopic investigations \cite{gorlitz2020spectroscopic}; and $\eta_{\text{orb}} = 0.65$, from a fit to the integrated areas of the C and D peaks presented in Fig. 1(d) of the main text. Combining these estimates, we obtain a quantum efficiency of $\eta_{QE}>51(8)\%$. We emphasize that this estimate represents a lower bound, as a nonideal separation between the emitter and the Bragg reflector would result in destructive interference, thereby lowering the fitted cooperativity and by extension the $\eta_{QE}$ estimate.

To model the power dependence of the reflection contrast in Fig. 4c of the main text, we use the expression  $\mathcal{R}(s) = \mathcal{R}(0)/(1+s)$ \cite{auffeves2007giant, javadi2015single}, where $\mathcal{R}(0)$ is given by Eq. \ref{eq:RNorm} and $s$ is the saturation parameter.